\documentclass[twocolumn,trackchanges]{aastex7}
\usepackage{lineno}
\linenumbers
\usepackage{natbib}
\usepackage{color}
\usepackage{booktabs}
\usepackage{mathtools}
\usepackage{overpic}

\usepackage{comment}

\DeclarePairedDelimiterX\braket[2]{\langle}{\rangle}{#1\,\delimsize\vert\,\mathopen{}#2}

\def    \apjl  		{\rm {ApJL}}

\def    \apjl  		{\rm {ApJL}}

\def	\cm		{\,{\rm {cm}}}
\def	\K		{\,{\rm K}}
\def	\g		{\,{\rm {g}}}
\def	\mum	{\,{\mu \rm{m}}}

\def \bea {\begin{eqnarray}}
	\def \ena {\end{eqnarray}}

\def	\ba	{\boldsymbol{a}}
\def	\be	{\boldsymbol{e}}
\def	\bB	{\boldsymbol{B}}

\def	\bJ	{\boldsymbol{J}} 
\def	\bk	{\boldsymbol{k}}

\def    \bmu    {{\hbox{\boldsym\char'026}}}	

\def	\bphi	{\boldsymbol{\phi}}
\def	\bxi	{\boldsymbol{\xi}}

\def	\bQ	{\boldsymbol{Q}} 

\def	\cm	{\,{\rm cm}}

\def	\max	{\,{\rm max}}
\def	\d	{{\rm d}}

\def	\D	{{\rm D}}
\def	\eff	{{\rm eff}}

\def	\ehat	{\hat{\bf e}}
\def	\erg	{\,{\rm erg}}

\def	\g	{\,{\rm g}}
\def	\gas	{\,{\rm gas}}
\def	\G	{{\rm G}}

\def	\H	{{\rm H}}

\def	\N	{{\rm N}}

\def	\pc	{\,{\rm pc}}

\def	\s	{\,{\rm s}}

\def	\au	{\,{\rm au}}

\def	\d 	{\rm d}
\def	\B 	{\rm B}

\def	\V	{{\rm V}}
\def	\Bar	{{\rm Bar}}
\def	\rad	{{\rm rad}}
\def	\yr	{{\rm yr}}

\def	\xhat		{\hat{\boldsymbol{x}}}
\def	\yhat   	{\hat{\boldsymbol{y}}}
\def	\zhat		{\hat{\boldsymbol{z}}}

\def	\ehat		{\hat{\boldsymbol{e}}}

\def    \gas     	{{\rm gas}}

\font\mib=cmmib10

\def\bOmega{\hbox{\mib\char"0A}}
\def\bmu{\hbox{\mib\char"16}}

\def\bGamma{\hbox{\mib\char"00}}

\makeatletter
\newcommand*{\rom}[1]{\expandafter\@slowromancap\romannumeral #1@}

\begin{document}

\title{Toward A General Theory of Grain Alignment and Disruption by Radiative Torques and Magnetic Relaxation}

\author{Thiem Hoang}
\affiliation{Korea Astronomy and Space Science Institute, Daejeon 34055, Republic of Korea}
\affiliation{Department of Astronomy and Space Science, University of Science and Technology, 217 Gajeong-ro, Yuseong-gu, Daejeon, 34113, Republic of Korea}

\email{thiemhoang@kasi.re.kr}  
\date{Draft version \today} 

\begin{abstract}
We generalize the Magnetically enhanced RAdiative Torque (MRAT) theory for astrophysical environments described by a dimensionless parameter $U/(n_{1}T_{2})$ with $U$ the radiation strength, $n_{1}=n_{\rm H}/(10\cm^{-3})$ the normalized hydrogen density, and $T_{2}=T_{\rm gas}/100{\rm K}$ the normalized gas temperature. We derive the critical magnetic relaxation $\delta_{\rm mag,cri}$ required to produce high-J attractors for different RAT models and local conditions, and find that $\delta_{\rm mag,cri}$ must increase with increasing $U/(n_{1}T_{2})$. We numerically study grain alignment and rotational disruption by the MRAT mechanism, considering both gas collisions and magnetic fluctuations. We find that, for the collision-dominated (CD) regime ($U/(n_{1}T_{2})\leq 1$), collisional and magnetic excitations slowly transport large grains from low-J to high-J attractors, leading to perfect {\it slow alignment} within $\sim 10-100$ damping times due to MRAT. However, for the radiation-dominated (RD) regime ($U/(n_{1}T_{2})>1$), only a fraction of grains can be fast aligned at high-J attractors by MRAT, and the majority of grains are trapped at low-J rotation due to strong RATs, a new effect termed {\it RAT trapping}. For extreme conditions of $U/(n_{1}T_{2})>10^{4}$, the efficiency of magnetic relaxation on grain alignment is suppressed, and grains only have fast alignment and disruption purely determined by RATs. We quantified the fraction of grains with fast alignment at high-J attractors, $f_{\rm high-J}^{\rm fast}$, for different RAT models, magnetic relaxation, and $U/(n_{1}T_{2})$, and found that the maximum $f_{\rm high-J}^{\rm fast}$ can reach $45\%$ by MRATs and $22\%$ by RATs. We discuss the implications of our results for describing alignment and disruption of astrophysical dust.

\end{abstract}


\section{Introduction}
The alignment of non-spherical dust grains with ambient magnetic fields causes the polarization of starlight \citep{Hall.1949,Hiltner.1949} and thermal dust emission \citep{Planck.2015}. The polarization of light induced by dust grain alignment (hereafter dust polarization) is widely used to trace the two-dimensional (2D) magnetic fields (B-fields) projected in the plane of the sky, from the diffuse interstellar medium (ISM) and molecular clouds to star- and planet-forming regions \citep{Pattle.2023ASPC,Tsukamoto.2023}. The polarization degree offers valuable information on dust properties (size, shape, and composition), and the physical processes responsible for grain alignment \citep{HoangBao.2024}. Recently, new techniques were introduced for probing three-dimensional (3D) B-fields by combining 2D B-fields traced by the dust polarization angles with the inclination angle constrained by the polarization degree \citep{HoangBao.2024,BaoHoang.2024}.

Grain alignment driven radiative torques (RATs), first noticed by \cite{Dolginov:1976p2480}, and then numerically demonstrated by \cite{DraineWein.1996}, is the standard theory for describing grain alignment in the ISM \citep{DraineWein.1997,Andersson.2015,LAH.2015}. The development of an analytical model (AMO) of RATs by \cite{LazHoang.2007} enabled a new era of quantitative grain alignment \citep{HoangLaz.2008,HoangLaz.2009a,HoangLaz.2009b}. The unified theory of grain alignment, namely magnetically enhanced RAT alignment (MRAT), was developed by \cite{LazHoang.2008,HoangLaz.2016}, which unifies the effects of RATs and magnetic relaxation due to iron inclusions embedded within dust grains. For an ensemble of different grain shapes and magnetic properties, there exists only a fraction of grains that can be aligned at an attractor with angular momentum greater than the thermal value (so-called high-J attractors), denoted by $f_{\rm high-J}^{\rm ens}$ \citep{HoangLaz.2016}. For the typical diffuse ISM, \cite{HoangLaz.2016} found that, if $\delta_{\rm mag}>10$, high-J attractors become universal, regardless of the RAT models (e.g., grain shapes and composition), and $f_{\rm high-J}^{\rm ens}=1$. However, for $\delta_{\rm mag}\ll 1$, high-J attractors are only determined by RATs, and $f_{\rm high-J}^{\rm ens}\sim 0.2-0.6$ \citep{Herranen.2021}.

For a given grain shape and magnetic property that exhibit high-J attractors, if the grains have initially random orientation with the magnetic field, only a fraction of the grains can be rapidly aligned at high-J attractor within a gas damping time (so-called {\it fast alignment}). The rest of the grains are driven toward an attractor at thermal angular momentum (aka. low-J attractor). For the typical ISM with the gas density $n_{\H}=30\cm^{-3}$, gas temperature $T_{\gas}=100\K$, and standard interstellar radiation field (ISRF), numerical simulations in \citep{HoangLaz.2008,HoangLaz.2016} identified the crucial role of {\it collisional excitations} by gaseous random collisions in scattering grains out of the low-J attractors and slowly transporting them to high-J attractors, which can lead to the {\it perfect alignment} of grains within several gaseous damping times, so-called {\it slow alignment} (see also \citealt{LazHoang.2021}). However, for grains without high-J attractors, grain alignment is negligible due to the strong randomization by gas collisions \citep{HoangLaz.2008,HoangLaz.2016}. As a result, the net alignment degree of grains can be determined by $f_{\rm high-J}^{\rm ens}$. The MRAT theory is successfully applied to the ISM, molecular clouds, and star-forming regions (SFRs) \citep{Gianghoang.2024}, and incorporated in the publicly available POLARIS code to model dust polarization \citep{Reissl.2016,Giangetal.2024}.

A new important physical effect arising from suprathermal rotation of grains by RATs is rotational disruption induced by centrifugal force, aka. radiative torque disruption (RATD) \citep{Hoang.2019nas,Hoang.2019}. Therefore, radiative torques govern both dust grain alignment and dust evolution, which have broader implications for astrophysics because dust is involved in various important physical and chemical processes, as well as acting as a momentum carrier for radiation pressure feedback \citep{Draine.2011book}. Various observational evidence for RATD mechanism were reported in different environments with strong radiation fields (see \citealt{Tramhoang.2022} for a review), including the lines of sight toward supernovae \citep{Gianghoang.2020}, the torus around active galactic nuclei (AGN; \citealt{Gianghoang.2021}), the envelope of evolved stars \citep{Truong.2022}, photodissociation regions \citep{Tramlee.2021,Tramhoang.2021,Ngoc.2024apj}. In particular, in the time-domain era, the efficiency of RATD is essential for accurate modeling of the intrinsic luminosity of transients as well as understanding the dust evolution caused by transient feedback. For a given grain shape and composition, if grains initially have random orientation and one turn on the radiation flash, the fraction of grains with fast alignment at high-J attractors is described $f_{\rm high-J}^{\rm fast}$ \citep{LazHoang.2021}. Therefore, the RATD efficiency can be described by $f_{\rm disr}=f_{\rm high-J}^{\rm ens}\times f_{\rm high-J}^{\rm fast}$. \cite{LazHoang.2021} calculated $f_{\rm high-J}^{\rm fast}$ for the standard ISM, implying the increases of $f_{\rm high-J}^{\rm fast}$ with the magnetic relaxation $\delta_{\rm mag}$ and can reach $50\%$. Nevertheless, the effects of magnetic relaxation on enhancing the value of $f_{\rm high-J}^{\rm ens}$ and $f_{\rm high-J}^{\rm fast}$ have not been studied for strong radiation fields. Moreover, the efficiency of collisional excitations in transporting grains from low-J to high-J attractors, which increases the alignment and disruption, has not been quantified for the environments of strong radiation field.

The main goal of this paper is to develop a general theory of the MRAT alignment for various astrophysical environments, and to quantify the efficiency of both {\it fast and slow} alignment and disruption driven by the MRAT mechanism using numerical simulations. The rest of the paper is structured as follows. In Section \ref{sec:review}, we review the basic physical processes and characteristic timescales for grain alignment by RATs and magnetic relaxation. Section \ref{sec:newtheory} describes the general MRAT theory, the lower and upper limits for the magnetic relaxation effect, and numerical methods to quantify grain alignment. In Section \ref{sec:results}, we show the numerical results of grain alignment for the different models of RATs, magnetic properties, and radiation fields. An extended discussion is presented in Section \ref{sec:discuss}, and a summary of our main results is presented in Section \ref{sec:summary}. 

\section{Review of Grain Alignment and Disruption Physics}\label{sec:review}
Here, we first review the fundamental physical processes involved in the grain alignment process to be used for reference. For grains with iron inclusions considered in this paper, internal relaxation is fast compared to the other dynamical timescales due to Barnett relaxation \citep{Dolginov.1976} and inelastic relaxation \citep{LazEfroim.1999}, which induces the internal alignment of the grain axis of maximum inertia with its angular momentum (see more details in \citealt{Hoang.2021,Hoangetal.2022}). Thus, we only discuss physical processes inducing external alignment of grain angular momentum ($\bJ$) with the magnetic field ($\bB$), including magnetic relaxation and radiative torques.

\subsection{Rotational Damping}
Grain rotation is damped by collisions with gas atoms/molecules, followed by thermal evaporation (see, e.g., \citealt{Roberge.1993}). For an oblate spheroidal grain of mass density $\rho$, the characteristic damping timescale due to gas collisions is given by
\bea
\tau_{\rm gas}&=&\frac{3}{4\sqrt{\pi}}\frac{I_{\|}}{n_{\H}m_{\H}
v_{T}a^{4}\Gamma_{\|}},\nonumber\\
&\simeq&2.2\times 10^{5} \hat{\rho}\hat{s}a_{-5}\left(\frac{100\K}{T_{\gas}}\right)^{1/2}\left(\frac{10
\cm^{-3}}{n_{\H}}\right)\left(\frac{1}{\Gamma_{\|}}\right) \yr,~~~~\label{eq:taugas}
\ena
where $n_{\H}$ is the hydrogen number density, $T_{\gas}$ is the gas temperature, $v_{T}=\left(2k_{B}T_{\rm gas}/m_{\H}\right)^{1/2}$ is the hydrogen thermal velocity, $a=a_{-5}10^{-5}\cm$ is the length of the semi-major axis, and $s=b/a<1$ is the ratio of the semi-minor to semi-major axes with $\hat{s}=s/0.5$, $I_{\|}=8\pi \rho s a^{5}/15$ is the moment of inertia along the principal axis of maximum inertia moment (also denoted as $I_{1}$), and $\Gamma_{\|}$ is a geometrical parameter of order unity (\citealt{Roberge.1993}). 

Dust grains are heated by stellar radiation and cool down by infrared emission. Infrared emission reduces the grain's angular momentum. The grain rotational damping rate due to infrared emission is described by $\tau_{\rm IR}^{-1}=F_{\rm IR}\tau_{\rm gas}^{-1}$ with $F_{\rm IR}$ the dimensionless parameter given by (see \citealt{Draine.1998}):
\bea
F_{\rm IR}\simeq 1.2U^{2/3}\left(\frac{10^{-5}\cm}{a_{\eff}}\right)
\left(\frac{10 \cm^{-3}}{n_{\H}}\right)\left(\frac{100 \K}{T_{\gas}}\right)^{1/2},\label{eq:FIR}
\ena 
where $U= u_{\rm rad}/u_{\rm MMP}$ with $u_{\rm rad}$ the energy density of the radiation field is the radiation strength and $u_{\rm MMP}=8.64\times 10^{-13}\erg\cm^{-3}$ is the energy density of the standard ISRF in the solar neighborhood from \cite{Mathis.1983}, and $a_{\eff}$ is the effective size of an irregular grain defined as the radius of an equivalent sphere of the same volume $V_{\rm gr}=4\pi a_{\eff}^{3}/3$. For numerical estimates of the characteristic timescales, we use the oblate spheroidal shape for convenience, so $a_{\eff}=as^{1/3}$, where $s=1/2$ is fixed in this paper.

The total damping rate for grain rotation by gas collisions and infrared emission is given by
\bea
\tau_{\rm damp}^{-1}=\tau_{\rm gas}^{-1} + \tau_{\rm IR}^{-1}=\tau_{\rm gas}^{-1}(1+F_{\rm IR}).\label{eq:tau_damp}
\ena

The total damping torque by gas collisions and IR emission can be described by 
\bea
\bGamma_{\rm damp}=-\frac{\bJ}{\tau_{\rm damp}}.\label{eq:Gamma_damp}
\ena

\subsection{Barnett magnetic moment and Larmor precession}
\subsubsection{Dust Grain Magnetism}
Iron is among the most abundant elements in the universe. Spectroscopic observations show that more than $95\%$ of iron is depleted from the gas \citep{Jenkins.2009} and thus embedded into dust grains. Iron atoms, if distributed diffusely within a dust grain, produce ordinary paramagnetic material (PM). The zero-frequency susceptibility $\chi(0)$ of such a paramagnetic dust is given by the Curie's law:
	\bea
	\chi_{\rm PM}(0)&=&\frac{n_{p}\mu^{2}}{3k_{\B}T_{\d}},\label{eq:curielaw_PM}
	\ena
	where the effective magnetic moment per iron atom $\mu$ reads
	\bea
	\mu^{2}\equiv p^{2}\mu_{\B}^{2} =g_{e}^{2}\mu_{B}^{2}\left[{j(j+1)}\right],\label{eq:mu}
	\ena
	with $j$ being the angular momentum quantum number of electrons in the outer partially filled shell, and $p\approx 5.5$ is taken for silicate (see \citealt{Draine.1996}).
	
	Plugging the typical numerical values into Equation (\ref{eq:curielaw_PM}), we obtain
	\bea
	\chi_{\rm PM}(0)\simeq 0.03f_{p}\hat{n}_{23}\left(\frac{p}{5.5}\right)^{2}\left(\frac{20\K}{T_{d}}\right),\label{eq:chi_PM}
	\ena
	where $\hat{n}_{23}=n/10^{23}\cm^{-3}$ is the atomic density of material, $f_{p}$ is the fraction of paramagnetic (Fe) atoms in the dust grain which is $f_{p}=1/7$ for silicate of structure MgFeSiO$_{4}$.

When grains contains embedded iron clusters, they become superparamagnetic material (SPM), and their magnetic susceptibility is significantly enhanced from PM grains. Let $N_{\rm cl}$ be the number of iron atoms per inclusion (cluster). The SPM magnetic susceptibility is given by
\bea
\chi_{\rm SPM}(0)=\frac{n_{\rm cl}m^{2}}{3kT_{d}},\label{eq:curilaw_SPM}
\ena
where $n_{\rm cl}$ is the volume density of iron clusters, and $m=N_{\rm cl}\mu_{0}=N_{\rm cl}p\mu_{B}$ is the average magnetic moment per iron clusters and $m_{0}$ is the average magnetic moment of Fe.

Let $\phi_{sp}$ be the volume filling factor of iron inclusions of the same size. One obtains $n_{\rm cl}=\mathcal{N}/V_{\rm gr}$ with $\mathcal{N}$ the total number of iron clusters given by
\bea
\mathcal{N}&&=\frac{\phi_{sp}V_{\rm gr}}{N_{\rm cl}V_{\rm Fe}},\nonumber\\
&&\simeq 3.5\times 10^{8}\phi_{sp}N_{\rm cl}^{-1}a_{-5}^{3}, \label{eq:Ncl}
\ena
where $V_{\rm Fe}=4\pi R_{\rm Fe}^{3}/4$ with $R_{\rm Fe}$ is the radius of iron atom \citealt{HoangLaz.2016}). 

By plugging $m$ and $n_{\rm cl}=\mathcal{N}/V$ into Equation (\ref{eq:curilaw_SPM}), we obtain:
\bea
\chi_{\rm SPM}(0)\approx 0.026N_{\rm cl}\phi_{\rm sp}\hat{p}^{2}\left(\frac{20\K}{T_{d}}\right),\label{eq:chi_SPM}
\ena
which is a factor $N_{\rm cl}$ greater than the PM susceptibility given by Equation (\ref{eq:chi_PM}).

\subsubsection{Grain Magnetic Moment from Barnett effect and Larmor Precession}

A rotating PM/SPM grain acquires a magnetic moment through the Barnett effect (\citealt{Barnett.1915}). The instantaneous Barnett magnetic moment of a grain of volume $V_{\rm gr}$ rotating with the angular velocity $\bOmega$ is equal to
\bea
\bmu_{\rm Bar}=\frac{\chi(0)V_{\rm gr}\bOmega}{\gamma_{g}}=-\frac{\chi(0)V_{\rm gr}}{g_{e}\hbar\mu_{B}}\bOmega,\label{eq:muBar}
\ena
where $\chi(0)$ is given by Equation (\ref{eq:chi_PM}) for PM and (\ref{eq:chi_SPM}) for SPM grains, $\gamma_{g}=-g_{e}\mu_{B}/\hbar\approx -e/(m_{e}c)$ is the gyromagnetic ratio of an electron, $g_{e}\approx 2$ is the $g-$factor, and $\mu_{B}=e\hbar/2m_{e}c\approx 9.26\times 10^{-21} \erg \G^{-1}$ is the Bohr magneton. 

The interaction of the grain magnetic moment with an external static magnetic field $\bB$ results in a magnetic torque
\bea
\bGamma_{B}&&=[\bmu_{\rm Bar}\times\bB]\nonumber\\
&&=-|\mu_{\Bar}|B\sin\beta\hat{\bphi}\equiv I_{\|}\omega\sin\beta d\phi/dt \hat{\bphi},
\ena
where $\hat{\bphi}$ is the unit vector (see Figure \ref{fig:alignRAT}), resulting in the rapid Larmor precession of $\bJ\| \bOmega$ around $\bB$.

The period of such a Larmor precession denoted by $\tau_{\rm Lar}$, is given by
\bea
\tau_{\rm Lar}&&=\frac{2\pi}{|d\phi/dt|}\nonumber\\
&&
=\frac{2\pi I_{\|}g\mu_{B}}{\chi_{0}V\hbar B}\simeq 0.65 a_{-5}^{2}\hat{s}^{-2/3}\frac{\hat{\rho}}{\hat{\chi}
	\hat{B}} ~\yr,
\label{eq:tauB}
\ena
where $\hat{\chi}=\chi(0)/10^{-4}$, and $\hat{B}=B/10\mu$G is the normalized magnetic field strength. Comparing Equation (\ref{eq:tauB}) with Equations (\ref{eq:taugas}) and (\ref{eq:tau_DG_sup}), it is seen that the Larmor precession is much faster than the rotational damping and magnetic relaxation.

\subsection{Magnetic Relaxation}
Rotating magnetic (para-/superparamagnetic) grains experience the dissipation of the grain rotational energy into heat due to the lag, resulting in the alignment of $\bJ$ with the ambient field. This is described by a magnetic torque \citep{DavisGreenstein.1951,DraineWein.1997}:
\bea
\bGamma_{\rm mag}=-\frac{\bJ}{\tau_{\rm mag}}\left(\sin\xi\cos\xi\hat{\bxi} + \sin^{2}\xi \hat{\bJ}\right),\label{eq:Gamma_mag}
\ena
where $\hat{\bxi}$ is the unit vector describing the alignment between $\bJ$ and $\bB$ (see Figure \ref{fig:alignRAT}), $\tau_{\rm mag}$ is the characteristic time of the magnetic relaxation given by 
\bea
\tau_{\rm mag} &=& \frac{I_{\|}}{K(\omega)V_{\rm gr}B^{2}}=\frac{2\rho a^{2}s^{-2/3}}{5K(\omega)B^{2}},\nonumber\\
&\simeq & 6\times 10^{5}\hat{\rho}\hat{s}^{-2/3}a_{-5}^{2}\hat{B}^{-2}\hat{K}^{-1} \yr,\label{eq:tau_DG_sup}
\ena
where $V_{\rm gr}=4\pi s a^{3}/3$ is the grain volume, and $\hat{K}=K(\omega)/10^{-13}\s$ and
$K(\omega)=\chi_{2}(\omega)/\omega$ with $\chi_{2}(\omega)$ as the imaginary part of the complex magnetic susceptibility of the grain material.

To describe the aligning effect of magnetic relaxation relative to the disalignment by gas collisions, the dimensionless magnetic relaxation strength was introduced in \cite{HoangLaz.2016} as:
\bea
\delta_{\rm mag}&&=\frac{\tau_{\gas}}{\tau_{\rm mag}},\nonumber\\ 
&&\simeq 0.3\hat{\rho}{a}_{-5}\left(\frac{\hat{K}\hat{B}^{2}}{n_{1}T_{2}^{1/2}}\right),\label{eq:delta_mag}
\ena
where $n_{1}=n_{\H}/(10\cm^{-3})$ and $T_{2}=T_{\gas}/100\K$, where the physical parameters of the typical ISM are used. The exact value of $\delta_{\rm mag}$ depends on the grain size, magnetic properties and magnetic field strength as well as the local gas conditions (see \citealt{HoangLaz.2016} for more details). 

\subsection{Radiative Torques and the RAT Alignment}
\begin{figure}
	\includegraphics[width=0.5\textwidth]{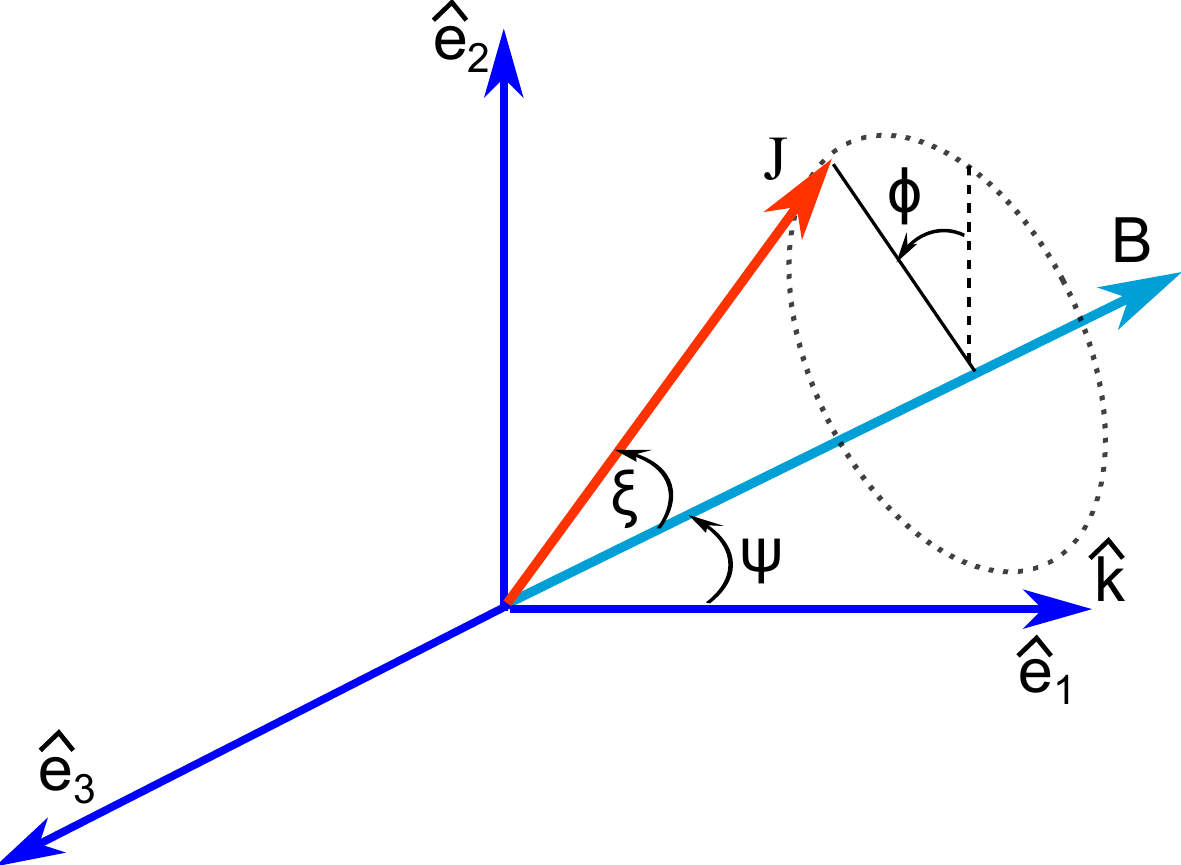}   
	\	\caption{Orientation of the grain angular momentum $\bJ$ in a lab coordinate system $\ehat_{1}\ehat_{2}\ehat_{3}$. The axis $\ehat_{1}$ is defined by the radiation direction $\bk$, the axis $\ehat_{2}$ is perpendicular to $\ehat_{1}$ and lies in the plane $\ehat_{1}$ and magnetic field $\bB$. The radiation direction makes an angle $\psi$ with the magnetic field. The alignment angle between $\bJ$ and $\bB$ is described by $\xi$, and fast Larmor precession of $\bJ$ around $\bB$ is described by the angle $\phi$.}
	\label{fig:alignRAT}
\end{figure}

Anisotropic radiation propagating along $\bk$ exerts a radiative torque on an irregular grain. Let $u_{\lambda}$ be the spectral energy density of the radiation field at wavelength $\lambda$ and $\gamma$ be its anisotropy. The energy density of the radiation field is $u_{\rad}=\int u_{\lambda}d\lambda$. The radiative torque arising from the interaction of the anisotropic radiation field with an irregular grain of size $a$ is then given by
\bea
{\bGamma}_{\rm RAT}&&=\int \left(\frac{\gamma \pi a_{\eff}^{2}
\lambda u_{\lambda}}{2\pi}\right){\bQ}_{\Gamma} d\lambda,\nonumber\\
&&=\left(\frac{\gamma a_{\eff}^{2}\bar{\lambda}u_{\rm rad}}{2}\right)\bar{\bQ}_{\Gamma} d\lambda,\label{eq:GammaRAT}
\ena
where ${\bQ}_{\Gamma}$ is the RAT efficiency which can be decomposed into three components $Q_{e1}, Q_{e2}$ and $Q_{e3}$ in the alignment reference frame defined by unit vectors $\ehat_{1},\ehat_{2},\ehat_{3}$ with $\hat{\be}_{1} \| \bk$, $\ehat_{2}\perp \ehat_{1}$ and $\ehat_{3}=\ehat_{1}\times \ehat_{2}$ (see Fig.\ref{fig:alignRAT}). Above, $\bar{\lambda}$ and $\bar{\bQ}$ are the average wavelength and RAT efficiency over the local radiation spectrum, and $u_{\rm rad}=Uu_{\rm MMP}$ is the radiation energy density of the local radiation field.

The magnitude of RAT efficiency, $Q_{\Gamma}$, in general, depends on the radiation field, grain shape, size and grain orientation relative to $\bk$ \citep{DraineWein.1996,LazHoang.2007}. Due to the uncertainty in interstellar grain shapes, an analytical model of RATs was introduced by \cite{LazHoang.2007}, and they demonstrated that their analytical torques exhibits the similar functional forms and basic features as calculated by Discrete Dipole Approximation code (DDSCAT) (\citealt{DraineFlatau.1994}) for irregular shapes. Recently, the AMO is well supported with numerical calculations for an ensemble of shapes \citep{Herranen:2019kj}. However, the relative amplitude of $Q_{e1}(\Theta)$ and $Q_{e2}(\Theta)$ changes from one shape to another and also varies with the wavelength. Therefore, we can the AMO and describe the RATs of different grain shapes and compositions through the ratio of the magnitudes of these torque components \citep{LazHoang.2007}
\begin{equation}
q^{\rm max}=\frac{Q_{e1}^{\rm max}}{Q_{e2}^{\rm max}}.\label{eq:qmax}
\end{equation}

The magnitude of RAT efficiency for a given grain size and a wavelength can be approximately described by the empirical scaling law from \citep{LazHoang.2007}, which is comparable to the recent results by \cite{Herranen:2019kj} and \cite{Jager.2024}. Therefore, the combination of AMO and the RAT magnitude is sufficient to describe RATs for different grain shapes and sizes, and we assume that for dust grains \citep{HoangLaz.2016} and in this paper.

Following \cite{LazHoang.2007}, RATs can cause three different effects on grain dynamics, including the precession of the grain angular momentum around the radiation direction $\bk$, spinup/spindown of the grain rotation, and the alignment of $\bJ$ and $\bk$. Rotating paramagnetic and superparamagnetic grains acquire a magnetic moment due to Barnett effect \citep{Barnett.1915,Dolginov.1976}, resulting in the Larmor precession of the grain angular momentum around the ambient magnetic field $\bB$. For PM/SPM grains, the Larmor precession is dominant over the radiative precession around $\bk$, which establishes the $\bB$ as an axis of the grain alignment. If the radiative precession is faster than the Larmor precession, the alignment axis is $\bk$. Figure \ref{fig:alignRAT} describes the orientation of the grain angular momentum with $\bB$ due to RATs from a radiation beam along $\bk$. 

It is convenient to study the grain orientation in the lab system using spherical coordinates, which is completely determined by three variables: the angle $\xi$ between the angular momentum vector $\bJ$ and the magnetic field direction $\bB$, the Larmor precession angle $\phi$ of ${\bJ}$ around $\bB$ and the value of the angular momentum $J$. To describe the effect of grain alignment by RATs in the lab coordinate system, one decomposes RATs into the net torque component along $\bJ$ responsible for spinup/spindown, denoted by $H(\xi,\phi,\psi)$, the torque component perpendicular to $\bJ$ directed along $\hat{\bxi}$ responsible for alignment/disalignment of $\bJ$ with $\bB$, denoted by $F(\xi,\phi,\psi)$, and the torque component directed along $\hat{\bphi}$ causing the radiative precession, $G(\xi,\phi,\psi)$. Therefore, Equation (\ref{eq:GammaRAT}) can be rewritten as
\bea
\bGamma_{\rm RAT}=M\left[H(\xi,\phi,\psi)\hat{\bJ} + F(\xi,\phi,\psi)\hat{\bxi}
+ G(\xi,\phi,\psi)\hat{\bphi}\right].\label{eq:GammaRAT_sphere} 
\ena
where \bea
M=\frac{\gamma 	\bar{\lambda} u_{\rm rad}a_{\eff}^{2}}{2}.
\ena
	
For more details, see previous papers \citep{DraineWein.1997,LazHoang.2007,HoangLaz.2008,HoangLaz.2016,Hoang.2022}. 

\begin{figure*}
	\centering
	\begin{overpic}[width=0.45\textwidth]{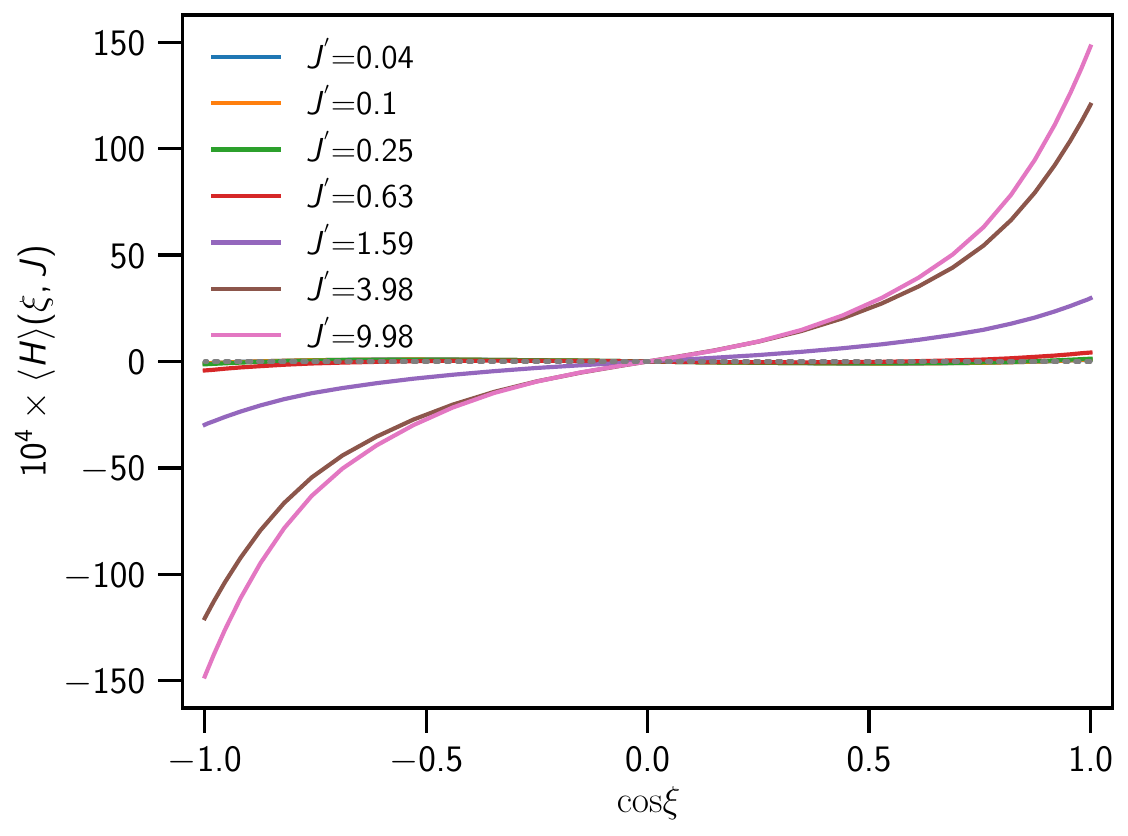}  
		\put(80,65){\small \textbf{(a)} $\langle H\rangle$}
	\end{overpic} 
	\begin{overpic}[width=0.45\textwidth]{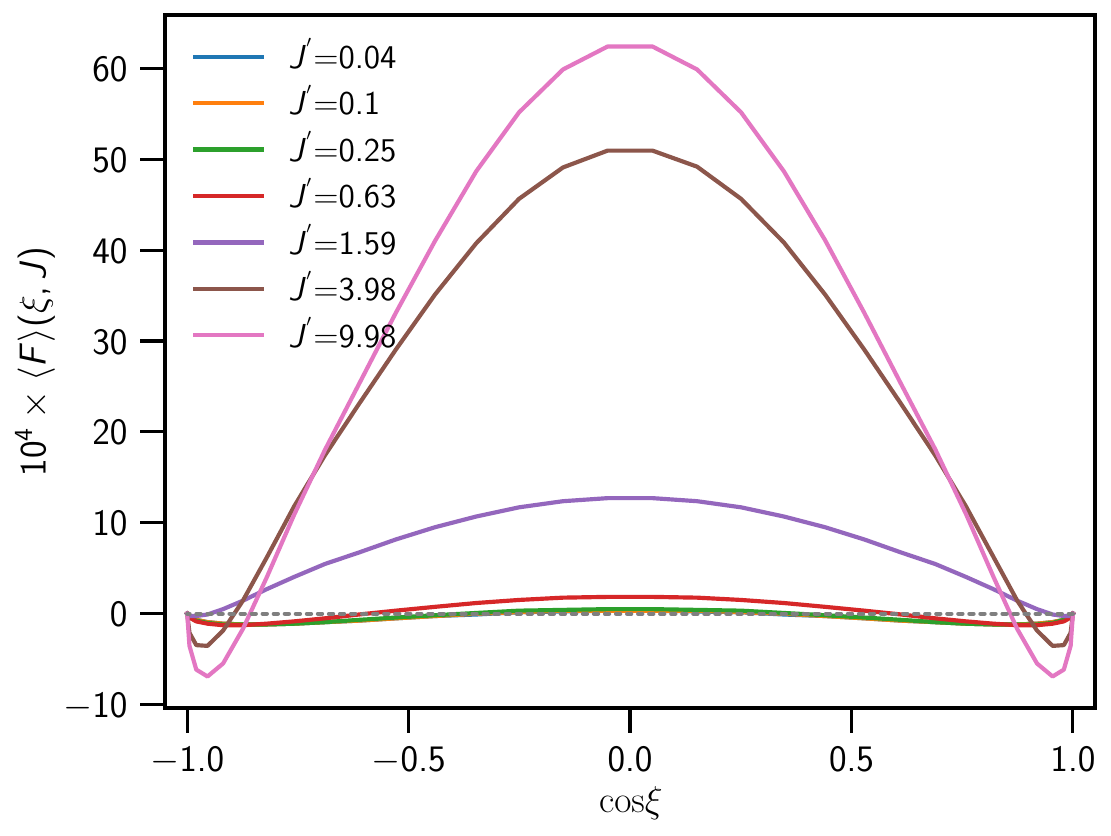}   
		\put(80,65){\small \textbf{(b)} $\langle F\rangle$}
		\put(97,20){\circle{2}}
		\put(79,20){\circle{2}}
		\put(34,20){\circle{2}}
		\put(17,20){\circle{2}}
		\put(33,15){B}
		\put(95,23){A}
	\end{overpic}
	\caption{The averaged aligning (panel (a)) and spinup (panel (b)) torque components for the different grain angular momentum, assuming the RAT model of $q^{\rm max}=3$ and the radiation parallel to the magnetic field, $\psi=0^{\circ}$. The intersection between the dotted line and $\langle F\rangle$ determines the stationary points marked by circles for the case $J'=0.63$.}
	\label{fig:FHmean}
\end{figure*}

Figure \ref{fig:FHmean} shows the averaged torque components, $\langle H\rangle$ and $\langle F\rangle$, for a RAT model with $q^{\rm max}=3$, obtained from averaging $H(\xi,\phi,\psi)$ and $F(\xi,\phi,\psi)$ over fast Larmor precession (angle $\phi$) and thermal fluctuations of the grain axis of maximum inertia $\ba_{1}$ with $\bJ$, as function of $\cos\xi$ for the case $q^{\rm max}=3$ and $\psi=0^{\circ}$, assuming the different values of $J$ (see \citealt{HoangLaz.2008,HoangLaz.2016} for details). For this RAT model, $\langle H\rangle>0$ for $\cos\xi>0$ and $\langle H\rangle <0$ for $\cos\xi<0$. There are four stationary points determined by $\langle F\rangle=0$ (marked by circles), among which $\cos\xi=1$ is an attractor with $\langle H\rangle>0$. Under the sole effect of RATs, grain alignment typically occurs at low-J and high-J attractors; however, the existence of high-J attractors is not universal and depends on grain shapes and radiation fields \citep{LazHoang.2007}.

\subsection{Radiative Torque Disruption (RATD)}
\label{sec:rot_disr}
An effectively spherical grain of radius $a_{\eff}$ spinning at angular velocity $\Omega$ experiences a centrifugal stress $S=\rho a_{\eff}^{2}\Omega^{2}/4$ acting on the plane through the grain center. Let $S_{\rm max}$ be the tensile strength of the grain material. The critical angular speed above which the grain is disrupted is determined by setting $S=S_{\rm max}$, yielding,
\begin{eqnarray}
	\Omega_{\rm disr}=\frac{2}{a_{\eff}}\left(\frac{S_{\rm max}}{\rho}\right)^{1/2}\simeq \frac{3.65\times 10^{8}}{a_{\eff,-5}}\left(\frac{S_{\rm max,7}}{\hat{\rho}}\right)^{1/2}~\rm rad s^{-1},\label{eq:w_disr}~~~~
\end{eqnarray}
where $a_{\eff,-5}=a_{\eff}/10^{-5}\cm$, and $S_{\rm max,7}=S_{\rm max}/(10^{7}\rm erg \cm^{-3})$.

RATs can rapidly spin up grains to suprathermal rotation with angular velocity greater than $\Omega_{\rm disr}$, resulting in instantaneous destruction of the grain into small fragments, a new effect termed RAdiative Torque Disruption (RATD) \citep{Hoang.2019nas,Hoang.2019}. Therefore, the RATD can occur whenever the grain's angular velocity exceeds the critical value. The minimum size for the RATD is given by (see Appendix \ref{apdx:RATD}; \citealt{Hoang.2019nas,Hoang.2019}):
\bea
a_{\rm disr}\simeq 0.22\bar{\lambda}_{0.5}S_{\max,7}^{1/3.4}
(1+F_{\rm IR})^{1/1.7}\left(\frac{n_{1}T_{2}^{1/2}}{\gamma_{-1}U}\right)^{1/1.7}\mum,~~~\label{eq:adisr}
\ena
where $\bar{\lambda}_{0.5}=\bar{\lambda}/0.5\mum$ and $\gamma_{-1}=\gamma/0.1$.

\section{General Theory of the MRAT Alignment}\label{sec:newtheory}
\subsection{Basic Equations of Grain Rotational Dynamics}
As discussed in Section \ref{sec:review}, the external alignment of grains with the magnetic field is mainly governed by rotational damping processes, magnetic relaxation, and RATs. Numerical calculations of grain alignment by RATs are studied by following the evolution of grain angular momentum in the phase space (\citealt{DraineWein.1997,LazHoang.2007,HoangLan.2016}). 

The evolution of the grain angular momentum in the lab coordinate system defined by the magnetic field and radiation direction (see Figure \ref{fig:alignRAT}) is described by the equation of motion \citep{DraineWein.1997,LazHoang.2007}
\bea
\frac{d\bJ}{dt} =\bGamma_{\rm RAT}+ \bGamma_{B}+ \bGamma_{\rm damp}+\bGamma_{\rm mag},\label{eq:dJvec_dt}
\ena
where the RAT torque $\bGamma_{\rm RAT}$ is given by Equation (\ref{eq:GammaRAT}), $\bGamma_{\rm damp}$ by Equation (\ref{eq:Gamma_damp}), and $\bGamma_{\rm mag}$ by Equation (\ref{eq:Gamma_mag}). 

In the spherical coordinate system with unit basis vectors $(\hat{\bJ}=\bJ/|\bJ|, \hat{\xi},\hat{\phi})$, $\bJ$ is described by three variables $J,\xi,\phi$ (see Figure \ref{fig:alignRAT}), and the time-evolution of $\bJ$ is represented as \citep{Weingartner.2003,HoangLaz:2008gb}
\bea
\frac{d\bJ}{dt} = \frac{dJ}{dt}\hat{\bJ} + \frac{Jd\xi}{dt}\hat{\bxi} + \frac{J\sin\xi d\phi}{dt}\hat{\bphi}.\label{eq:dJdt_sphere}
\ena

After averaging Equations (\ref{eq:dJvec_dt}) and (\ref{eq:dJdt_sphere}) over the angle $\phi$ due to the fast Larmor precession, it reduces to two equations that determine the alignment of $\bJ$ with the magnetic field:
\bea
\frac{d\xi}{dt}&&=\frac{M \langle F\rangle(\xi, \psi)}{J} -\frac{\sin\xi\cos\xi}{\tau_{\rm mag}} ,\label{eq:dxidt}\\
\frac{dJ}{dt}&&= M \langle H\rangle(\xi,\psi)-\frac{J}{\tau_{\rm damp}}-\frac{J}{\tau_{\rm mag}}\sin^{2}\xi,\label{eq:dJdt}
\ena
where $\langle F\rangle(\xi,\psi)$ and $\langle H(\xi,\psi)\rangle$ are the aligning and spin-up torque components averaged over the precession angle $\psi$ and thermal fluctuations (see e.g., Figure \ref{fig:FHmean}), and $\tau_{\rm damp}$ is the rotational damping time of grains due to both gas collisions and IR emission. These equations are different from previous studies \citep{HoangLaz.2016,LazHoang.2021} in which only gas damping is taken into account.

As in previous studies \citep{HoangLaz.2008,HoangLaz.2016}, we use the dimensionless parameter, $J'=J/J_{T}$ with $J_{T}=I_{\|}\omega_{T}$ and  $\omega_{T}=(kT_{\gas}/I_{\|})^{1/2}\simeq 10^{5}s^{-1/2}a_{-5}^{-5/2}T_{2}^{1/2}$ as thermal angular momentum along the grain symmetry axis. The dimensionless time is defined by $t'=t/\tau_{\rm damp}$, which is different from previous work in which $\tau_{\rm gas}$ is used due to the subdominance of IR emission in the standard ISM. Equations (\ref{eq:dxidt}) and (\ref{eq:dJdt}) become
\bea
\frac{d\xi}{dt'}&&=\left(\frac{M\tau_{\rm damp}}{J_{T}}\right)\frac{\langle F\rangle(\xi, \psi)}{J'} - \frac{\tau_{\rm damp}}{\tau_{\rm mag}}\sin\xi\cos\xi, \label{eq:dxidt_norm}\\ 
\frac{dJ'}{dt'}&&=\left(\frac{M\tau_{\rm damp}}{J_{T}}\right) \langle H\rangle(\xi,\psi)-J'\left(1 + \frac{\tau_{\rm damp}}{\tau_{\rm mag}}\sin^{2}\xi\right).~~~~~\label{eq:dJdt_norm}
\ena 

Let us denote the prefactor in the first terms of the above equation by
\bea
\delta_{\rm RAT}\equiv \frac{M\tau_{\rm damp}}{J_{T}}\simeq 6.2\times 10^{4}a_{-5}^{1/2} \frac{U}{n_{1}T_{2}},\label{eq:delta_RAT}
\ena
which characterizes the relative importance of RATs and the damping by gas collisions and IR emission. Therefore, the term $U/(n_{1}T_{2})$ can be used to characterize the difference of the local environment from the standard ISM with $n_{\H}=10\cm^{-3}$ and $T_{\rm gas}=100\K$.

Using the standard definition of the magnetic relaxation strength from Equation (\ref{eq:delta_mag}), one then writes
\bea
\frac{\tau_{\rm damp}}{\tau_{\rm mag}}=\frac{\delta_{\rm mag}}{1+F_{\rm IR}},\label{eq:delta_mag_damp}
\ena
which implies the reduction of magnetic relaxation by a factor $1+F_{\rm IR}$.

In the steady state, the maximum angular momentum spun-up by RATs can be obtained by setting $d\xi/dt'=0, dJ/dt'=0$, which yields
\bea
J_{\rm RAT}^{'{\rm max}}=\delta_{\rm RAT}\langle H\rangle.\label{eq:Jmax-RAT}
\ena

The above equations of motion can be rewritten as
\bea
\frac{d\xi}{dt'}&=&\frac{\delta_{\rm RAT}\langle F\rangle(\xi, \psi)}{J'} - \frac{\delta_{\rm mag}}{1+F_{\rm IR}}\sin\xi\cos\xi ,\label{eq:dxidt_new}\\ 
\frac{dJ'}{dt'}&=&\delta_{\rm RAT}\langle H\rangle(\xi,\psi)-J'\left(1 + \frac{\delta_{\rm mag}}{1+F_{\rm IR}}\sin^{2}\xi\right),\label{eq:dJdt_new}
\ena 
which reveal that grain alignment is fully determined by three parameters, $\delta_{\rm RAT}, \delta_{\rm mag}$, and RAT efficiency $\langle F\rangle,\langle H\rangle$. Therefore, Equations (\ref{eq:dxidt_new}) and (\ref{eq:dJdt_new}) can be used to simulate the alignment process of grains subject to a wide range of environments by varying the parameters $\delta_{\rm RAT}$.


\subsection{Critical Magnetic Relaxation for Producing High-J Attractors in the MRAT Mechanism}
\subsubsection{Critical Magnetic Relaxation for the MRAT}
Previous studies \citep{HoangLaz.2016,Herranen.2021} derive the critical relaxation required for producing high-J attractors by MRATs for the ISM with the standard ISRF for which the IR damping is subdominant. Here, we derive the criteria for high-J attractors when taking into account the effect of IR damping, which is significant for strong radiation fields.

Following the method in \cite{HoangLaz.2016}, we determine the criteria for the stationary point at $\sin\xi_{s}=0, J'_{s}=\delta_{\rm RAT}\langle{\rm H}(\xi_{s})\rangle$ for which $(d\xi/dt'=0,dJ'/dt'=0)$ (see Eq.\ref{eq:dJdt_new}) to be an attractor as follows:
\bea
\left. \frac{1}{\langle H\rangle}\frac{d\langle F\rangle}{d\xi}\right|_{\sin\xi_{s}=0} -\frac{\delta_{\rm mag}}{1+F_{\rm IR}}<0,~~\langle{H}\rangle>0,\label{eq:attractor}
\ena
which simply indicates that to produce an attractor point, the rate of increasing $\xi$ from $\xi_{s}$ by RATs (i.e., disaligning case by RATs) must be lower than the rate of decreasing $\xi$ (aligning) by magnetic relaxation $\delta_{\rm mag}$. 

The above equation yields the critical magnetic relaxation for high-J attractors:
\bea
\delta_{\rm mag}>\left(1+F_{\rm IR}\right)\left[\frac{1}{\langle H\rangle}\frac{d\langle F\rangle}{d\xi} \right]_{\sin\xi_{s}=0},~\langle H\rangle>0.\label{eq:delta_mag_cri}
\ena

When the radiation direction is directed along the magnetic field ($\psi=0$), the above conditions yield the critical strength required for high-J attractors \citep{HoangLaz.2016}:
\bea
\delta_{\rm mag,cri}=(1+F_{\rm IR})\left[\frac{2-q^{\rm max}}{q^{\rm max}}\right].\label{eq:delta_mag_cri1}
\ena

One notes that the above condition is always satisfied if $q^{\rm max}>2$. However, for $q^{\rm max}<2$, RATs do not produce high-J attractors, and the magnetic relaxation {is required to produce high-J attractors}. Therefore, the critical magnetic relaxation has an additional term $1+F_{\rm IR}$ different from \cite{HoangLaz.2016}. In typical and weak radiation fields, $F_{\rm IR}\ll 1$, and it is not important. But in a strong radiation field, $F_{\rm IR}\gg 1$, and it influences the efficiency of magnetic relaxation on grain alignment. Since $\delta_{\rm mag}$ has an upper limit determined by embedded iron inclusions and the magnetic field strength ($B$), the magnetic relaxation effect will become ineffective in a strong radiation field.

\begin{figure*}
	\centering
	\begin{overpic}[width=0.45\textwidth]{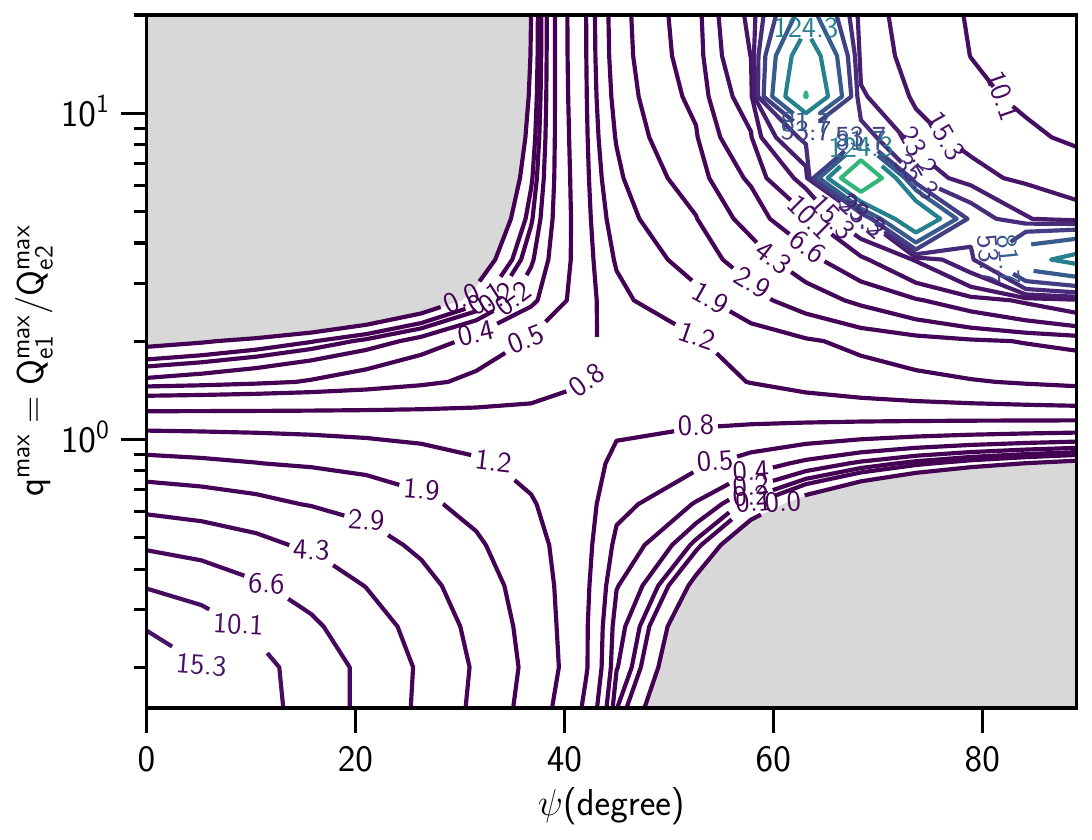}
			\put(14,70){\small \textbf{(a)} $U = 1$}
            \put(15,60){\small \textbf{High-J by RAT}}
            \put(65,20){\small \textbf{High-J by RAT}}
            \put(35,25){\rotatebox{35}{\fontsize{10}{6} \textbf{High-J by MRAT}}}            
	\end{overpic}
	\begin{overpic}[width=0.45\textwidth]{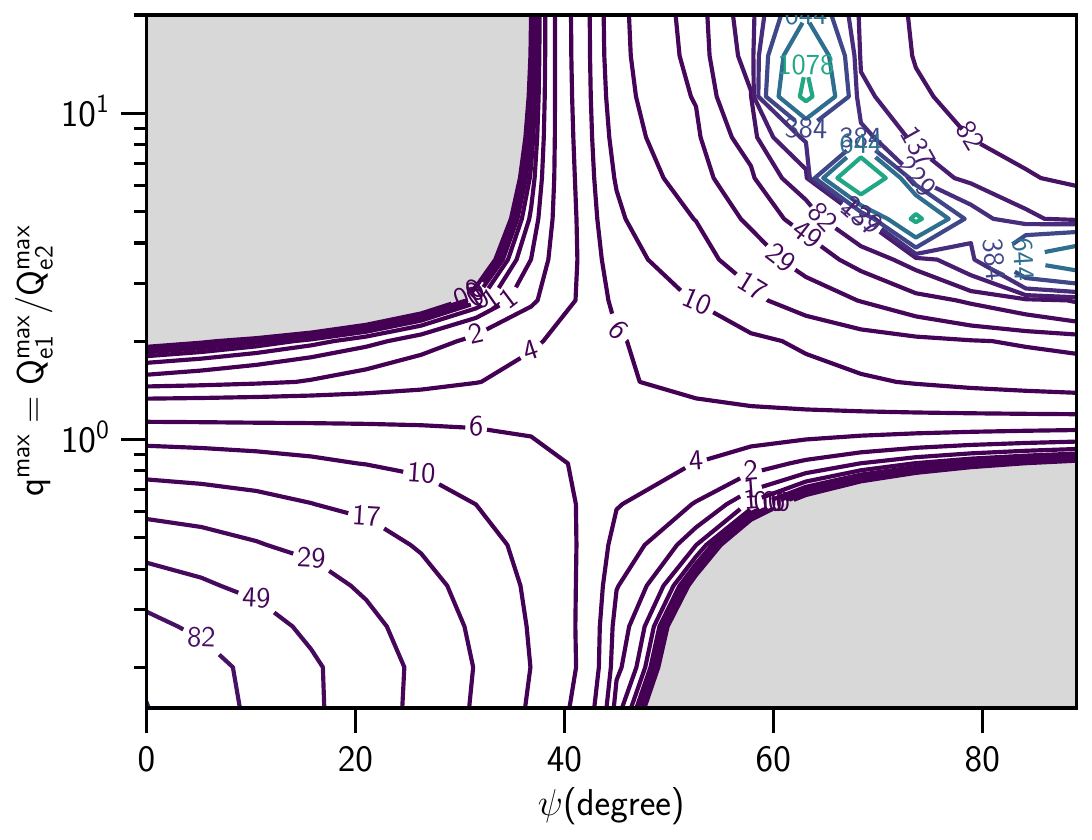}
			\put(14,70){\small \textbf{(b)} $U = 10^{2}$}
            \put(15,60){\small \textbf{High-J by RAT}}
            \put(65,20){\small \textbf{High-J by RAT}}
            \put(35,25){\rotatebox{35}{\fontsize{10}{6} \textbf{High-J by MRAT}}}
	\end{overpic}
	\begin{overpic}[width=0.45\textwidth]{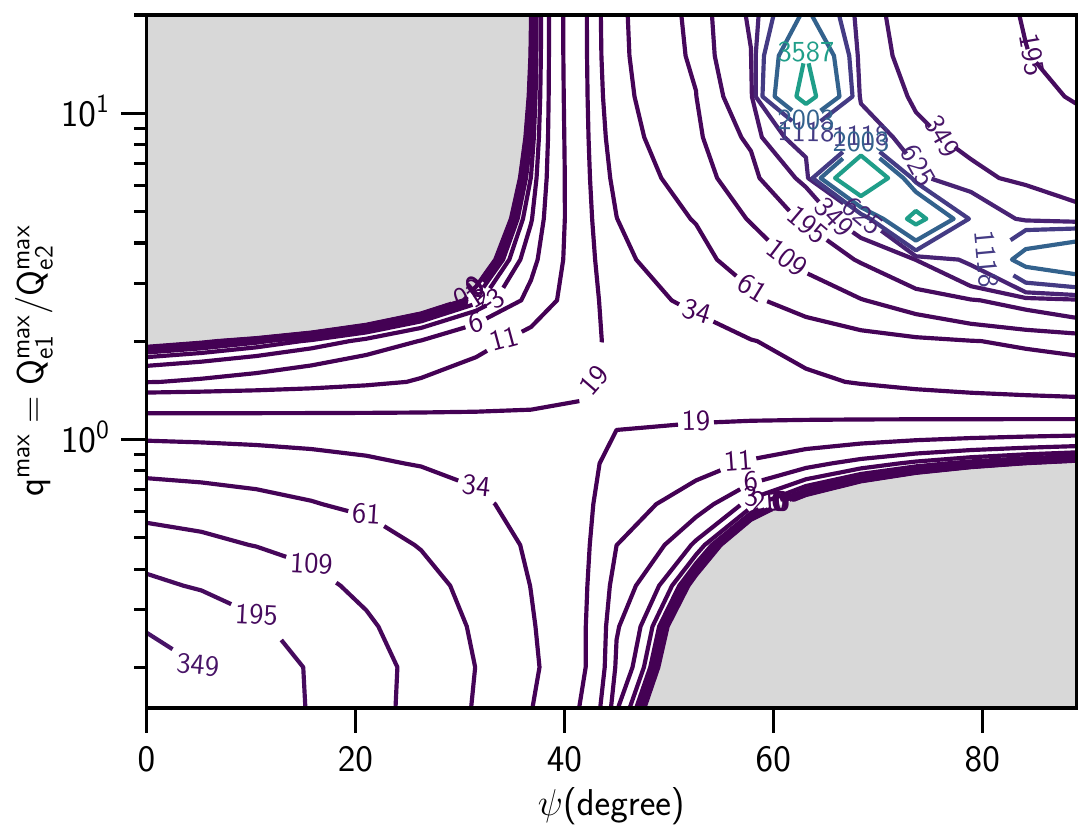}
			\put(14,70){\small \textbf{(c)} $U = 10^{3}$}
            \put(15,60){\small \textbf{High-J by RAT}}
            \put(65,20){\small \textbf{High-J by RAT}}
            \put(35,25){\rotatebox{35}{\fontsize{10}{6} \textbf{High-J by MRAT}}}
	\end{overpic}
	\begin{overpic}[width=0.45\textwidth]{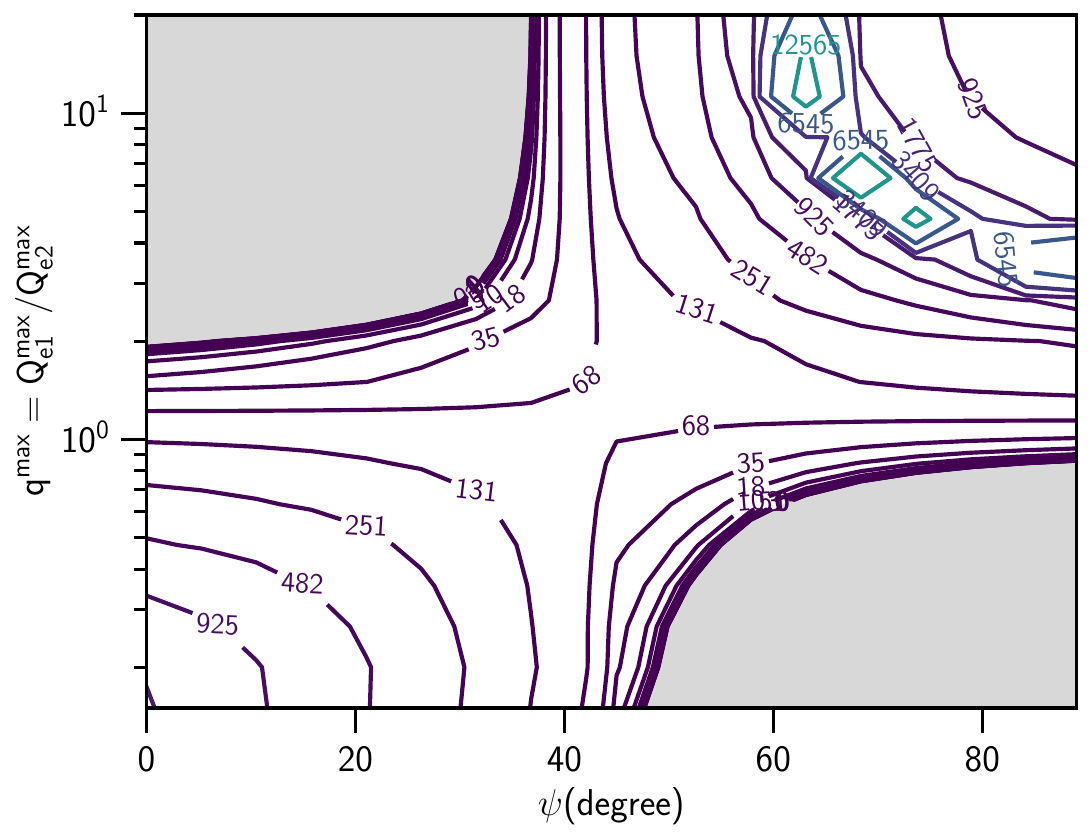}
			\put(14,70){\small \textbf{(d)} $U = 10^{4}$}
            \put(15,60){\small \textbf{High-J by RAT}}
            \put(65,20){\small \textbf{High-J by RAT}}
            \put(35,25){\rotatebox{35}{\fontsize{10}{6} \textbf{High-J by MRAT}}}
	\end{overpic}
	\caption{The map of the critical magnetic relaxation ($\delta_{\rm mag, cri}$) required to produce high-J attractors for the different RAT models ($q^{\rm max}$) and the angle between radiation direction with the magnetic field ($\psi$) in the different radiation fields of $U=1-10^{4}$ (panels (a)-(d)). The {gray shaded} areas denote the parameter space where high-J attractors can be produced by RATs only, no need for magnetic relaxation. The contour levels show some values of $\delta_{\rm mag, cri}$. The critical magnetic relaxation strength must be increased to have universal high-J attractors in stronger radiation fields.}
	\label{fig:deltacri_lowJ_highJ}
\end{figure*}

Equation (\ref{eq:delta_mag_cri1}) provides the critical magnetic relaxation required for producing high-J attractors by the MRAT when the radiation is parallel to the magnetic field. For a general angle $\psi$, we numerically solve Equation (\ref{eq:delta_mag_cri}) to $\delta_{\rm mag,cri}$, as in \cite{HoangLaz.2016}. Figure \ref{fig:deltacri_lowJ_highJ} shows the results for the standard ISRF (upper left) and intense radiation fields with $U=10^{2}, 10^{3}, 10^{4}$, assuming the gas density $n_{\rm H}=10\cm^{-3}$ and $T_{\rm gas}=100\K$. The critical magnetic relaxation must increase with increasing radiation strength in order to create high-J attractors, except in the parameter space where RATs alone can produce high-J attractors (gray shaded areas). As a result, for a given $\delta_{\rm mag}$ determined by the grain magnetic properties and local gas conditions, the efficiency of magnetic relaxation on grain alignment is weaker for stronger radiation fields.

\subsubsection{Upper Threshold of the MRAT}

We can determine the threshold of the radiation strength for which magnetic relaxation is still effective, so that grain alignment is described by MRATs as follows. For a strong radiation field of $U/(n_{1}T_{2})>1$, $F_{\rm IR}\sim U^{2/3}/n_{1}\gg1$, and Equation (\ref{eq:delta_mag_cri}) can be rewritten as
\bea
\delta_{\rm mag,cri}&&= \frac{2-q^{\rm max}}{q^{\rm max}}F_{\rm IR}\\
&&=\frac{2-q^{\rm max}}{q^{\rm max}}\left(\frac{1.2U^{2/3}}{a_{\eff,-5}}\right)\left(\frac{1}{n_{1}T_{2}^{1/2}}\right).\label{eq:delta_mag_cri2}
\ena
which depends on local radiation and gas density as $U^{2/3}/n_{1}$.

For grains with iron inclusions, following \cite{HoangLaz.2016}, one has
\bea
\delta_{\rm mag,SPM}=0.3\hat{\rho}{a}_{\eff,-5}\left(\frac{10^{5}N_{cl,5}\hat{B}^{2}}{n_{1}T_{2}^{1/2}}\right),\label{eq:delta_SPM_max}
\ena
where $N_{\rm cl}=10^{5}N_{cl,5}$ is the number of iron atoms per cluster, and the $10^{5}$ is the upper limit of iron clusters for SPM grains \citep{JonesSpitzer.1967,Yang.2021}.

Setting $\delta_{\rm mag,SPM}=\delta_{\rm mag,cri}$, one obtains the upper limit for the radiation strength below which magnetic relaxation is still effective:
\bea
U^{\rm MRAT}_{\rm max} \simeq 5.2\times 10^{6}a_{\eff,-5}^{3}N_{cl,5}^{3/2}\hat{B}^{3}\left(\frac{q^{\rm max}}{2-q^{\rm max}}\right)^{3/2},
\label{eq:Umax}
\ena
which corresponds to the dust temperature
\bea
T^{\rm MRAT}_{\rm max}\simeq 217.2a_{\eff,-5}^{1/2}N_{cl,5}^{1/4}\hat{B}^{1/2}\left(\frac{q^{\rm max}}{2-q^{\rm max}}\right)^{1/4}.\label{eq:Td_max}
\ena

Therefore, the MRAT alignment is only effective for the radiation fields of $U<U^{\rm MRAT}_{\rm max}$ or dust temperature $T_{d}<T^{\rm MRAT}_{\rm max}$. Above this limit, grain alignment and disruption is only determined by RATs.

\subsection{Effects of Collisional and Magnetic Excitations on the MRAT Alignment}
The effect of gas collisions in the framework of paramagnetic alignment was studied 
by many authors (\citealt{Cugnon.1983,1997MNRAS.288..609L}) using the Fokker-Planck equations. The Langevin equation (LE) approach was used to numerically study this problem in \cite{Roberge.1993} and \cite{RobergeLaz.1999}. It was later applied to study the effect of collisional excitations in the framework of RAT alignment \citep{HoangLaz.2008}, spinning dust emission (\citealt{Hoang.2011}; \citealt{Hoang.2014b}), and magnetic dust emission (\citealt{HoangLaz.2016b}). Here, we follow a similar approach in \cite{HoangLaz.2016}.

Let ($\xhat,\yhat,\zhat$) be the inertial frame of reference where $\zhat$ is defined by the magnetic field direction. An increment of grain angular momentum after a time interval $dt$ due to various random interactions can be described by the LE (\citealt{Roberge.1993}) 
\bea 
d J_{i}=A_{i}(t)dt+B_{ij}(J,t)dw_{j},~ i=x,y,z,\label{eq:LE}
\ena
where $d\omega_{j}$ are the Wiener coefficients, and $A_{i}, B_{ij}$ are the diffusion coefficients defined as 
\bea
A_{i}&=\langle \Delta J_{i} \rangle,~ i=x,y,z, \label{eq:Acoef}\\ 
(BB^{T})_{ij}&=\langle \Delta J_{i} \Delta J_{j}\rangle,i,j=x,y,z,\label{eq:Bcoef}
\ena
where $B^{T}$ is the transposal matrix of $B$ (see Appendix B in \citealt{HoangLaz.2016}). 

The collisional excitation coefficients in the dimensionless units of $J'\equiv J/I_{\|}\omega_{T}$ and $t'\equiv t/\tau_{\rm damp}$ due to gas collisions are then given by
\bea
B'_{{\rm coll},i}&&=B_{{\rm coll},i}\times \left(\frac{\tau_{\rm damp}}{2I_{\|}kT_{\gas}} \right),\nonumber\\
&&=B_{{\rm coll},i}\times \left(\frac{\tau_{\rm gas}}{2I_{\|}kT_{\gas}} \right) \frac{1}{1+F_{\rm IR}}{\rm for~ i=x,y,z},\label{eq:Bcoll}
\ena
where $B_{\rm coll,i}$ are given by Equations in Appendix B of \cite{HoangLaz.2016}, but the difference is $\tau_{\rm gas}$ is replaced by $\tau_{\rm damp}$. 

The diffusion coefficients for magnetic excitations are described by (see Appendix B in \citealt{HoangLaz.2016})
\bea
B'_{\rm mag,xx}=B'_{\rm mag,yy}&&=\left(\frac{\delta_{\rm mag}}{1+F_{\rm IR}}\right)\left(\frac{T_{d}}{T_{\rm gas}}\right),\nonumber\\
B'_{{\rm mag},zz}&&=0,\label{eq:Bmag}
\ena 
for which we denote $\tilde{\delta}_{\rm mag}=\delta_{\rm mag}/(1+F_{\rm IR})$.

The total excitation coefficients become $B'_{ii}={B'}_{ii,\rm coll} + {B'}_{ii,\rm mag}$ where $B_{ii}$ with $i=x,y,z$ denotes the diagonal matrix components. The magnetic fluctuations increase with $\delta_{\rm mag}$, but both collisional and magnetic excitations are reduced significantly in strong radiation fields with $F_{\rm IR}>1$.

In dimensionless units, Equation (\ref{eq:LE}) becomes 
\bea
dJ'_{i}=A'_{i}dt'+\sqrt{B'_{ii}}dw'_{i} \mbox{~for~} i= x,~y,~z,\label{eq:dJp_dt}
\ena
where $\langle dw_{i}^{'2}\rangle=dt'$ and
\bea
A'_{i}&&=-{J'_{i}}\left[1+\tilde{\delta}_{\rm mag}(1-\delta_{iz})\right],\nonumber\\
B'_{ii}&&= B'_{\rm coll,i}+\tilde{\delta}_{\rm mag}(1-\delta_{iz})\left(\frac{T_{\d}}{T_{\gas}}\right),\label{eq:Bii}
\ena
with $\delta_{iz}=1$ for $i=z$ and $\delta_{iz}=0$ for $i= x,y$.

\subsection{Numerical Simulations of Grain Alignment and Disruption}
As in \cite{HoangLaz.2008,HoangLaz.2016}, to account for the randomization effect by gas collisions and magnetic fluctuations on grain alignment, we employ the hybrid approach. For each time step, we first solve the LEs subject to gas-grain collisions and magnetic fluctuations by solving for three components $J_{x}, J_{y}, J_{z}$ using the 4th order Runge-Kutta integrator implemented in \cite{HoangLaz.2016} and compute the current values of $\tilde{J}=\sqrt{J_{x}^{2}+J_{y}^{2}+J_{z}^{2}}$, $\cos\tilde{\xi}=J_{z}/\tilde{J}$. Then, we use $\tilde{J}$ and $\tilde{\xi}$ as input parameters to solve Equations (\ref{eq:dxidt}) and (\ref{eq:dJdt}) for new values of $\xi, J$ as a result of regular RATs only, where the terms involving the gas and magnetic damping are removed because those effects are already accounted for by the LEs. For estimates of rotational damping and excitation coefficients by gas collisions and magnetic fluctuations, we consider the oblate spheroidal shape for grains as in \cite{HoangLaz.2016}.

To check whether grains converge to some attractor point, we introduce the convergence criteria as follows:
\bea
\frac{|J'_{i+1}-J'_{i}|}{J'_{i}} \le \epsilon_{J},\
\frac{|\cos(\xi_{i+1})-\cos(\xi_{i})|}{|\cos(\xi_{i})|} \le \epsilon_{\xi},
\ena
where $i$ is the timestep, and we adopt $\epsilon_{J}=10^{-6}$ and $\epsilon_{\xi}=10^{-6}$.

\subsubsection{Calculations of Grain Alignment Degrees}
The alignment of grain axes with the magnetic field includes the internal alignment of the grain axis of maximum inertia with the angular momentum ($\bJ$) and external alignment of ${\bJ}$ with the magnetic field. Let $Q_{J}=(3\cos^{2}\xi-1)/2$ be the measure of the external alignment of ${\bJ}$ with the magnetic field, and $Q_{X}=(3\cos^{2}\theta-1)/2$ be the internal alignment of the grain axis of maximum inertia with the angular momentum.

At each time step, we calculate $Q_{J}$ by averaging over the ensemble of $N_{\rm gr}$ grains:
\begin{eqnarray}
    Q_{J}=\frac{1}{N_{\rm gr}}\sum_{i=1}^{N_{\rm gr}} \frac{(3\cos^{2}\xi_{i}-1)}{2}.\label{eq:QJ}
\end{eqnarray}

The Rayleigh reduction factor \citep{Greenberg.1968}, which describes the alignment of the grain principal axis with the magnetic field \citep{RobergeLaz.1999}, is calculated as
\begin{eqnarray}
 R=\frac{1}{N_{\rm gr}}\sum_{i=1}^{N_{\rm gr}} Q_{J}(\xi_{i})Q_{X}(J_{i}),\label{eq:Ralign}
\end{eqnarray}
and $Q_X$ characterizes the internal alignment of the grain axis of maximum inertia ($\ba_{1}$) with respect to ${\bJ}$. For an oblate grain with fast internal relaxation,
\bea
Q_{X}(J_{i})=\int_{0}^{\pi} q_{X}Z\exp\left({-J_{i}^{2}[1+(h-1)\sin^{2}\theta]}\right)\sin\theta d\theta,~~~~
\label{eq:QX}
\ena
where $h=I_{\|}/I_{\perp}$ with $I_{\|,\perp}$ the principal inertia moments parallel and perpendicular to the grain symmetry axis, and the distribution of the angle $\theta$ between $\ba_{1}$ and $\bJ$ is described by the local thermal equilibrium function, which is valid for grains with fast internal relaxation due to Barnett relaxation and inelastic relaxation (see \citealt{HoangLaz.2016}).\footnote{In this paper, fast internal relaxation is assumed for simplicity \citep{HoangLaz.2009b}, and this assumption may not be valid for very large grains in dense environments \citep{Hoangetal.2022}.}

\subsubsection{Model Parameters and Setup}
In dimensionless units ($J'$ and $t')$, the alignment of grains is determined by three parameters, including $\delta_{\rm RAT}, \delta_{\rm mag}$, and RATs ($\langle F\rangle,\langle H\rangle$). Therefore, Equations (\ref{eq:dxidt_new}) and (\ref{eq:dJdt_new}) can be used to simulate the alignment process of grains subject to a wide range of environments by varying the parameters $\delta_{\rm RAT}$. To focus on the effects of local physical conditions on grain alignment, we fix the effective grain size of $a_{\eff}=0.2\mum$ (i.e., comparable to the maximum size of grains in the diffuse ISM) and the angle between the radiation direction and magnetic field at $\psi=0^{\circ}$. We consider different RAT models of $q^{\rm max}=1,2,3$, corresponding to the cases of RAT alignment without high-J attractor ($f_{\rm high-J}^{\rm ens}=0)$, with high-J attractors ($f_{\rm high-J}^{\rm ens}=1$), respectively (see Figure \ref{fig:deltacri_lowJ_highJ}). We assume two typical models of magnetic relaxation with $\delta_{\rm mag}=1$ and $\delta_{\rm mag}=10^{3}$, corresponding to PM and SPM grains containing a high level of iron inclusions. To account for a wide range of astrophysical environments, we consider a range of values of $\delta_{\rm RAT}\propto U/(n_{1}T_{2})$ (see Eq. \ref{eq:delta_RAT}), characterized by $U/(n_{1}T_{2})\lesssim 1$ (i.e., the collision-dominated regions) and $U/(n_{1}T_{2})>1 $ (i.e., radiation-dominated regions). {Table \ref{tab:models_qmax2} summarizes the model parameters adopted for our numerical calculations, the damping time, and the timescales required for perfect alignment for $q^{\rm max}=2$ for the SPM ($\delta_{\rm mag}=10^{3}$) and PM ($\delta_{\rm mag}=1$) grains. 
	For the numerical values shown in this Table, $\tau_{\rm damp}=\tau_{\rm gas}(1+F_{\rm IR})$ for $U\le 1$ and $\tau_{\rm damp}\sim \tau_{\rm gas}/F_{\rm IR}$ for $U>1$. The damping time is shorter for stronger radiation fields due to the increase of the infrared damping $F_{\rm IR}$.} Note that \cite{HoangLaz.2016,LazHoang.2021} run simulations for the typical ISM with $n_{\H}=30\cm^{-3}$ and $U=1$, corresponding to $U/(n_{1}T_{2})\approx 0.33$.

In addition to the gas density, gas and dust temperatures are also important parameters for estimating the characteristic timescales. In the ISM and radiation regions, gas is hotter than the gas due to photoelectric heating, but in dense and cold environments, gas and dust are thermally coupled. For a given radiation field, the dust temperature can be calculated assuming thermal equilibrium, which is a good approximation for the dust grains that can be aligned. Very small grains have a non-equilibrium temperature due to stochastic heating but are not aligned. The grain temperature of silicate and carbonaceous materials is given by \citep{Draine.2011book}
\bea
T_{\rm sil} &&\simeq 16.4U^{1/6}a_{\eff,-5}^{-1/15}~\K,\\
T_{\rm carb}&&\simeq 22.3U^{1/6}a_{\eff,-5}^{-1/40}~\K,\label{eq:Tsil_Tcarb}
\ena
and the temperature for the mixed dust grain made of silicate and carbonaceous material is
\bea
T_{d}  = \left(0.625T_{\rm sil}^{4} + 0.375T_{\rm carb}^{4}\right)^{1/4},\label{eq:Td_eq}
\ena
where we assume the volume fraction of silicate and carbonaceous material to be $62.5\%$ and $37.5\%$, respectively. The gas temperature can be determined by $T_{d}$ when the ratio $T_{\rm gas}/T_{d}$ is known. For the study here, we assume the gas temperature is higher than the dust as in the standard ISM with $T_{\rm gas}/T_{d}\sim 5$.


{We focus on a population of test grains initially rotating at $J=3J_{\rm T}$ but having different orientations with the magnetic field (cf. \citealt{HoangLaz.2008,HoangLaz.2016}).} We consider an ensemble of $N_{\rm gr}=16$ test grains to better visualize the phase-map of their time-dependent alignment, especially in the presence of stochastic excitations. The total integration time is chosen to be long enough for the perfect grain alignment induced by collisional and magnetic excitations.

\section{Numerical Results}\label{sec:results}
We first run simulations of grain alignment for a fixed gas density $n_{\rm H}=10\cm^{-3}$ with different ratios of $U/(n_{1}T_{2})$. We also tried to run simulations for different gas densities, such as $n_{\rm H}= 10^{2}-10^{4}\cm^{-3}$ and found that the alignment results are similar to those of $n_{\H}=10\cm^{-3}$ given the same ratio $U/(n_{1}T_{2})$. Therefore, here, we show the numerical results for the typical density $n_{\rm H}=10\cm^{-3}$ for the different ratio of $U/(n_{1}T_{2})$. We first show the typical RAT model of $q^{\rm max}=2$ for which RATs can induce high-J attractors (see Figure \ref{fig:deltacri_lowJ_highJ}).

\subsection{Collision-dominated (CD) Regime: $U/(n_{1}T_{2})\leq 1$}
Here, we first show the results for the CD regime with $U/(n_{1}T_{2})\leq 1$. These CD regions encompass the standard ISM and star-forming environments, including molecular clouds and filaments, and extend to dense cores and protostellar regions.

\begin{figure*}
	\centering
	\begin{overpic}[width=0.32\linewidth]{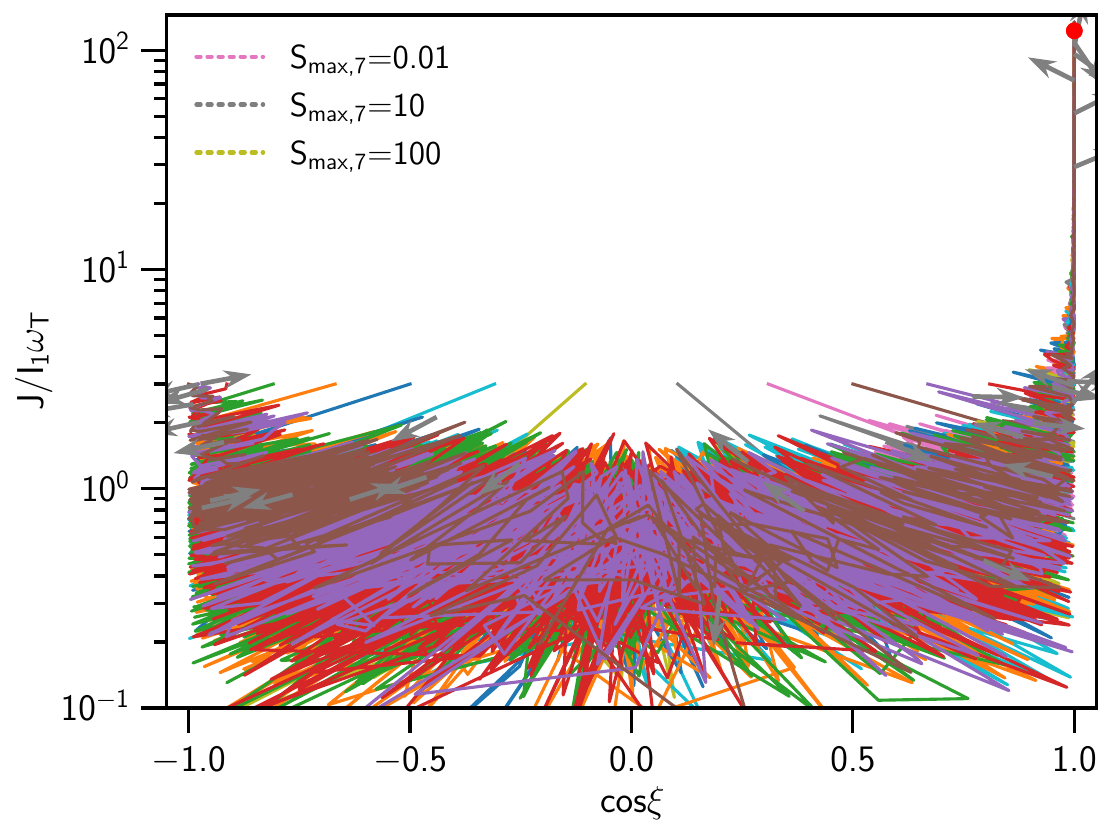}
		\put(65,75){\tiny \textbf{High-J Attractor}}		
		\put(40,65){\small \textbf{(a)} $U/(n_{1}T_{2}) = 0.16$}
	\end{overpic}
	\begin{overpic}[width=0.32\linewidth]{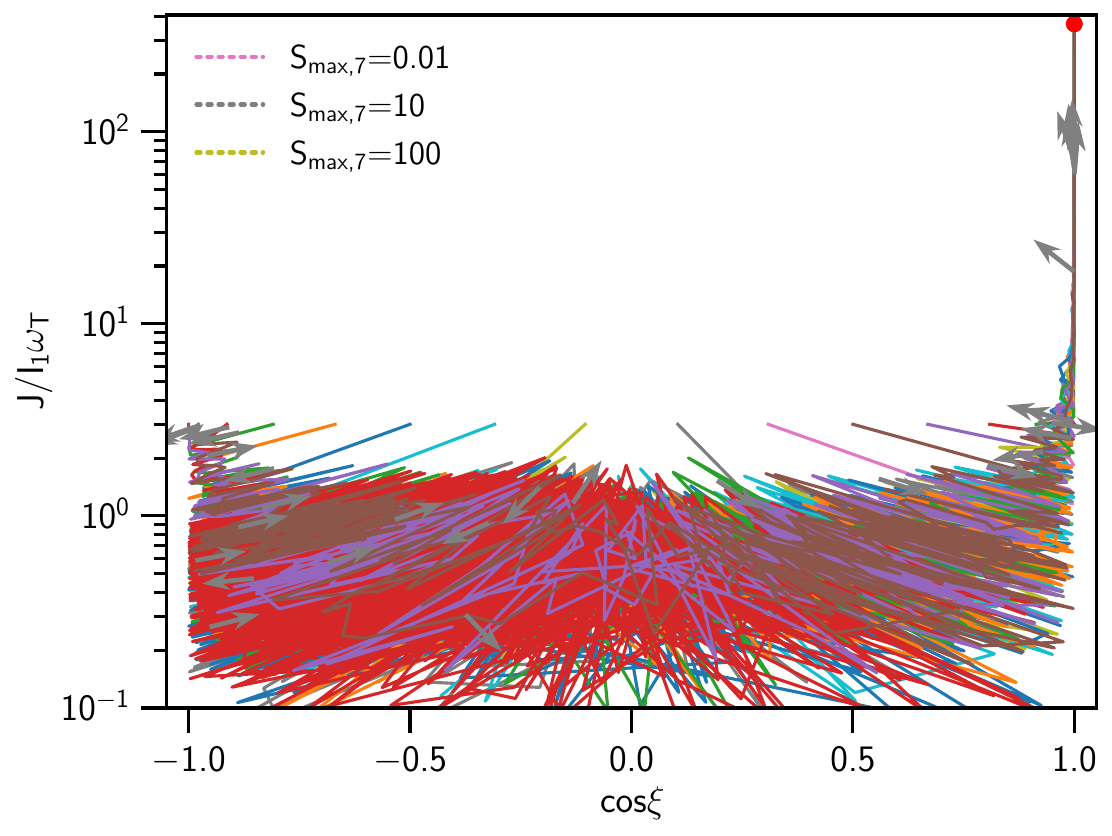}
				\put(65,75){\tiny \textbf{High-J Attractor}}
		\put(40,65){\small \textbf{(b)} $U/(n_{1}T_{2}) = 0.6$}
	\end{overpic}
	\begin{overpic}[width=0.32\linewidth]{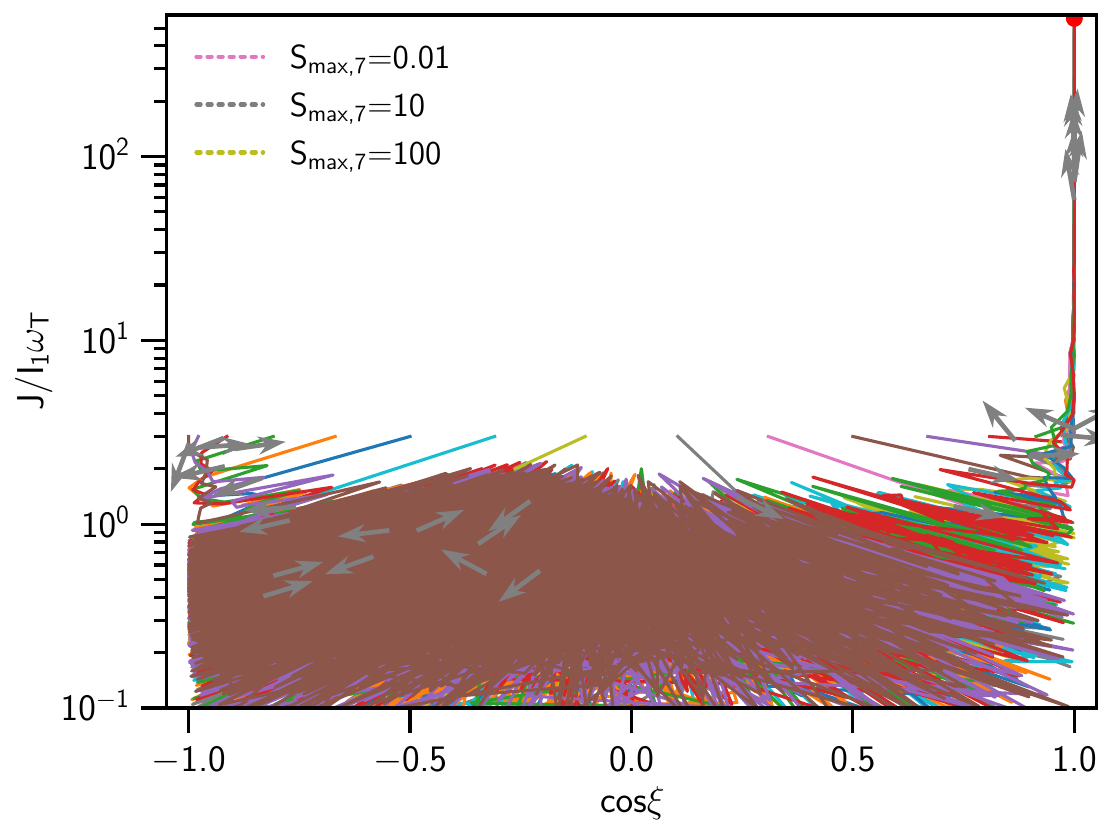}
				\put(65,75){\tiny \textbf{High-J Attractor}}
		\put(40,65){\small \textbf{(c)} $U/(n_{1}T_{2}) = 1$}
	\end{overpic}
\caption{Phase trajectory maps of grain alignment for the different radiation fields of $U/(n_{1}T_{2})\le 1$ (panels (a)-(c)), in the collision-dominated regime, assuming the RAT model of $q^{\rm max}=2$ and SPM grains with $\delta_{\rm mag}=10^{3}$. {Gray arrows indicate the instantaneous direction of grain orientation in the phase space, which are random for $J<J_{T}=I_{1}\omega_{T}$.} Some grains are rapidly spun up to the high-J attractor (red filled circles). The majority of grains are driven to low-J rotation, and subsequently, gas and magnetic excitations strongly disturb the grain orientation at low-J rotation and scatter them into stable high-J attractors. {The dotted lines showing the threshold for RATD for grains of different tensile strengths of $S_{\max,7}$ are not seen in these figures because RATD does not occur}.}
    \label{fig:CDR_map_qmax2_nH1e1_aef02}
\end{figure*}

\begin{figure*}
\centering
\begin{overpic}[width=0.32\textwidth]{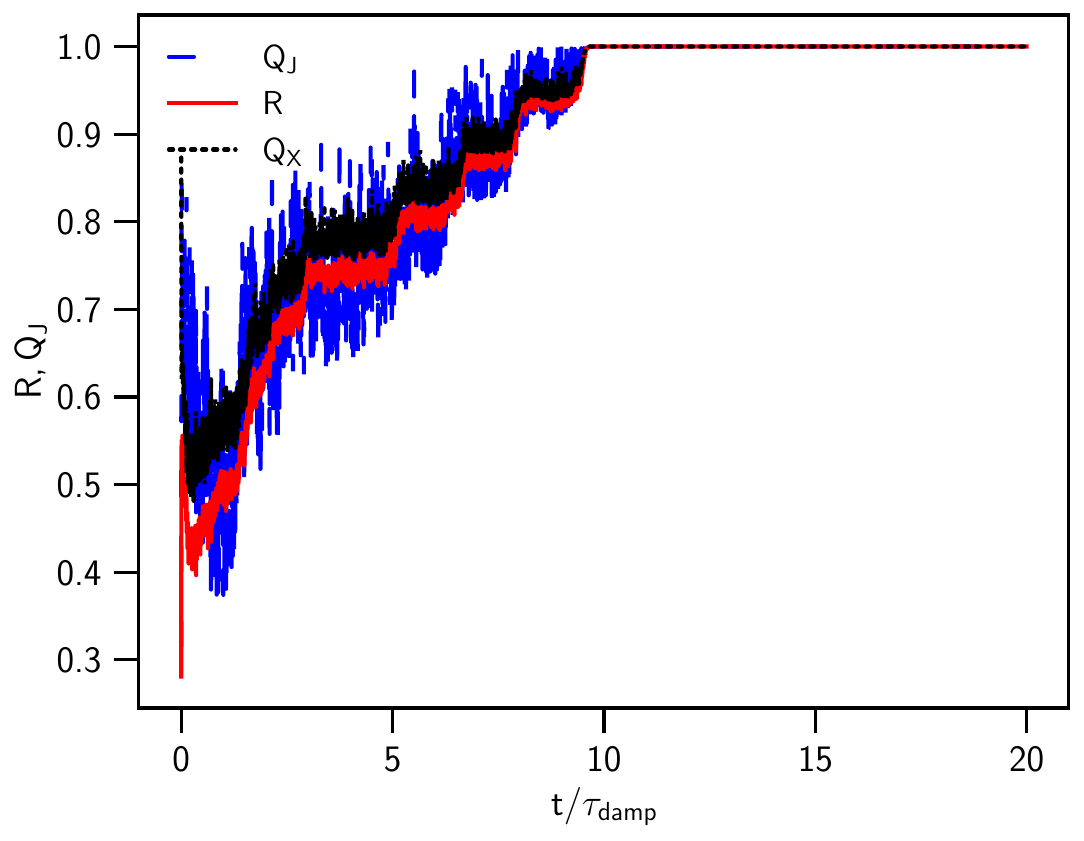}
		\put(40,20){\small \textbf{(a)} $U/(n_{1}T_{2}) = 0.16$}
\end{overpic}
\begin{overpic}[width=0.32\textwidth]{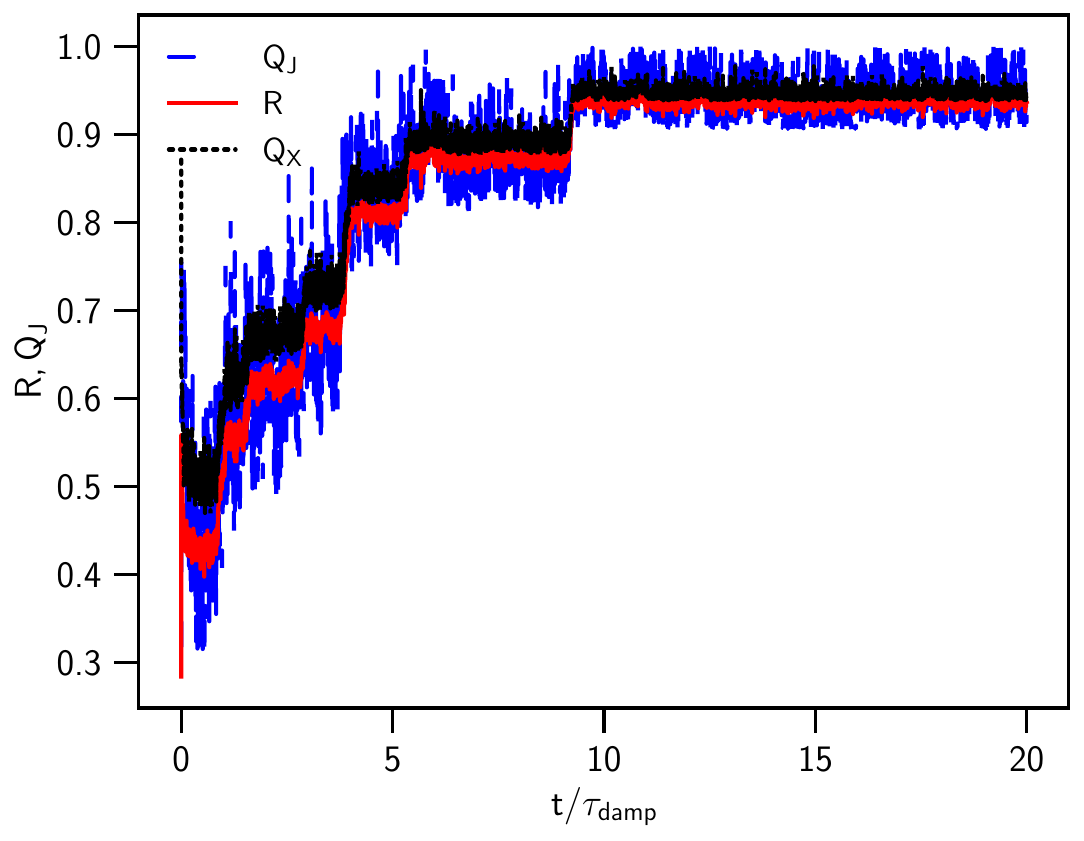}
		\put(40,20){\small \textbf{(b)} $U/(n_{1}T_{2}) = 0.6$}
\end{overpic}
\begin{overpic}[width=0.32\textwidth]{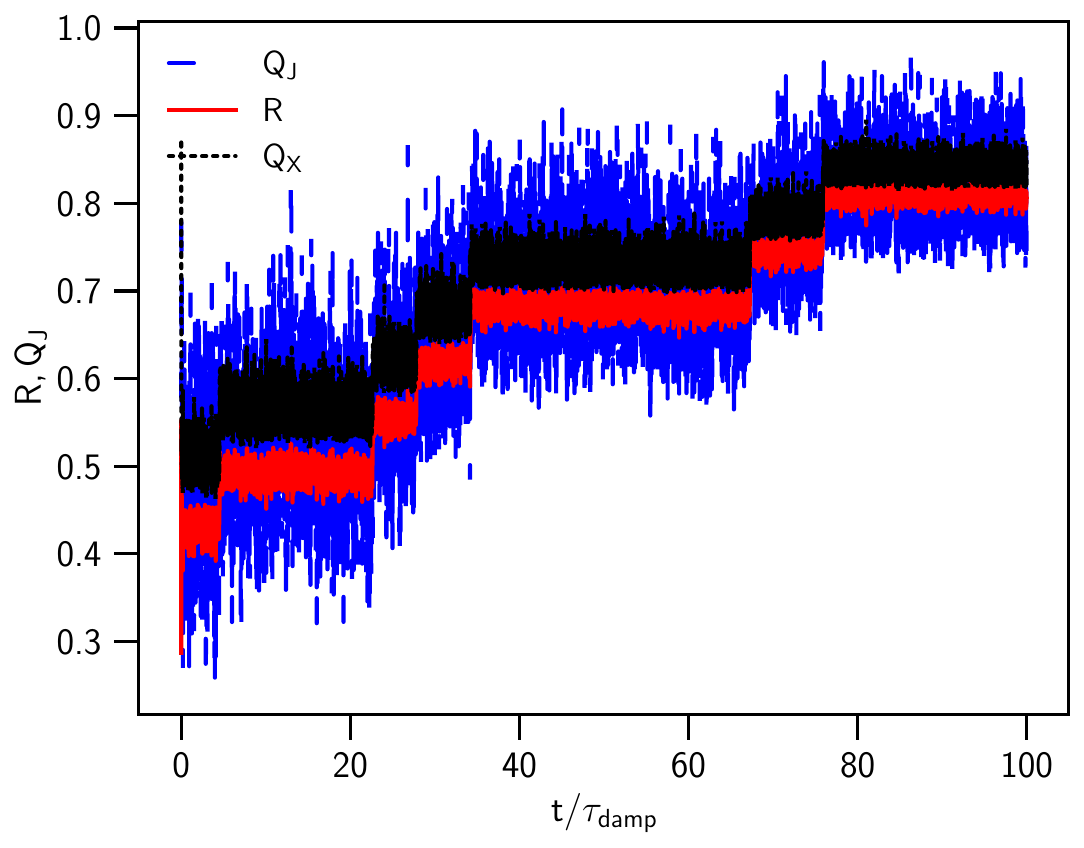}
		\put(40,20){\small \textbf{(c)} $U/(n_{1}T_{2}) = 1$}
\end{overpic}
\caption{Same as Figure \ref{fig:CDR_map_qmax2_nH1e1_aef02} but for the time-dependent alignment degrees $Q_{X}, Q_{J}$ and $R$. The alignment degrees increase with time due to the transport of grains to high-J attractors by gas and magnetic excitations and achieves perfect alignment of $R\sim 1$ after $\sim 10\tau_{\rm damp}$ for $U/(n_{1}T_{2})=0.16,0.6$ (panels (a),(b)) and near perfect alignment of $R\sim 0.8$ after $T_{\rm PA}\sim70\tau_{\rm damp}$ (panel (c)).}
\label{fig:CDR_RQJ_qmax2_nH1e1_aef02}
\end{figure*}

Figure \ref{fig:CDR_map_qmax2_nH1e1_aef02} shows the phase trajectory maps of grain alignment (i.e., alignment map) obtained from numerical simulations for the CD regime described by $U/(n_{1}T_{2})=0.16,0.6, 1$. {Initially, grains have the same angular momentum magnitude of $J=3J_{T}=3I_{1}\omega_{T}$ but different orientations with the magnetic field. Due to the effect of RATs and magnetic relaxation, a fraction of grains is rapidly driven to the high-J attractor (red circle), and the majority of grains is first driven to the low-J rotation where the grain orientation is strongly disturbed by gas collisions, as marked by gray arrows}. Subsequently, random excitations by gas collisions ($B_{\rm coll}$) and magnetic fluctuations ($B_{\rm mag}$) occasionally drive grains from low-J into high-J attractors, and all grains eventually reach high-J attractors with the perfect alignment. This effect was previously reported in \cite{HoangLaz.2008,HoangLaz.2016} for the standard diffuse ISM with $n_{\rm H}=30\cm^{-3},U=1$ or $U/(n_{1}T_{2})=0.33$. {In the phase trajectory map, we also show the threshold for RATD for grains of different tensile strengths of $S_{\max,7}=0.01,0.1,1$ by dotted horizontal lines to represent the RATD effect. For the CD regime, these RATD thresholds are not seen in this figure, implying RATD does not occur in the CD regime.}

Figure \ref{fig:CDR_RQJ_qmax2_nH1e1_aef02} shows the time-dependent alignment degrees calculated from numerical simulations for the model with the phase map shown in Figure \ref{fig:CDR_map_qmax2_nH1e1_aef02}. For the first two cases (panels (a) and (b)), the alignment degrees increase rapidly with time and achieve perfect alignment (PA) after $T_{\rm PA}\sim 10\tau_{\rm damp}\sim 10\tau_{\rm gas}$. However, for the later case of $U/(n_{1}T_{2})=1$, it requires a time of $\sim 70\tau_{\rm damp}$ to achieve significant alignment of $R=0.8$.

\begin{table*}[h]
	\centering
	\caption{Model parameters, the damping and perfect alignment timescales for $q^{\rm max}=2$ and $a_{\rm eff}=0.2\mum$.}
	\begin{tabular}{l l l l l l}
		$n_{\rm H}(\cm^{-3})$ & $U$ & $U/(n_{1}T_{2})$ & $\delta_{\rm mag}$  & $\tau_{\rm damp}(\yr)$ & $T_{\rm PA}(\tau_{\rm damp})$\cr
		\hline\hline\cr
		$10$ & 0.1 & 0.16 & $10^{3}$  & $4.3\times 10^{5}n_{1}^{-1}$ & $10$\cr
		$-$ & 0.5 & 0.6  & $-$ &  $3.1\times 10^{5}n_{1}^{-1}$ & $10$\cr
		$-$ & 1 & 1.0 & $-$ &  $2.5\times 10^{5}n_{1}^{-1}$ & $70$\cr
		\hline\\
		$10$ & 0.1 & 0.16 & $1$  & $4.3\times 10^{5}n_{1}^{-1}$ & $50$\cr
		$-$ & 0.5 & 0.6 &$-$  & $3.1\times 10^{5}n_{1}^{-1}$ & $70$\cr
		$-$ & 1 & 1.0 & $-$ &  $2.5\times 10^{5}n_{1}^{-1}$ & $80$\cr
		\hline
		$10$ & 2 & 1.9 & $10^{3}$ &  $2.9\times 10^{5}$ & NA \cr
		$-$ & 5 & 4.1 & $-$ &  $1.6\times 10^{5}$ & NA \cr
		$-$ & $10^{1}$ & 7.3 &$-$ & $9.9\times 10^{4}$ & NA \cr
		$-$ & $10^{2}$ & 49.5 &$-$ & $2.1\times 10^{4}$ & NA \cr
		$-$ & $10^{3}$ & 337 &$-$ & $4.6\times 10^{3}$ & NA \cr
		$-$ & $10^{4}$ & 2296 &$-$ & $9.9\times 10^{2}$ &   NA \cr
		$-$ & $10^{5}$ & 15640 &$-$ & $2.1\times 10^{2}$ &  NA \cr
		\hline\hline
	\end{tabular}
	\label{tab:models_qmax2}
\end{table*}

\subsection{Radiation-Dominated (RD) Regime: $U/(n_{1}T_{2})> 1$}
\begin{figure*}
\centering
\begin{overpic}[width=0.32\linewidth]{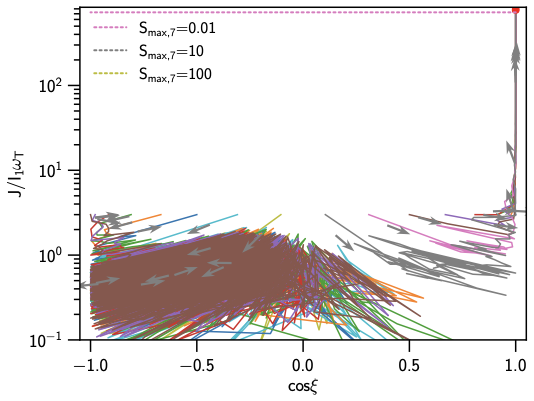}
		\put(65,75){\tiny \textbf{High-J Attractor}}
		\put(20,35){\tiny RAT Trapping}
\put(45,60){\small \textbf{(a)} $U/(n_{1}T_{2}) = 1.9$}
\end{overpic}
\begin{overpic}[width=0.32\linewidth]{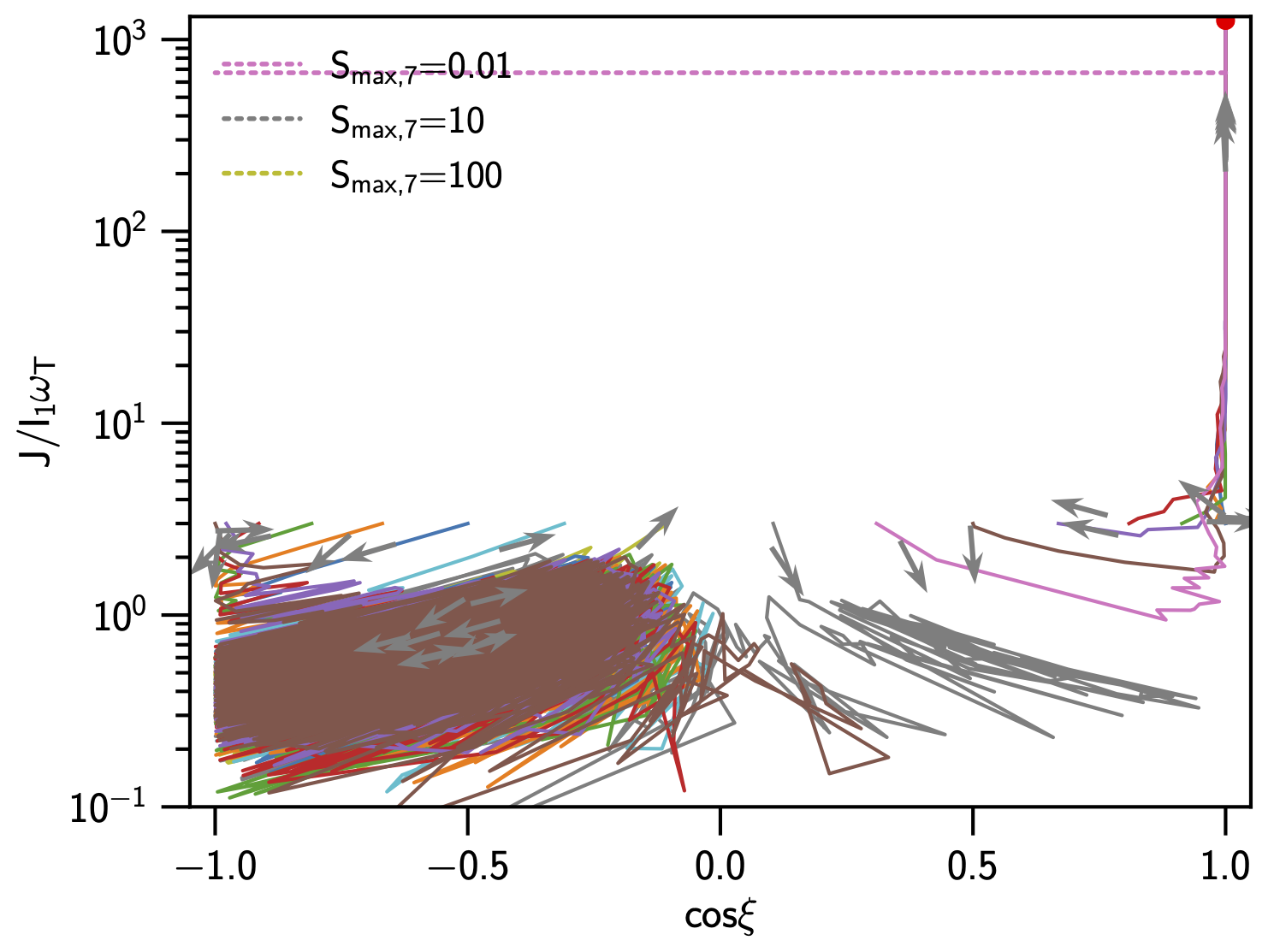}
		\put(65,75){\tiny \textbf{High-J Attractor}}
		\put(20,35){\tiny RAT Trapping}
\put(45,60){\small \textbf{(b)} $U/(n_{1}T_{2}) = 4.1$}
\end{overpic}
\begin{overpic}[width=0.32\linewidth]{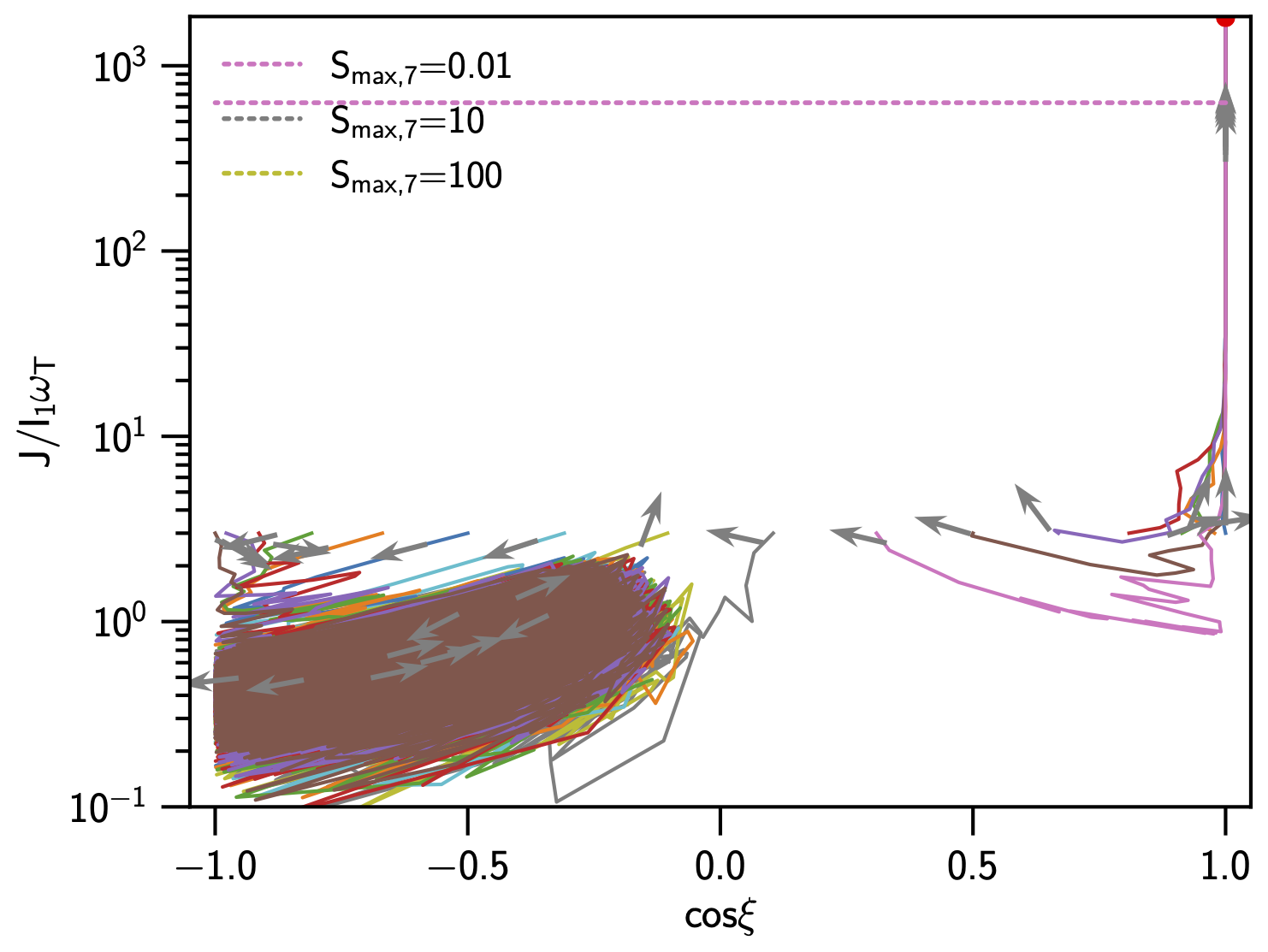}
		\put(65,75){\tiny \textbf{High-J Attractor}}
\put(45,60){\small \textbf{(c)} $U/(n_{1}T_{2}) = 7.3$}
\put(20,35){\tiny RAT Trapping}
\end{overpic}
\caption{Same as Figure \ref{fig:CDR_map_qmax2_nH1e1_aef02} but for the radiation-dominated regime with different $U/(n_{1}T_{2})$ (panels (a)-(c)). A fraction of grains are rapidly driven to the high-J attractor, and some grains are driven to the low-J rotation and trapped (confined) there by strong RATs. The confinement area is smaller for higher $U/(n_{1}T_{2})$. Collisional and magnetic excitations strongly disturb grain alignment but cannot scatter them into high-J rotation. Rotational disruption can occur for weak grains of $S_{\rm max,7}=0.01$ (pink dotted line), even when grains are not yet reaching a high-J attractor.}
\label{fig:RDR_map_qmax2_nH1e1_aef02}
\end{figure*}

Here we show the numerical results for the radiation-dominated regime of $U/(n_{1}T_{2})> 1$. Figure \ref{fig:RDR_map_qmax2_nH1e1_aef02} shows the phase trajectory maps of grain alignment for $U/(n_{1}T_{2})=1.9,4.1,7.3$ (panels (a)-(c)). As shown, a fraction of grains is rapidly driven to high-J attractors. The majority of grains is driven to low-J rotation during which collisional and magnetic excitations significantly disturb grain orientation. However, these excitation processes cannot transport grains to high-J attractors, and grains are radiatively trapped at the low-J rotation state. This is different from the CD regime observed in Figure \ref{fig:CDR_map_qmax2_nH1e1_aef02}. The trapping of grains at low-J rotation in this RD regime is the new effect, which we term RAdiative Torque Trapping (hereafter RATT).

The physics of the RATT effect can be understood using the alignment function of RATs as shown in Figure \ref{fig:FHmean} (panel (b)). There are two stationary points at $\cos\xi=\pm 1$ and $\cos\xi\sim \pm 0.6$ for $J'=0.63$. The stationary point $\cos\xi=1,J'\gg 1$ is a high-J attractor point (circle A), and $\cos\xi=-0.6,J'<1$ is a low-J attractor (circle B) produced by RATs (see Figure \ref{fig:FHmean}, panel (b)). When radiation field is strong, the low-J attractor is stabilized by RATs against gas random collisions, and grains are trapped at low-J attractors. However, for an average or weak radiation field, random collisions and magnetic fluctuations are dominant over RATs and can scatter grains out of this low-J attractor, slowly transporting them to the high-J attractor.

\begin{figure*}
\centering
\begin{overpic}[width=0.32\textwidth]{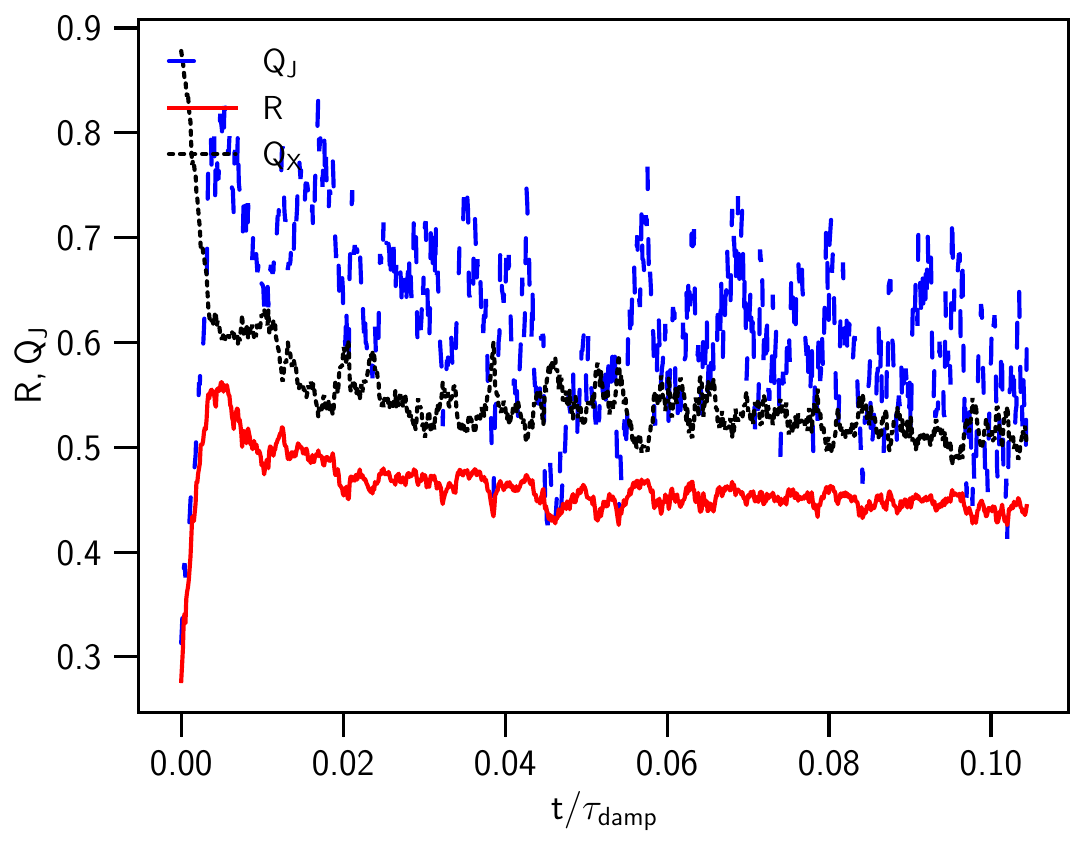}
\put(50,15){\small \textbf{(a)} $U/(n_{1}T_{2}) = 1.9$}
\end{overpic}
\begin{overpic}[width=0.32\textwidth]{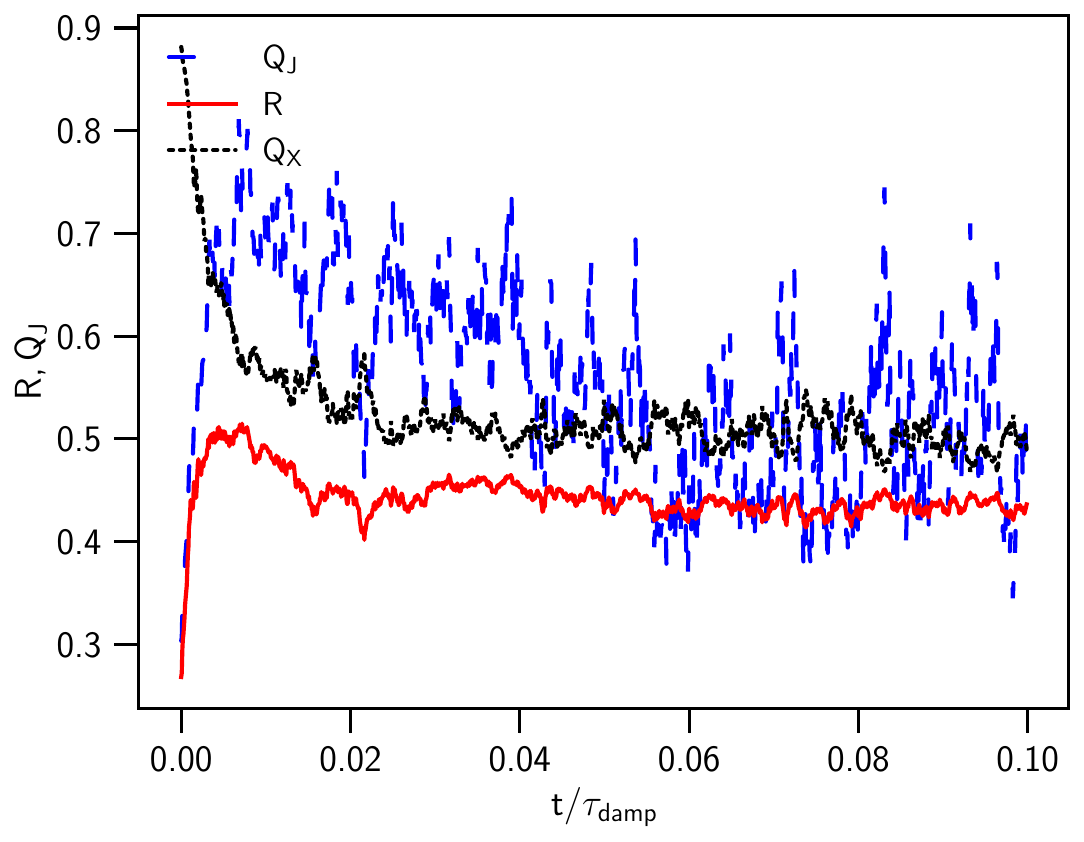}
\put(50,15){\small \textbf{(b)} $U/(n_{1}T_{2}) = 4.1$}
\end{overpic}
\begin{overpic}[width=0.32\textwidth]{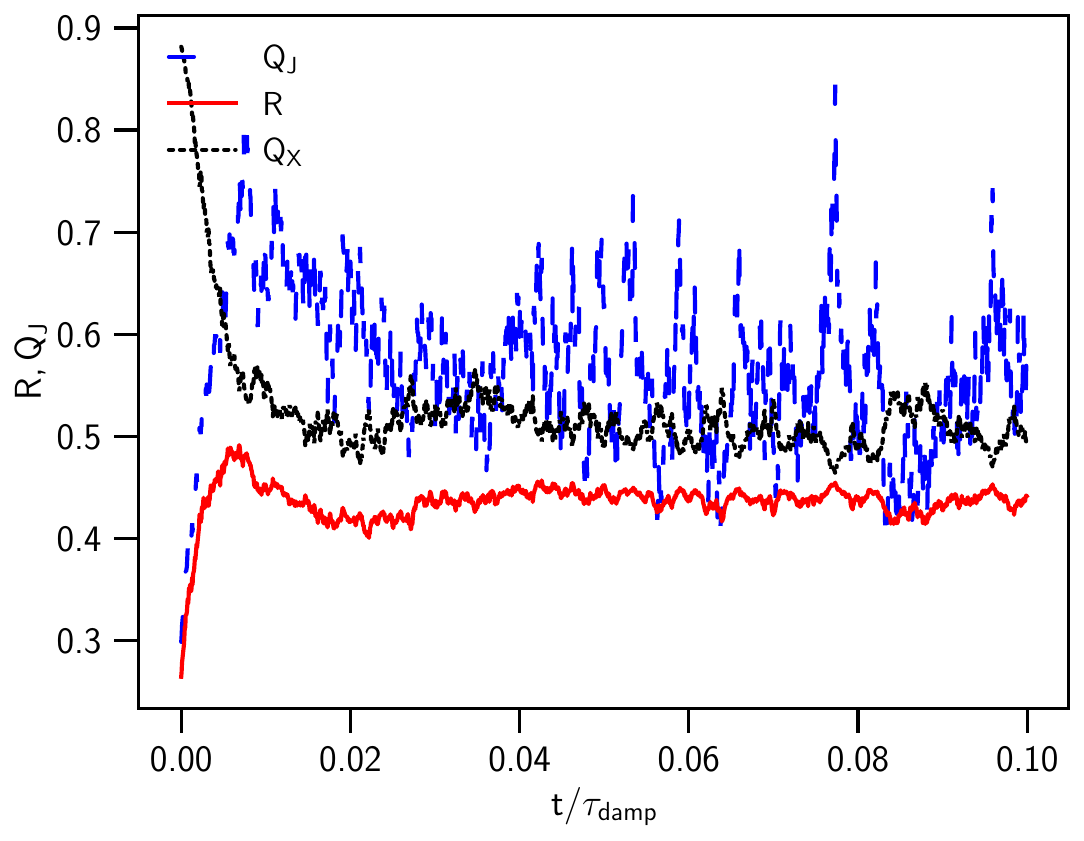}
\put(50,15){\small \textbf{(c)} $U/(n_{1}T_{2}) = 7.3$}
\end{overpic}
\caption{Same as Figure \ref{fig:RDR_map_qmax2_nH1e1_aef02} but for the time-dependent alignment degrees. Within a time $t\sim 0.01\tau_{\rm damp}$, the external and net alignment degrees ($Q_{J}, R$) increase rapidly due to external alignment by MRATs, but the internal alignment degree $Q_{X}$ decreases due to the damping of the grain angular momentum. Afterward, the alignment degrees decrease and slowly change after $t\sim 0.02\tau_{\rm damp}\sim 0.02\tau_{\rm gas}$ due to radiative torque trapping, although the external alignment is still fluctuating due to gas and magnetic excitation.}
\label{fig:RDR_RQJ_qmax2_nH1e1_aef02}
\end{figure*}

Figure \ref{fig:RDR_RQJ_qmax2_nH1e1_aef02} shows the time-dependent alignment degree calculated from numerical simulations for the model with the phase maps shown in Figure \ref{fig:RDR_map_qmax2_nH1e1_aef02}. The alignment degrees increase rapidly within $0.01-0.02\tau_{\rm damp}$ due to the fast alignment of grains to high-J attractors, and then they get saturated due to the RATT effect.

\subsection{Effects of Magnetic Relaxation and Excitations}
To study the effect of magnetic relaxation and excitations on the RAT alignment, we now show the results for the PM grains with $\delta_{\rm mag}=1$ to compare with the results for SPM grains in the previous section. We consider two different RAT models of $q^{\rm max}=2$ and $q^{\rm max}=1$ which have high-J attractor (former) and no high-J attractor (latter) as shown in Figure \ref{fig:deltacri_lowJ_highJ}.

\begin{figure*}
\centering
\begin{overpic}[width=0.32\linewidth]{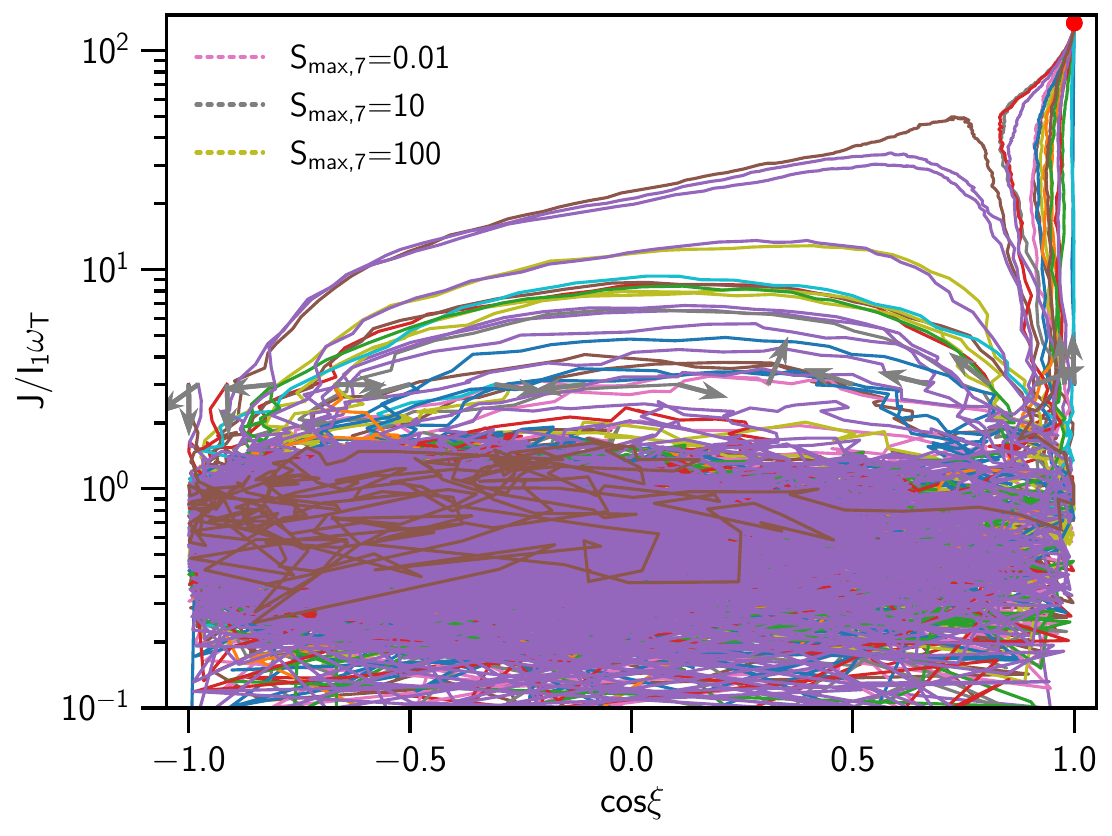}
			\put(65,75){\tiny \textbf{High-J Attractor}}
\put(15,75){\small \textbf{(a)} $U/(n_{1}T_{2}) = 0.16$}
\end{overpic}
\begin{overpic}[width=0.32\linewidth]{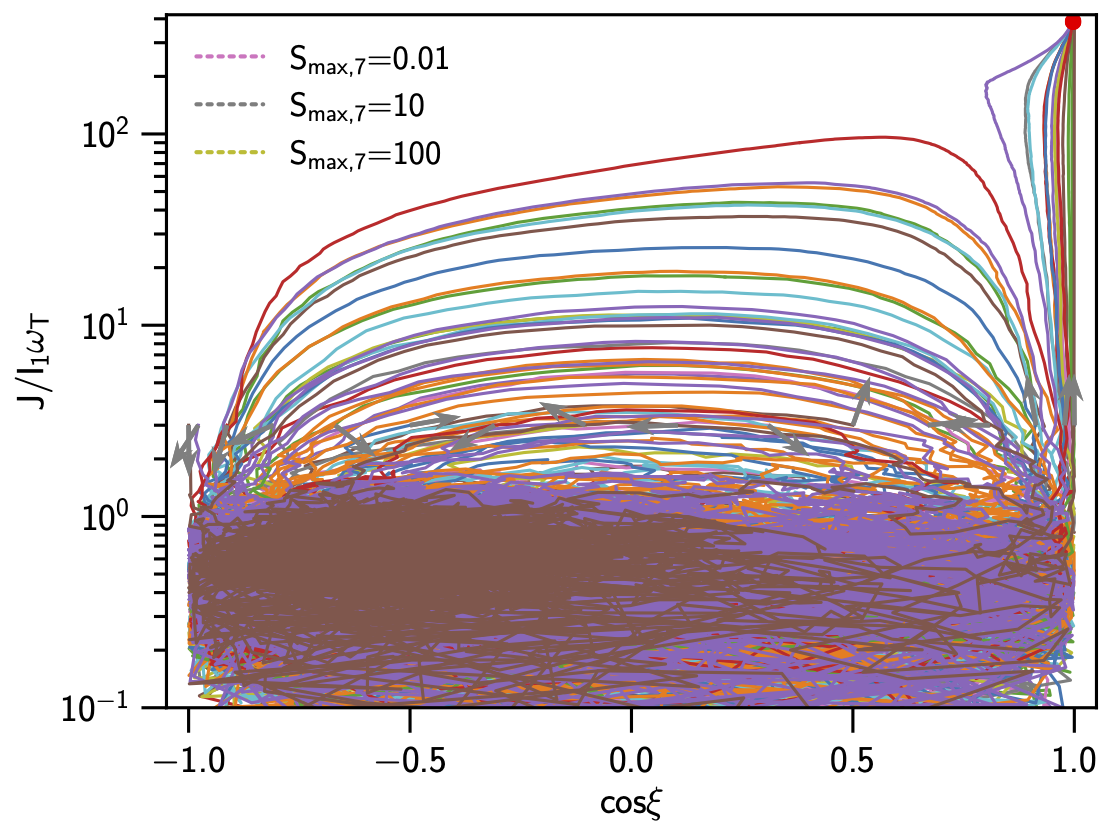}
			\put(65,75){\tiny \textbf{High-J Attractor}}
\put(15,75){\small \textbf{(b)} $U/(n_{1}T_{2}) = 0.6$}
\end{overpic}	
\begin{overpic}[width=0.32\linewidth]{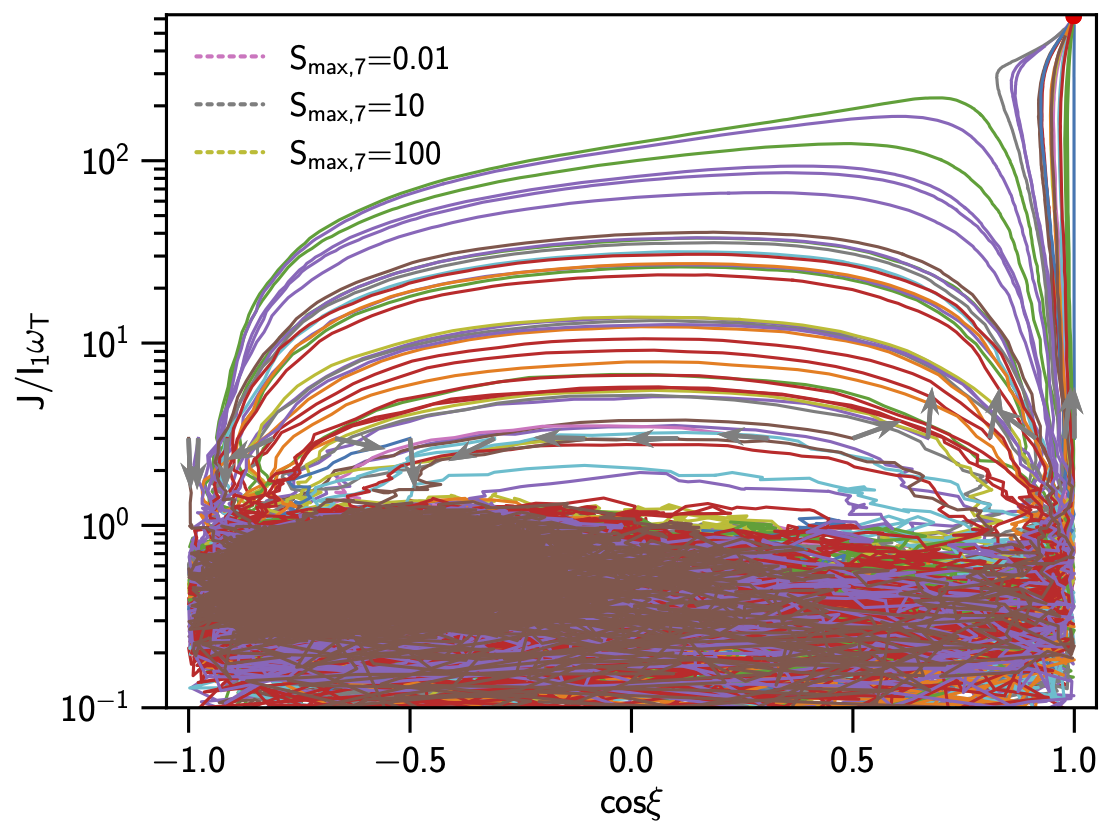} 
			\put(65,75){\tiny \textbf{High-J Attractor}}
\put(15,75){\small \textbf{(c)} $U/(n_{1}T_{2}) = 1$}
\end{overpic}       
\caption{Same as Figure \ref{fig:CDR_map_qmax2_nH1e1_aef02} but for PM grains with $\delta_{\rm mag}=1$. The phase trajectory during the low-J rotation is strongly disturbed by collisional and magnetic excitations, but the trajectory for suprathermal rotation of $J/I_{1}\omega_{T}>3$ is deterministic due to reduced magnetic relaxation. All grains eventually reach the high-J attractor (red circle) by MRATs due to gas and magnetic excitations.}
\label{fig:CDR_map_qmax2_nH1e1_aef02_deltam1}
\end{figure*}

\begin{figure*}
\centering
\begin{overpic}[width=0.32\textwidth]{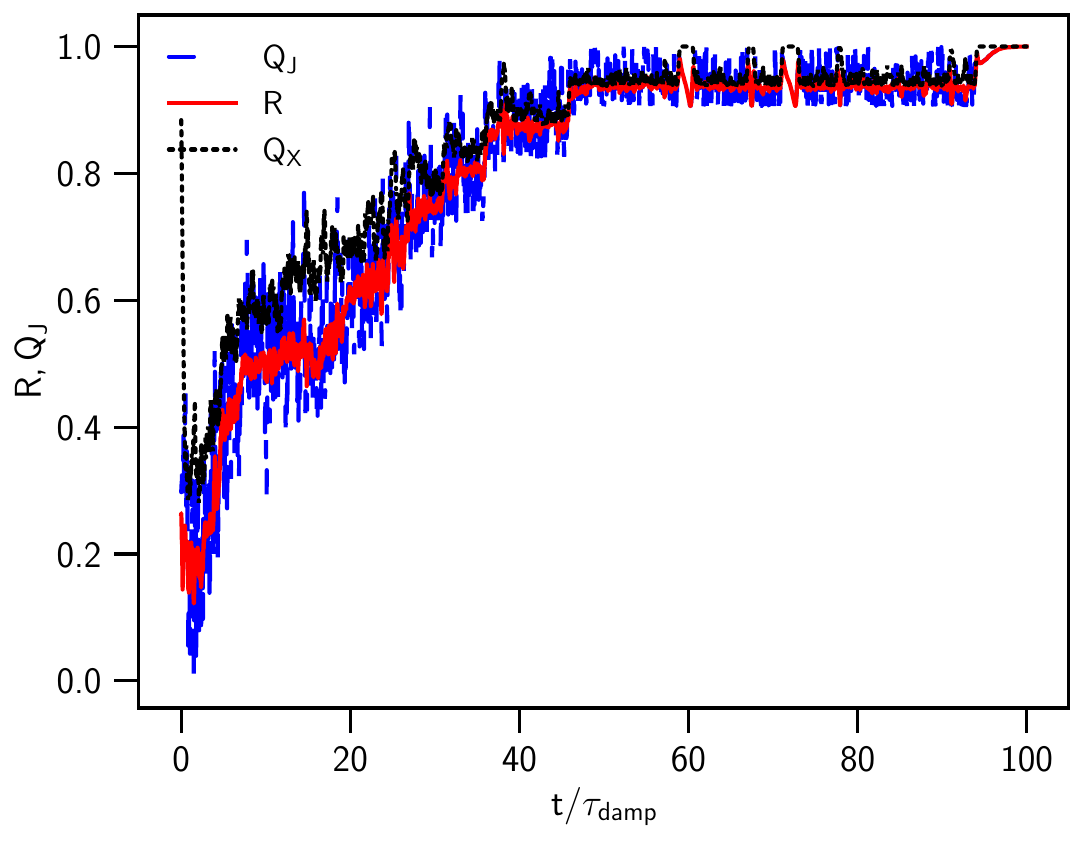}
\put(45,20){\small \textbf{(a)} $U/(n_{1}T_{2}) = 0.16$}
\end{overpic}	
\begin{overpic}[width=0.32\textwidth]{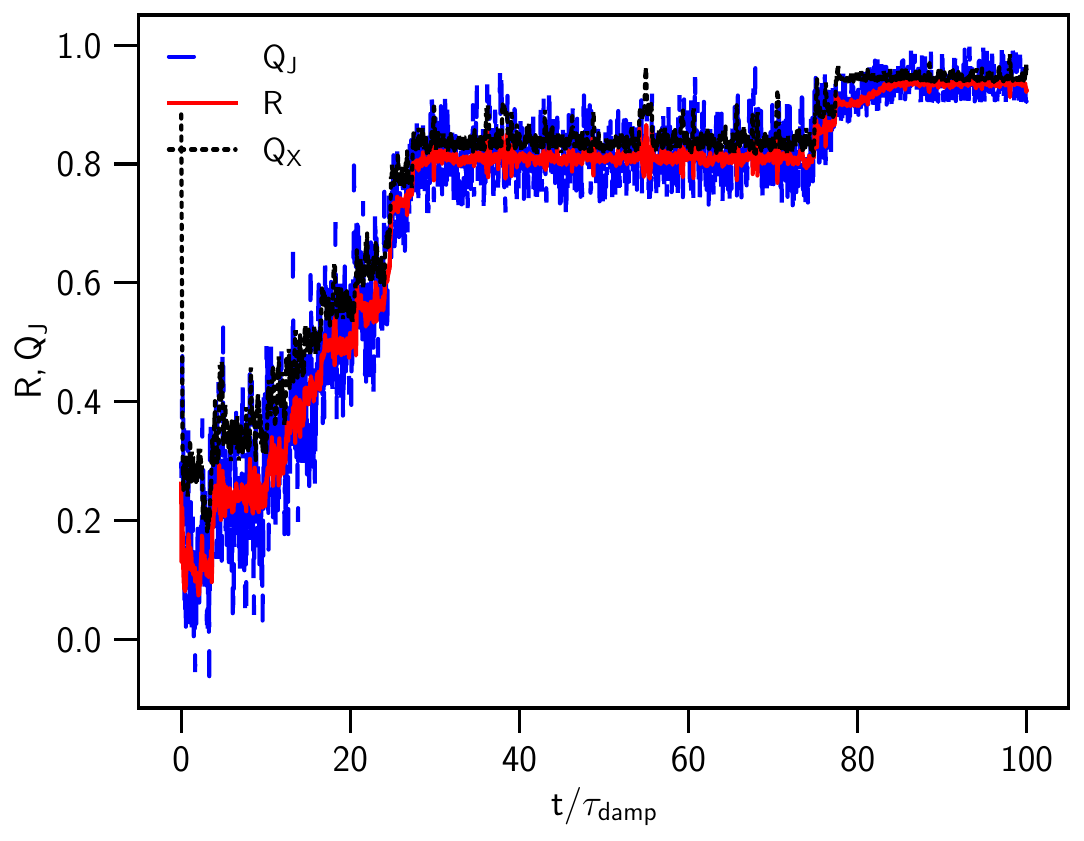}
\put(45,20){\small \textbf{(b)} $U/(n_{1}T_{2}) = 0.6$}
\end{overpic}	
\begin{overpic}[width=0.32\textwidth]{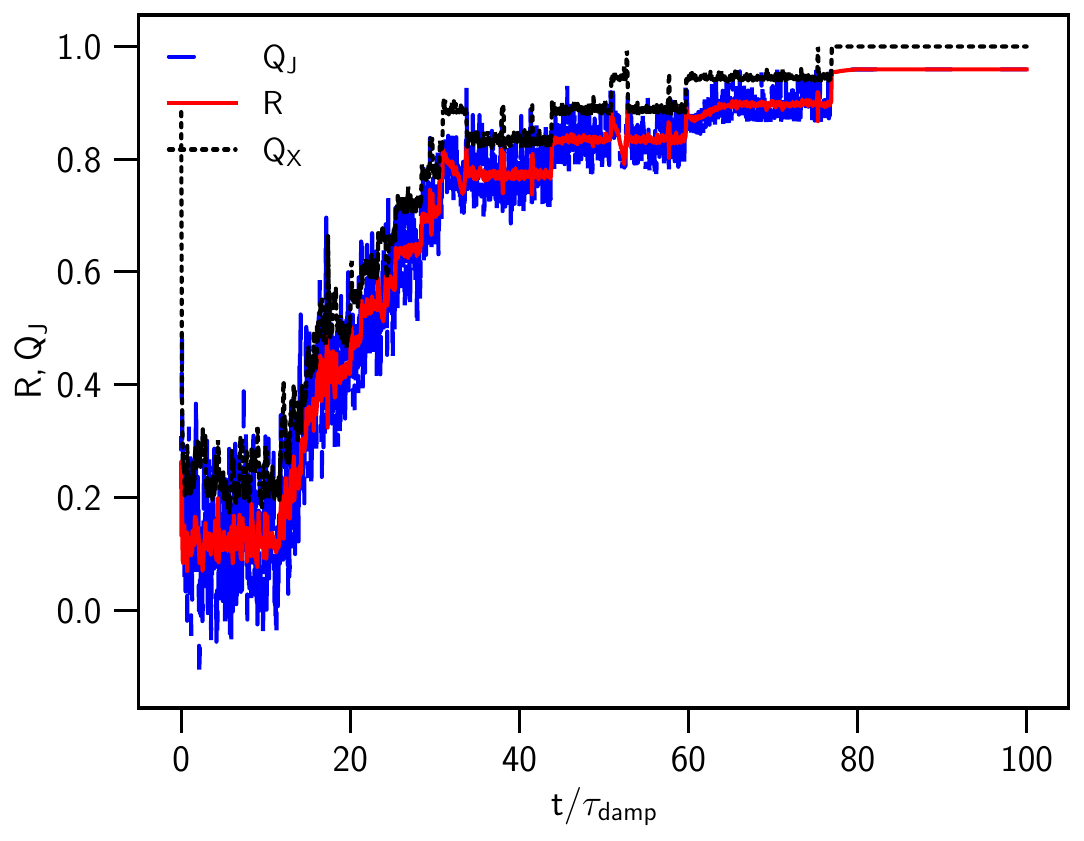}
\put(45,20){\small \textbf{(c)} $U/(n_{1}T_{2}) = 1$}
\end{overpic}	
\caption{Same as Figure \ref{fig:CDR_RQJ_qmax2_nH1e1_aef02} but for PM grains with $\delta_{\rm mag}=1$. The alignment degrees increase with time, but the time required to reach nearly perfect alignment is longer with $T_{\rm PA}\sim 40-70\tau_{\rm damp}$ due to the reduced magnetic excitations.}
\label{fig:CDR_RQJ_qmax2_nH1e1_aef02_deltam1}
\end{figure*}

Figures \ref{fig:CDR_map_qmax2_nH1e1_aef02_deltam1} and \ref{fig:CDR_RQJ_qmax2_nH1e1_aef02_deltam1} show the alignment map and time-dependent alignment degrees for the RAT model with a high-J attractor described by $q^{\rm max}=2$ and $\delta_{\rm mag}=1$ in the CD regime. Collisional and magnetic excitations are sufficient to significantly scatter grain orientations to transport them to high-J attractors compared to the CD regime. For the alignment degrees, one can see that the timescale for reaching the terminal alignment is longer for this model of $\delta_{\rm mag}=1$ is longer than that for SPM grains with $\delta_{\rm mag}=10^{3}$ shown in Figure \ref{fig:CDR_RQJ_qmax2_nH1e1_aef02}. This effect can arises from strong superparamagnetic excitations that enhances the scattering of grains out of the low-J rotation to high-J attractors.

\begin{figure*}
	\centering
	\begin{overpic}[width=0.32\linewidth]{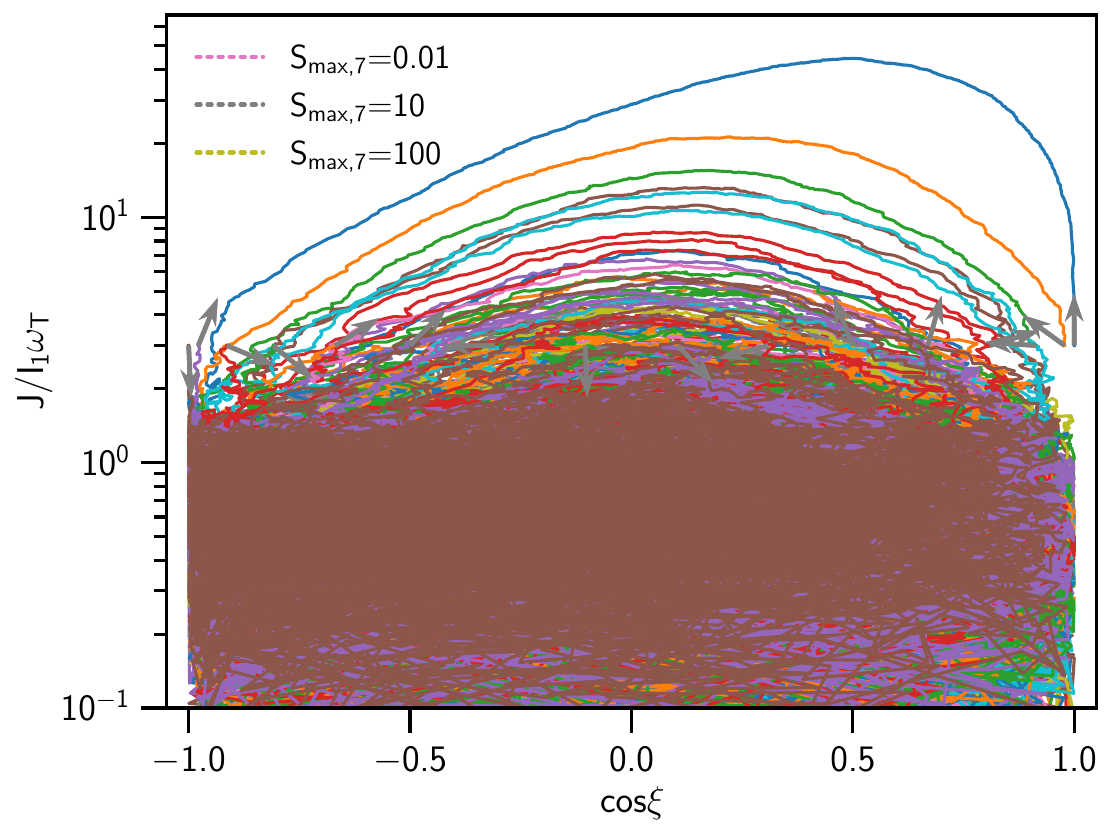}
		\put(15,75){\small \textbf{(a)} $U/(n_{1}T_{2}) = 0.16$}
		\end{overpic}	
	\begin{overpic}[width=0.32\linewidth]{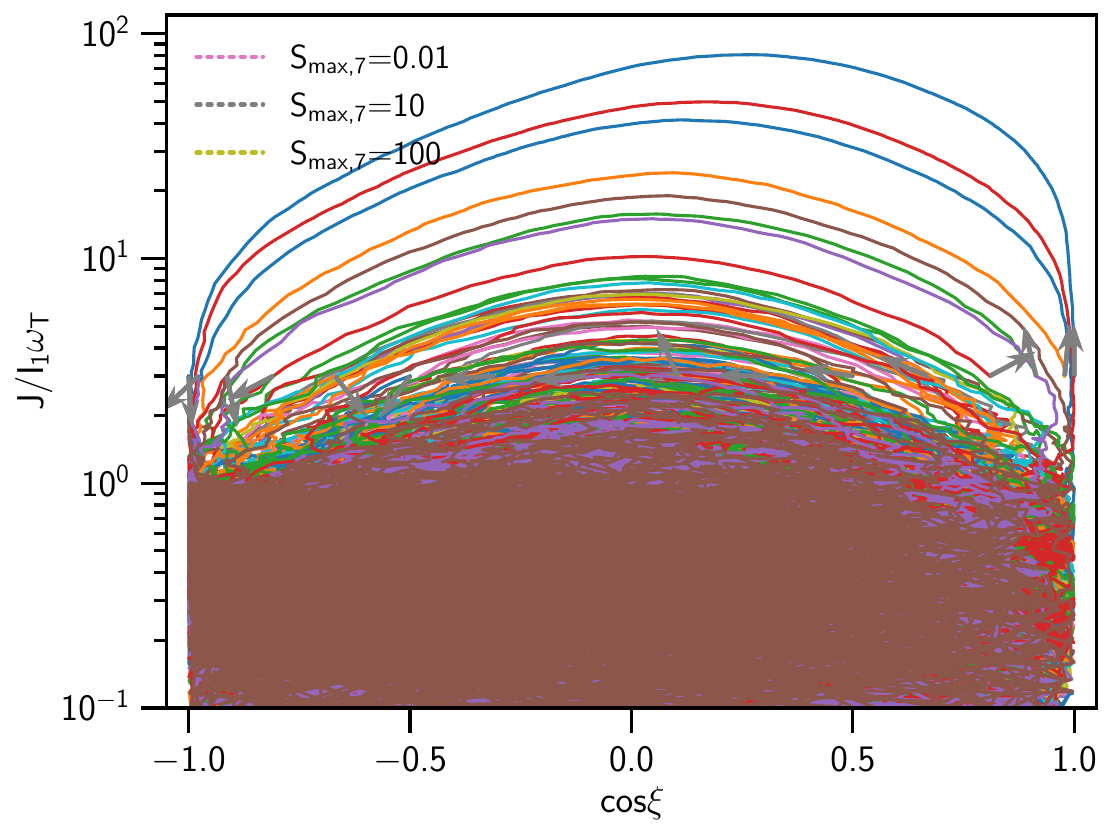}      
		\put(15,75){\small \textbf{(b)} $U/(n_{1}T_{2}) = 0.6$}
		\end{overpic}	
	\begin{overpic}[width=0.32\linewidth]{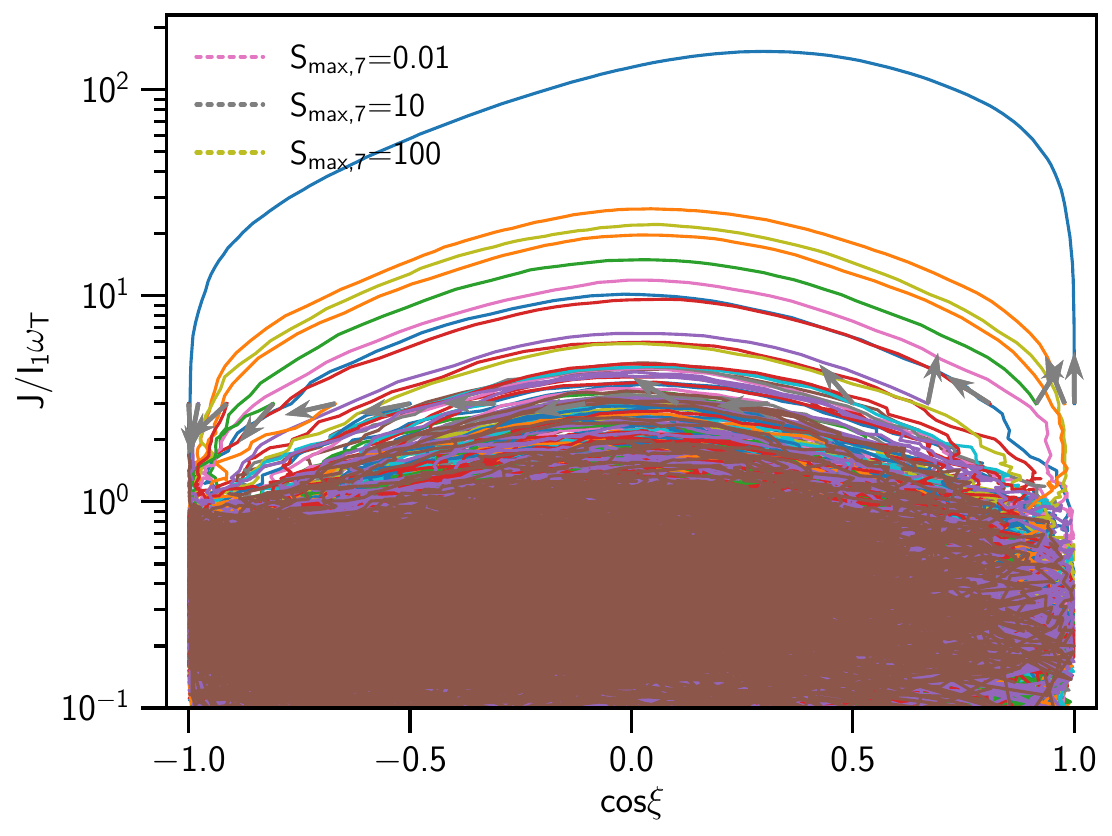}        
		\put(15,75){\small \textbf{(c)} $U/(n_{1}T_{2}) = 1$}
		\end{overpic}	
	\caption{Same as Figure \ref{fig:CDR_map_qmax2_nH1e1_aef02} but for PM grains with $\delta_{\rm mag}=1$ and the RAT model without high-J attractor of $q^{\rm max}=1$. The phase trajectory during the low-J rotation is significantly disturbed by collisional and magnetic excitations. All grains have cyclic trajectories during the suprathermal state of $J/I_{1}\omega_{T}>3$ and randomly during $J/I_{1}\omega_{T}<3$ due to gas and magnetic excitations.}
	\label{fig:CDR_map_qmax1_nH1e1_aef02_deltam1}
\end{figure*}

\begin{figure*}	
	\centering
	\begin{overpic}[width=0.32\textwidth]{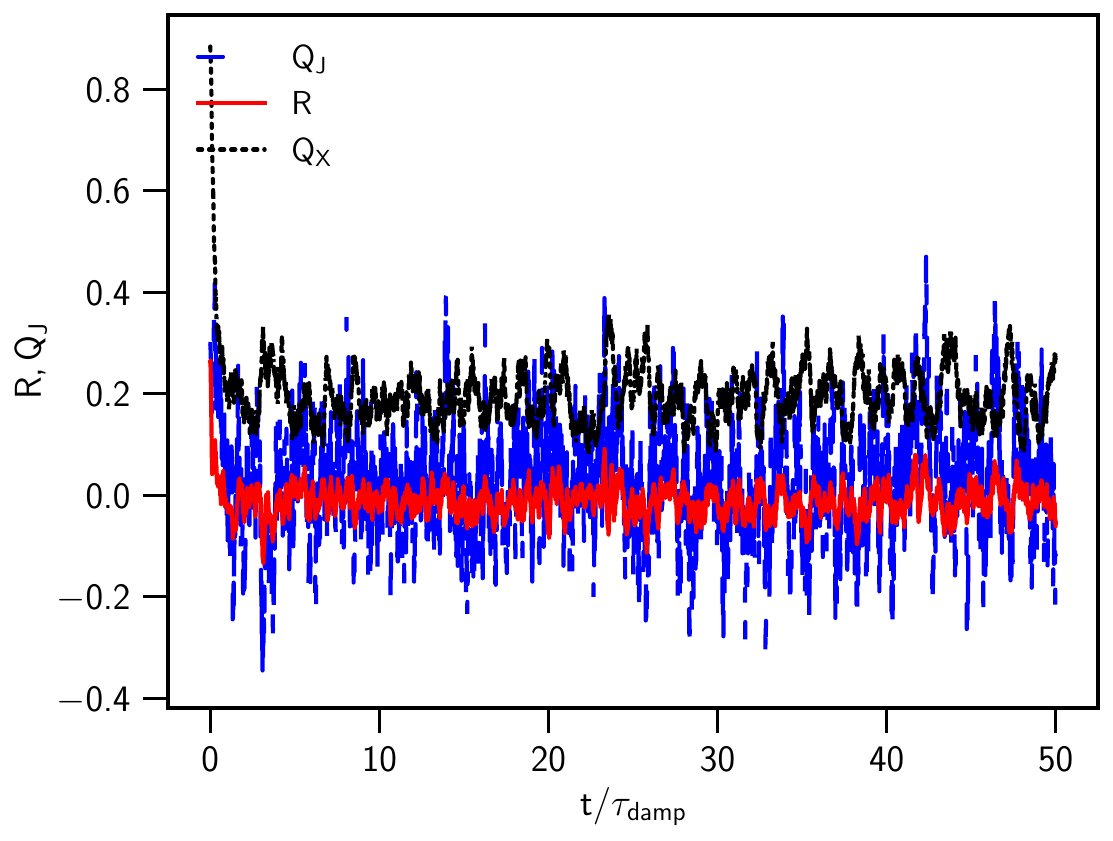}
	\put(45,65){\small \textbf{(a)} $U/(n_{1}T_{2}) = 0.16$}
	\end{overpic}	
	\begin{overpic}[width=0.32\textwidth]{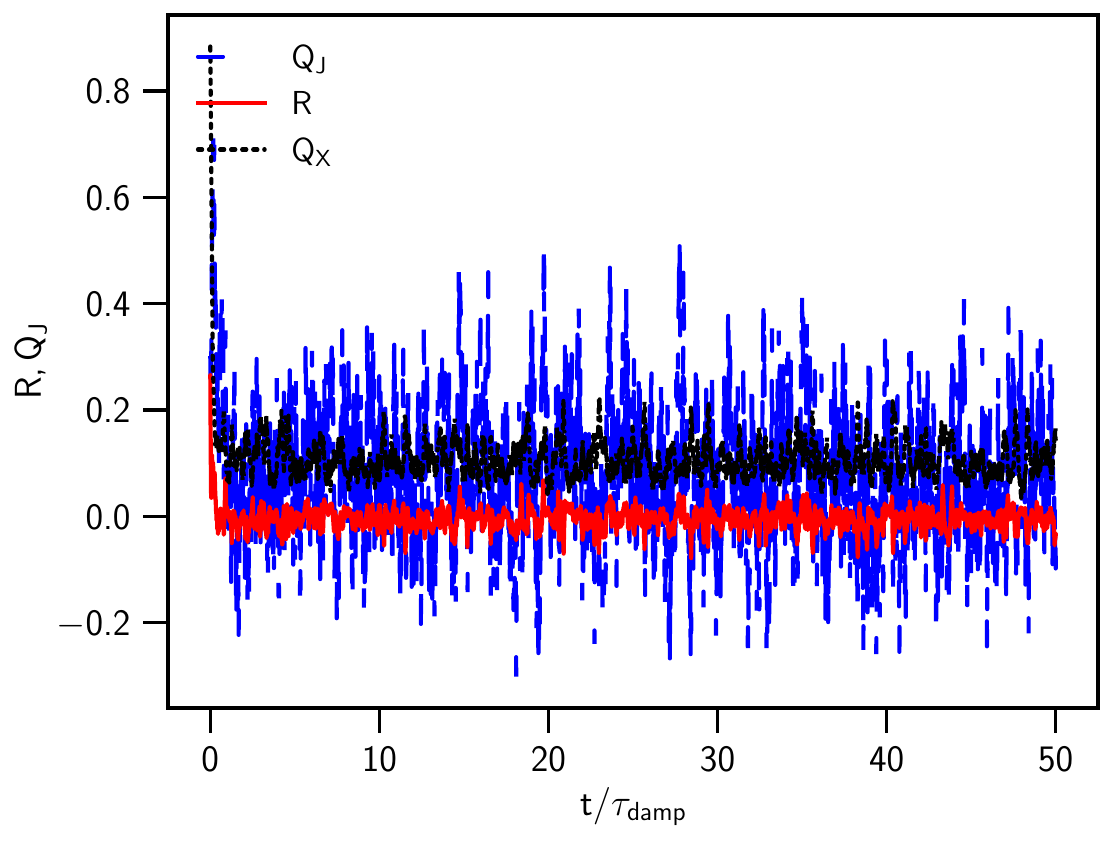}
	\put(45,65){\small \textbf{(b)} $U/(n_{1}T_{2}) = 0.6$}
	\end{overpic}	
	\begin{overpic}[width=0.32\textwidth]{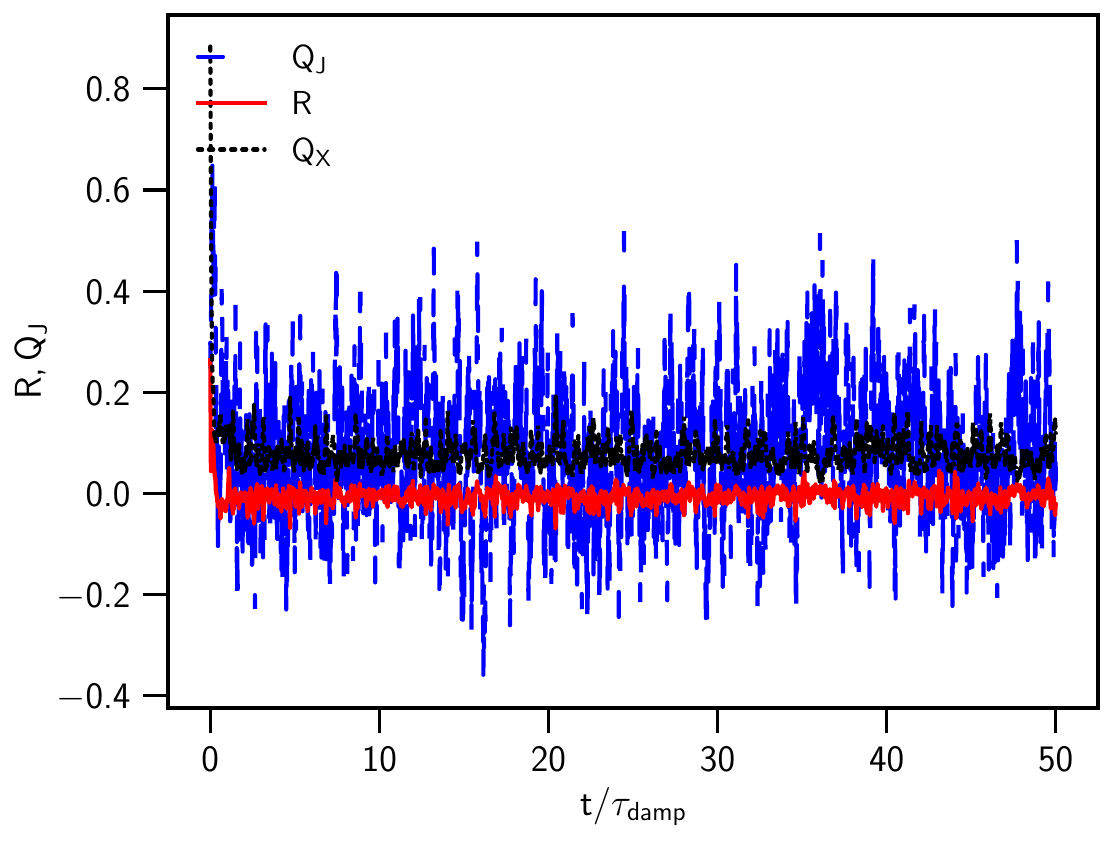}
	\put(45,65){\small \textbf{(c)} $U/(n_{1}T_{2}) = 1$}
	\end{overpic}	
	\caption{Same as Figure \ref{fig:CDR_RQJ_qmax2_nH1e1_aef02} but for PM grains with $\delta_{\rm mag}=1$ and $q^{\rm max}=1$. The net alignment degree $R$ is zero due to collisional randomization for the RAT model without high-J attractors, but the internal alignment degree is still finite with $Q_{X}\sim 0.2$.}
	\label{fig:CDR_RQJ_qmax1_nH1e1_aef02_deltam1}
\end{figure*}

Figures \ref{fig:CDR_map_qmax1_nH1e1_aef02_deltam1} and \ref{fig:CDR_RQJ_qmax1_nH1e1_aef02_deltam1} show the alignment map and time-dependent alignemnt degrees for the RAT model of $q^{\rm max}=1$ and PM grains of $\delta_{\rm mag}=1$ (without high-J attractors) in the CD regime. Collisional and magnetic excitations completely randomize grain alignment in the low-J rotation due to the lack of high-J attractors. For alignment degrees, one can see that the external alignment degree is $Q_{J}\sim 0$ and the net alignment of $R\sim 0$ due to strong gas randomization. Note that the degree of internal alignment ($Q_{X}$) is still considerable of $Q_{X}\sim 0.2$ due to the fast internal relaxation (by Barnett relaxation and inelastic relaxation), but $Q_{X}$ is lower for higher $U/(n_{1}T_{2})$ due to stronger RATT. This confirms our previous results on requiring high-J attractors for considerable alignment.

\subsection{Extreme radiation fields and reduced efficiency of magnetic relaxation on the RAT alignment}
Here we run the simulations for extreme radiation with $U=10^{2}-10^{5}$ and $q^{\rm max}=2$, assuming SPM grains of $\delta_{\rm mag}=10^{3}$. We note that these extreme radiation regimes with warm/hot dust are quite abundant in astrophysics, including circumstellar regions, AGN torus, planetary systems, and supernova ejecta.

\begin{figure*}
	\centering
	\begin{overpic}[width=0.45\linewidth]{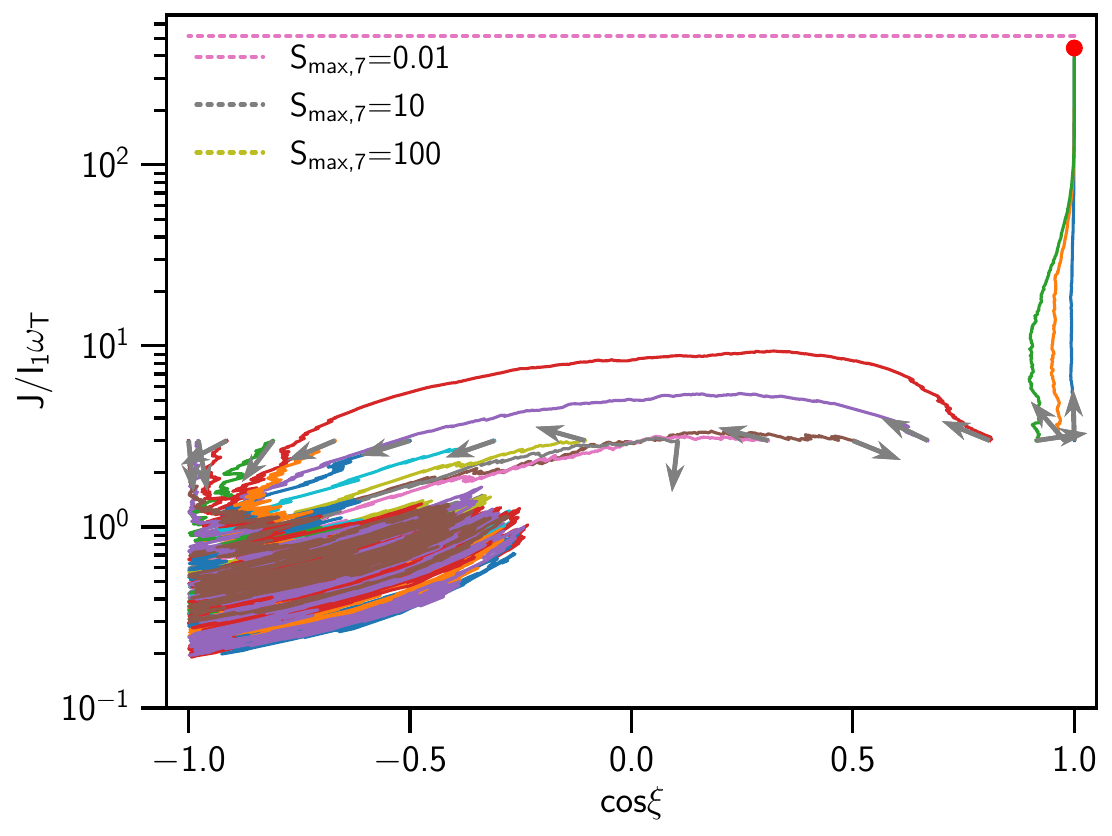}
			\put(75,75){\tiny \textbf{High-J Attractor}}
			\put(20,14){\tiny \textbf{RAT Trapping}}
		    \put(55,20){\small \textbf{(a)} $U/(n_{1}T_{2}) = 49.5$}
	\end{overpic}	
	\begin{overpic}[width=0.45\linewidth]{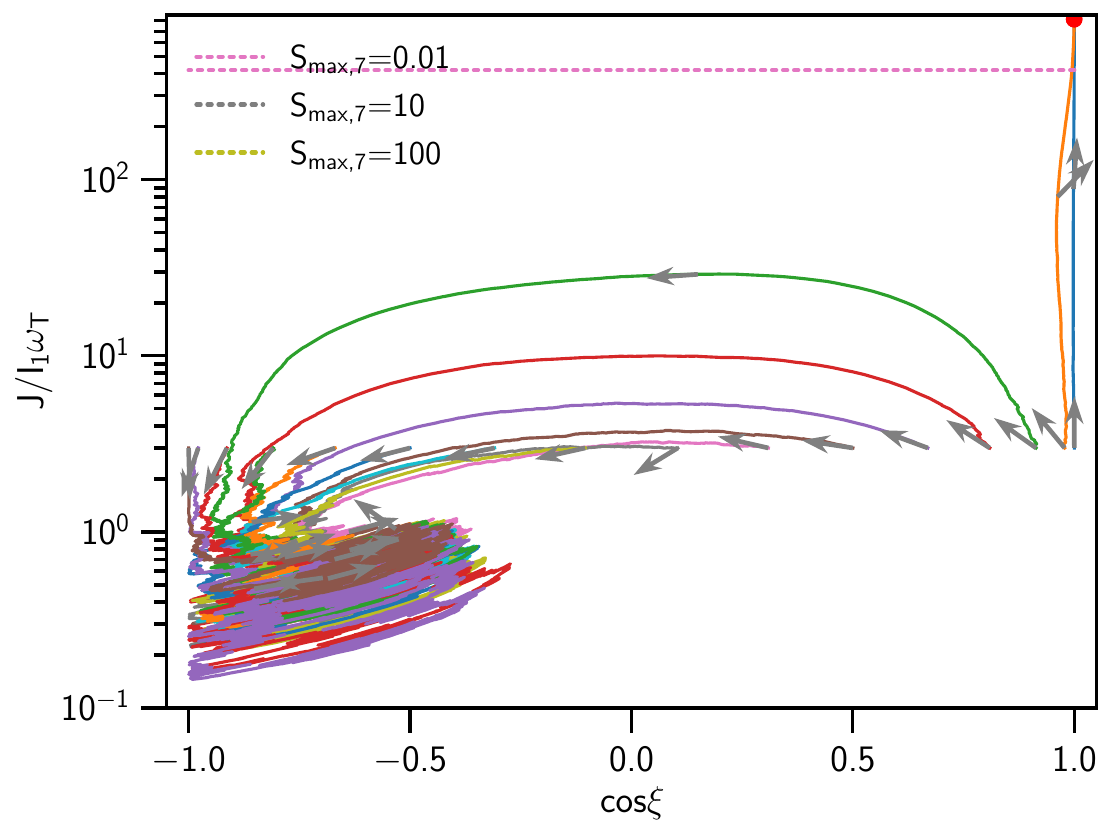}
			\put(75,75){\tiny \textbf{High-J Attractor}}
			\put(20,14){\tiny \textbf{RAT Trapping}}
			\put(55,20){\small \textbf{(b)} $U/(n_{1}T_{2}) = 337$}
		\end{overpic}	
	\begin{overpic}[width=0.45\linewidth]{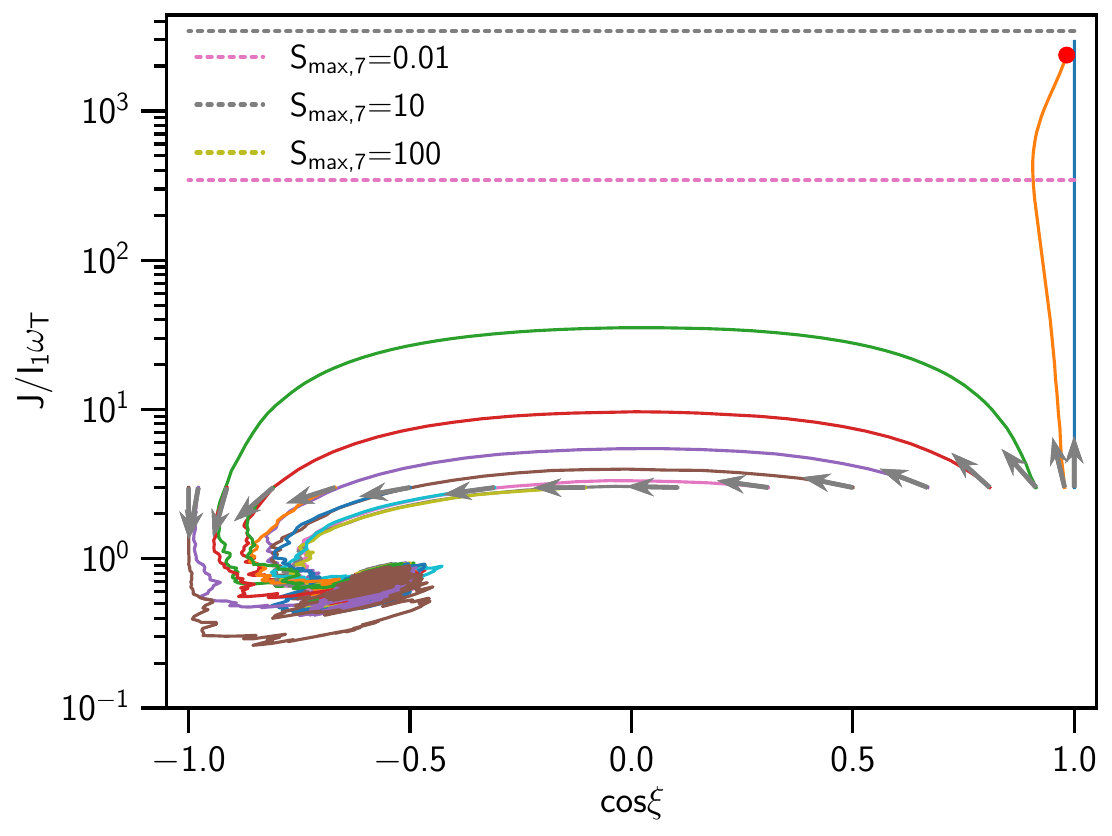} 
			\put(75,75){\tiny \textbf{High-J Attractor}}    
			\put(20,14){\tiny \textbf{RAT Trapping}}
			\put(55,20){\small \textbf{(c)} $U/(n_{1}T_{2}) = 2296$}
		\end{overpic}	
	\begin{overpic}[width=0.45\linewidth]{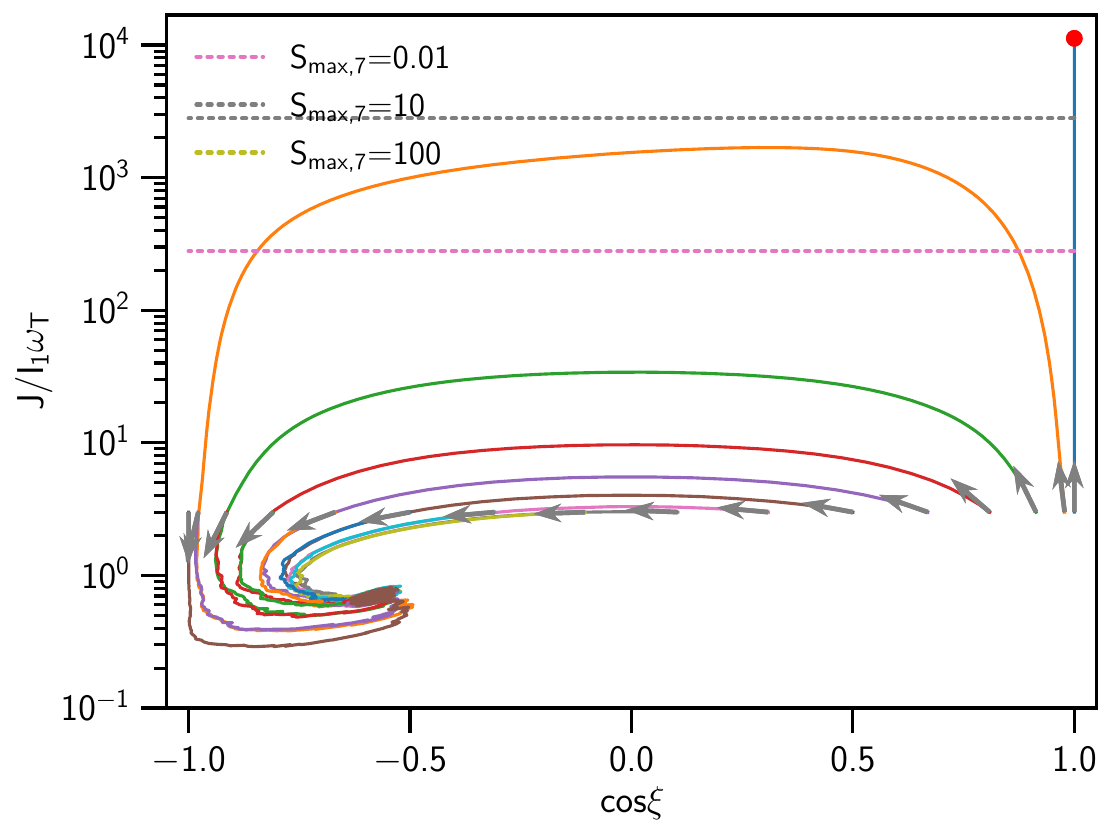} 
			\put(75,75){\tiny \textbf{High-J Attractor}}
			\put(20,14){\tiny \textbf{RAT Trapping}}
			\put(55,20){\small \textbf{(d)} $U/(n_{1}T_{2}) = 15640$}
		\end{overpic}	
	\caption{The phase trajectory map for the RAT model $q^{\rm max}=2$ and SPM grains with $\delta_{\rm mag}=10^{3}$ in extreme radiation fields with different $U/(n_{1}T_{2})$  (panels (a)-(d)). The fraction of grains with fast alignment at high-J attractor decreases with increasing $U/(n_{1}T_{2})$ due to reduced efficiency of magnetic relaxation. The trajectory of grains during low-J rotation is disturbed less, and the confinement area is narrower, for higher $U/(n_{1}T_{2})$ due to stronger RATT. Rotational disruption can occur even when grains are still on the way to the high-J attractor.}
	\label{fig:extreme_map_qmax2}
\end{figure*}

Figure \ref{fig:extreme_map_qmax2} shows the alignment maps for the extreme radiation regime with the different $U/(n_{1}T_{2})$. The fraction of grains with fast alignment at high-J attractor decreases with increasing $U/(n_{1}T_{2})$ due to reduced efficiency of magnetic relaxation on the RAT alignment, as predicted in Figure \ref{fig:deltacri_lowJ_highJ}. We note that for this model in the CD regime, all grains are aligned at high-J attractors, as shown in Figure \ref{fig:CDR_map_qmax2_nH1e1_aef02}. The trajectory of grains during low-J rotation is disturbed less for higher $U/(n_{1}T_{2})$ due to stronger RATT and reduction of magnetic excitations. One interesting result from Figure \ref{fig:extreme_map_qmax2} is that {some grains heading to high-J attractors can be disrupted by RATD, while the remaining grains that are trapped at low-J rotation can survive}.

\begin{figure*}	
\centering
	\begin{overpic}[width=0.45\textwidth]{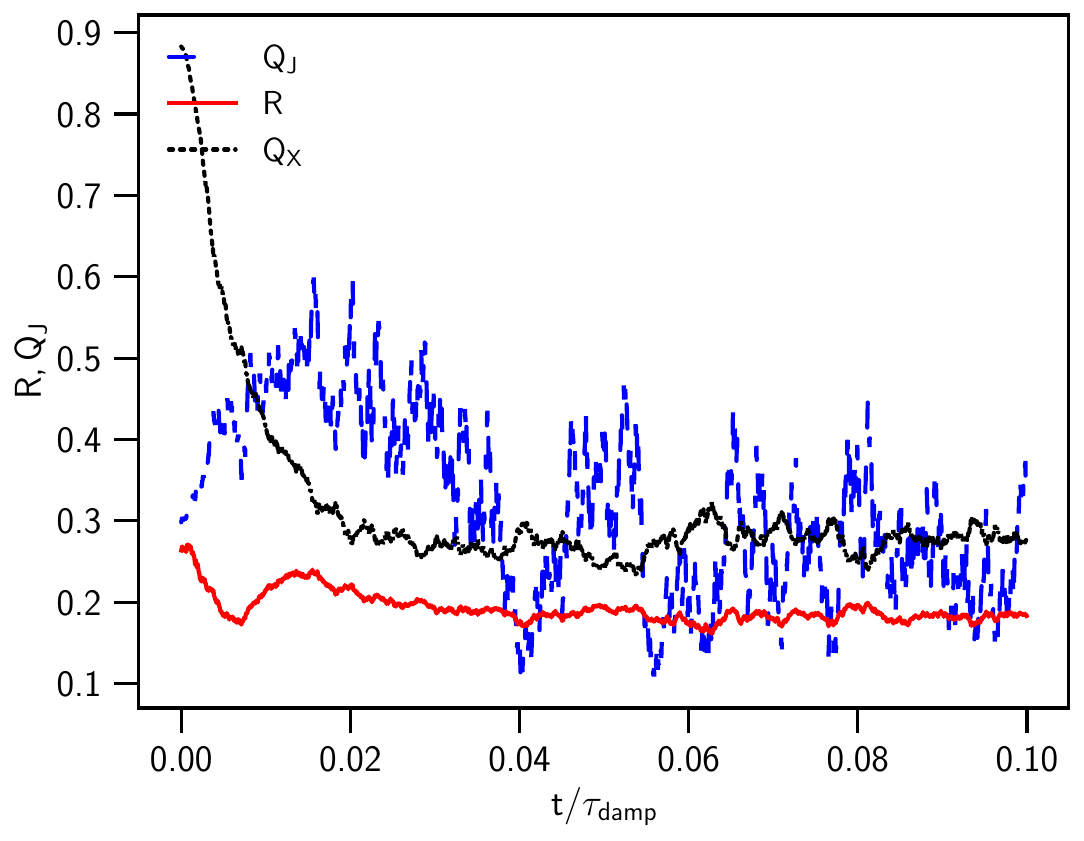}
			\put(45,65){\small \textbf{(a)} $U/(n_{1}T_{2}) = 49.5$}
	\end{overpic}	
\begin{overpic}[width=0.45\textwidth]{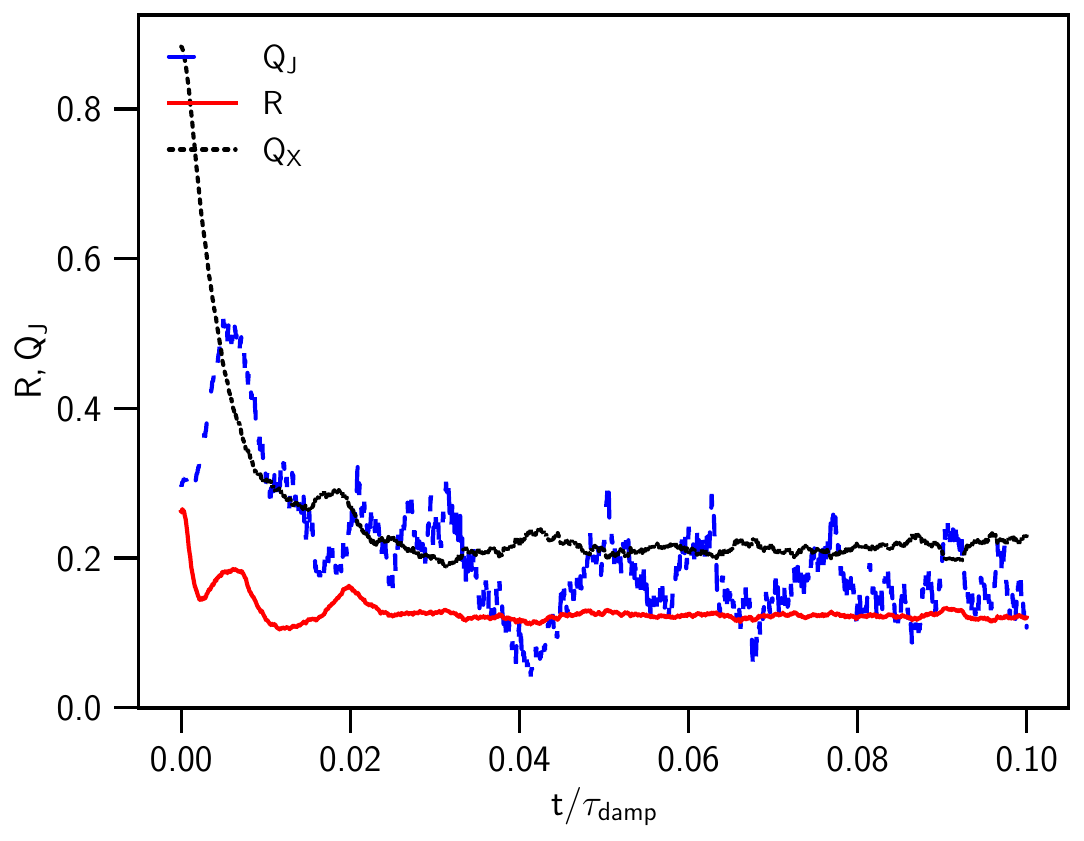}
			\put(45,65){\small \textbf{(b)} $U/(n_{1}T_{2}) = 337$}
	\end{overpic}	
\begin{overpic}[width=0.45\textwidth]{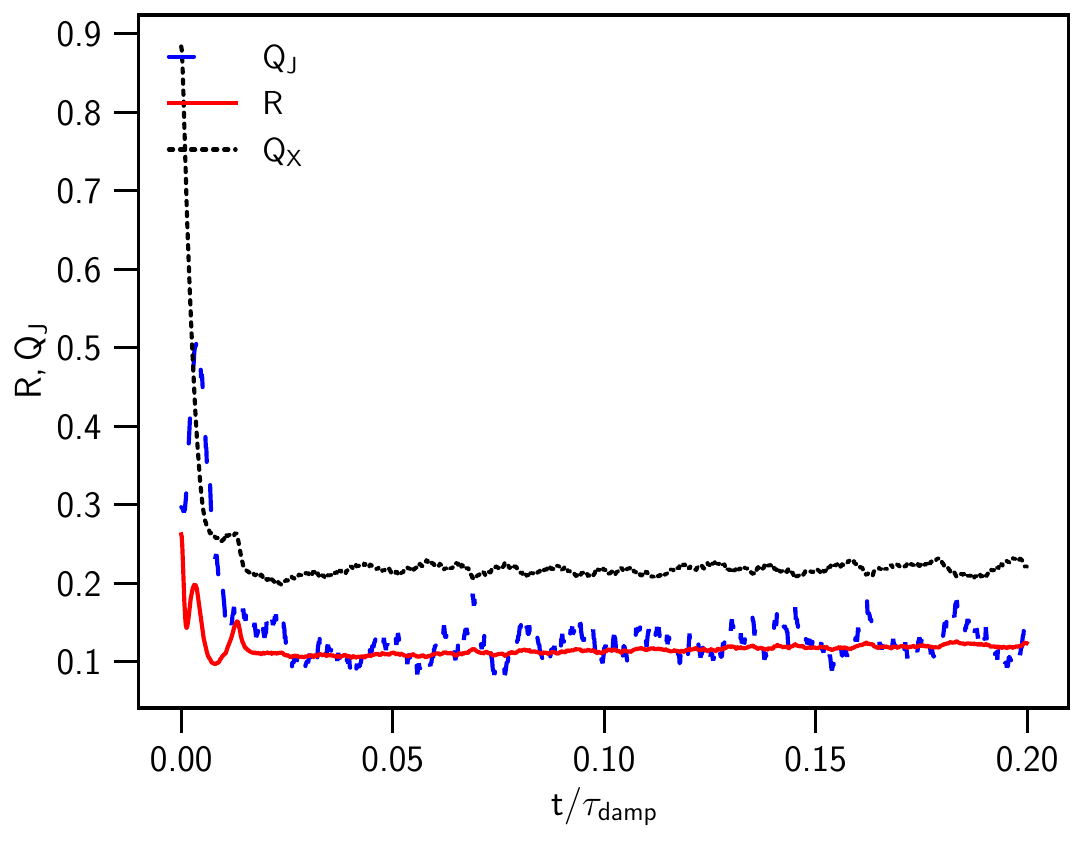}
			\put(45,65){\small \textbf{(c)} $U/(n_{1}T_{2}) = 2296$}
	\end{overpic}	
\begin{overpic}[width=0.45\textwidth]{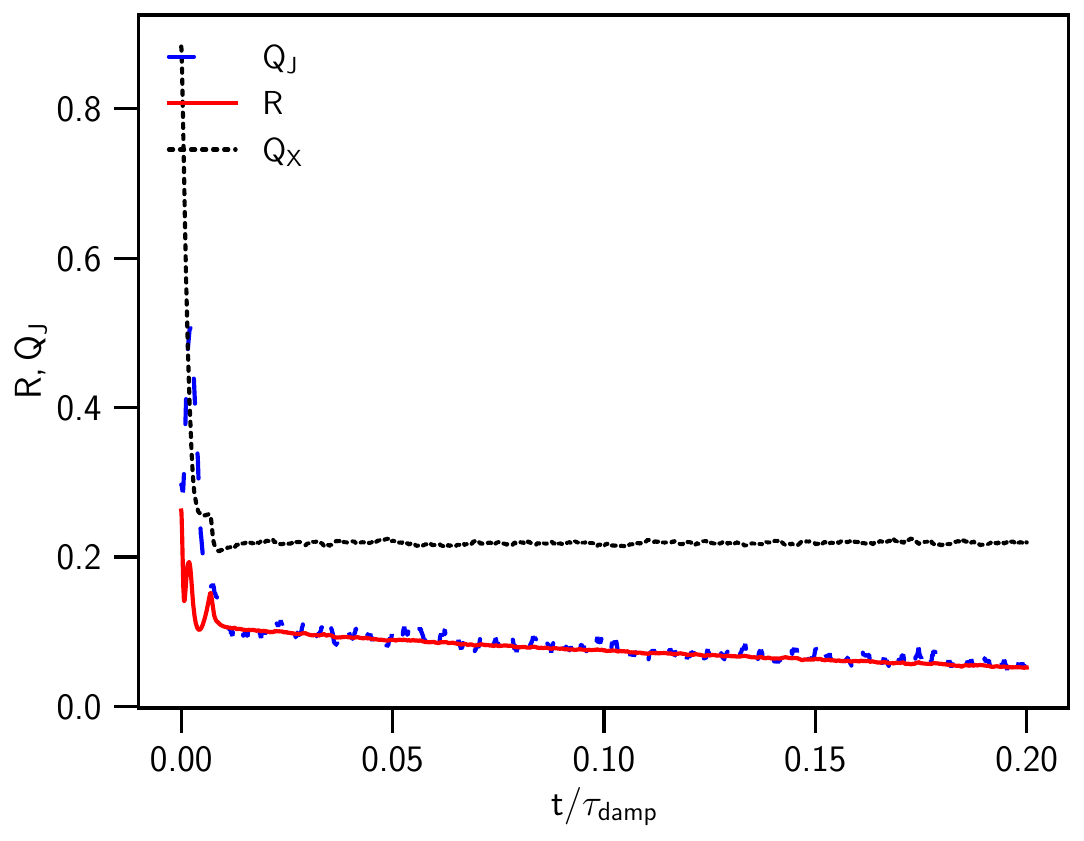}
			\put(45,65){\small \textbf{(d)} $U/(n_{1}T_{2}) = 15640$}
	\end{overpic}	
	\caption{Same as Figure \ref{fig:extreme_map_qmax2} but for for time-dependent alignment degrees. Grain alignment degrees decrease with increasing $U/(n_{1}T_{2})$ due to reduced efficiency of magnetic relaxation. Fluctuations of external alignment $Q_{J}$ (blue line) caused by gas and magnetic excitations at low-J rotation is much weaker for stronger radiation fields due to RATT effect.}
	\label{fig:extreme_RQJ_qmax2}
\end{figure*}

Figure \ref{fig:extreme_RQJ_qmax2} shows the time-dependent alignment degrees for the different $U/(n_{1}T_{2})$. The alignment degrees are much lower than in the RD regime shown in Figure \ref{fig:RDR_RQJ_qmax2_nH1e1_aef02}. Moreover, the grain alignment degrees decrease with increasing $U/(n_{1}T_{2})$ due to reduced efficiency of magnetic relaxation. For $U/(n_{1}T_{2})>10^{4}$, the alignment degree does not depend on the value of $\delta_{\rm mag}$ due to the suppression of magnetic relaxation and is purely determined by RATs. The fraction of grains aligned at high-J attractors depends only on $q^{\rm max}$. Collisional and magnetic excitations slightly scatter grain orientation out of the low-J attractor. We note that for this model in the CD regime, grains are perfectly aligned as shown in Figure \ref{fig:CDR_RQJ_qmax2_nH1e1_aef02}. 

We note that in the extreme radiation field, the alignment does not change significantly compared to the CD and RD regimes because only a fraction of grains with initial angle $\cos\xi\sim 1$ can be rapidly aligned with $\bB$. However, their angular momenta increase with time due to further spin-up by RATs over time. Therefore, the disruption can occur within $10^{-2}\tau_{\rm damp}$, which results in the rapid change in the grain size distribution and composition.

\subsection{The Fraction of Grains with Fast Alignment and Fast Disruption at High-J Attractors}
To determine the fraction of grains that can rapidly align at high-J attractors in less than one damping time, we run simulations for an integration time of $T_{\rm obs}=\tau_{\rm damp}/2$ for different RAT models, magnetic relaxation, and radiation strengths, assuming a ensemble of $N_{\rm gr}=32$ grains with initially random orientations. We consider $q^{\rm max}=0.5-4$, which is the most probable range of RATs from an ensemble of random shapes in \cite{Herranen:2019kj}

\begin{figure*}
\centering
\begin{overpic}[width=0.45\textwidth]{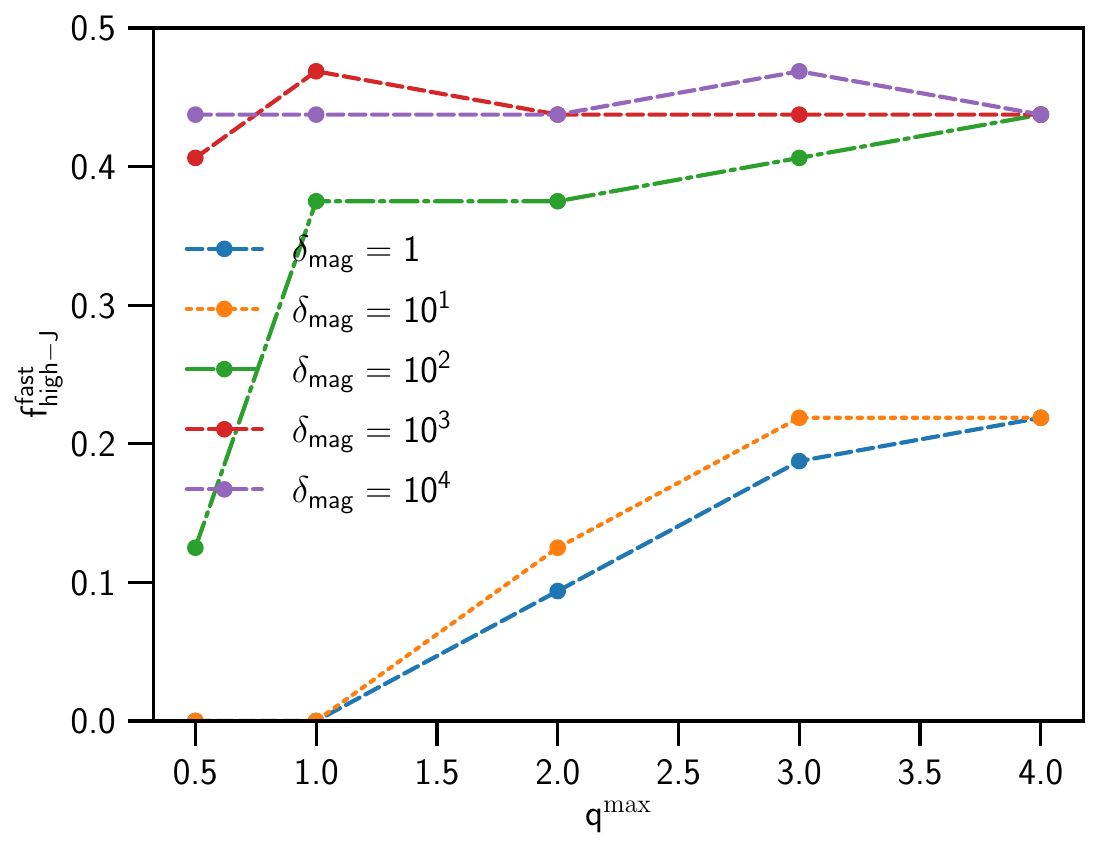} 
\put(55,15){\small \textbf{(a)} $U/(n_{1}T_{2}) = 1$}
\put(65,62){\small MRAT Alignment}
\put(65,27){\small RAT Alignment}
\end{overpic}	
\begin{overpic}[width=0.45\textwidth]{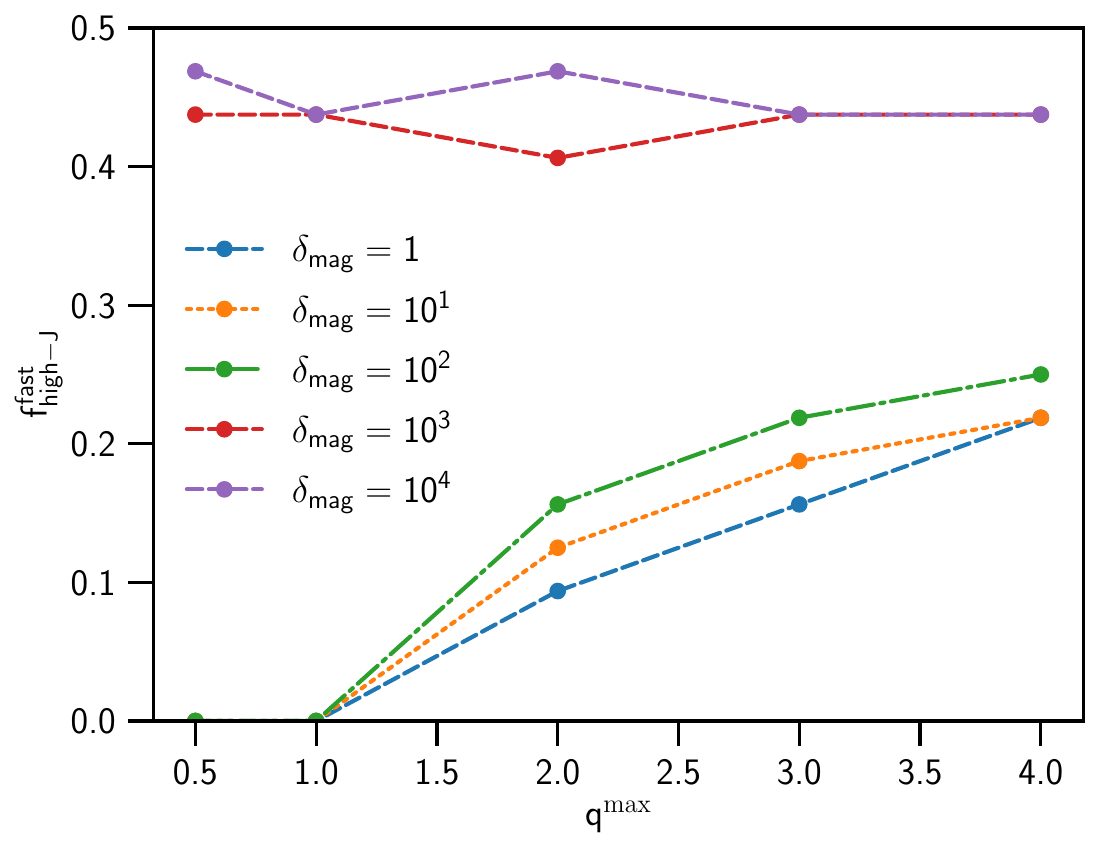}   
\put(55,15){\small \textbf{(b)} $U/(n_{1}T_{2}) = 7.3$}
\put(65,62){\small MRAT Alignment}
\put(65,27){\small RAT Alignment}
\end{overpic}	
\begin{overpic}[width=0.45\textwidth]{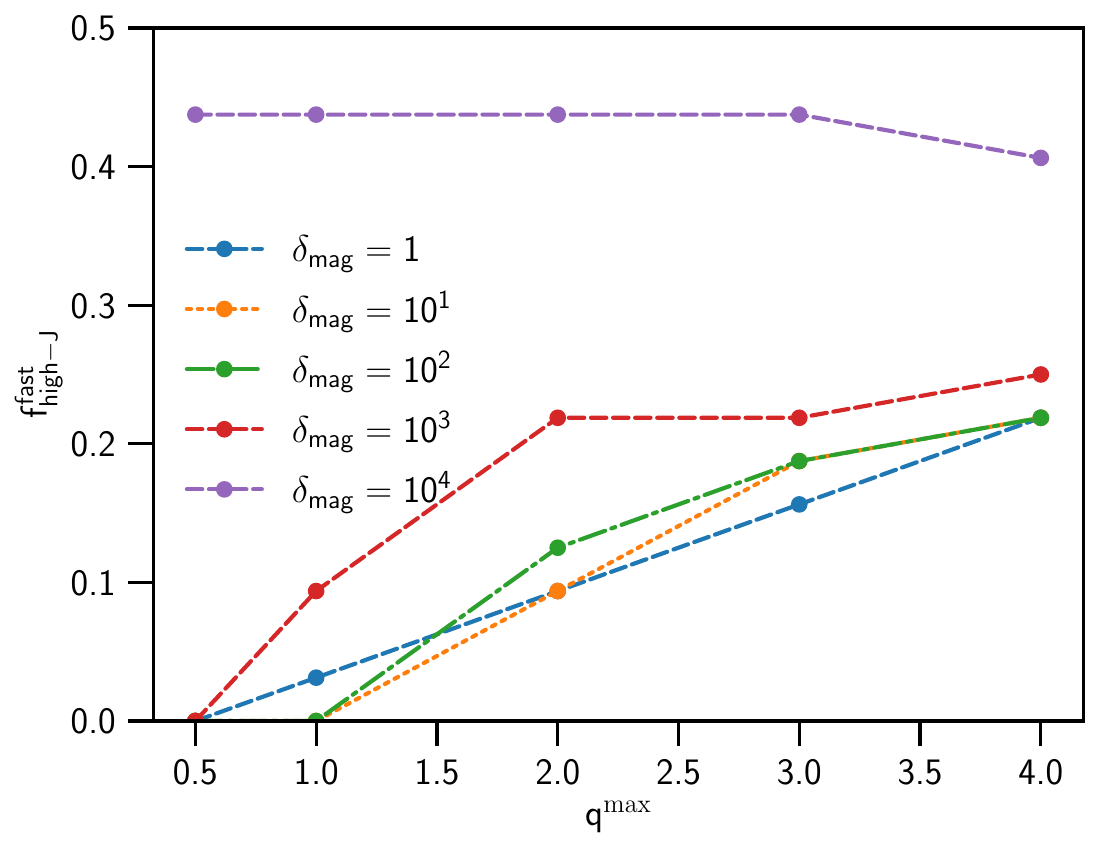}   
\put(55,15){\small \textbf{(c)} $U/(n_{1}T_{2}) = 49.5$}
\put(65,62){\small MRAT Alignment}
\put(65,27){\small RAT Alignment}
\end{overpic}	
\begin{overpic}[width=0.45\textwidth]{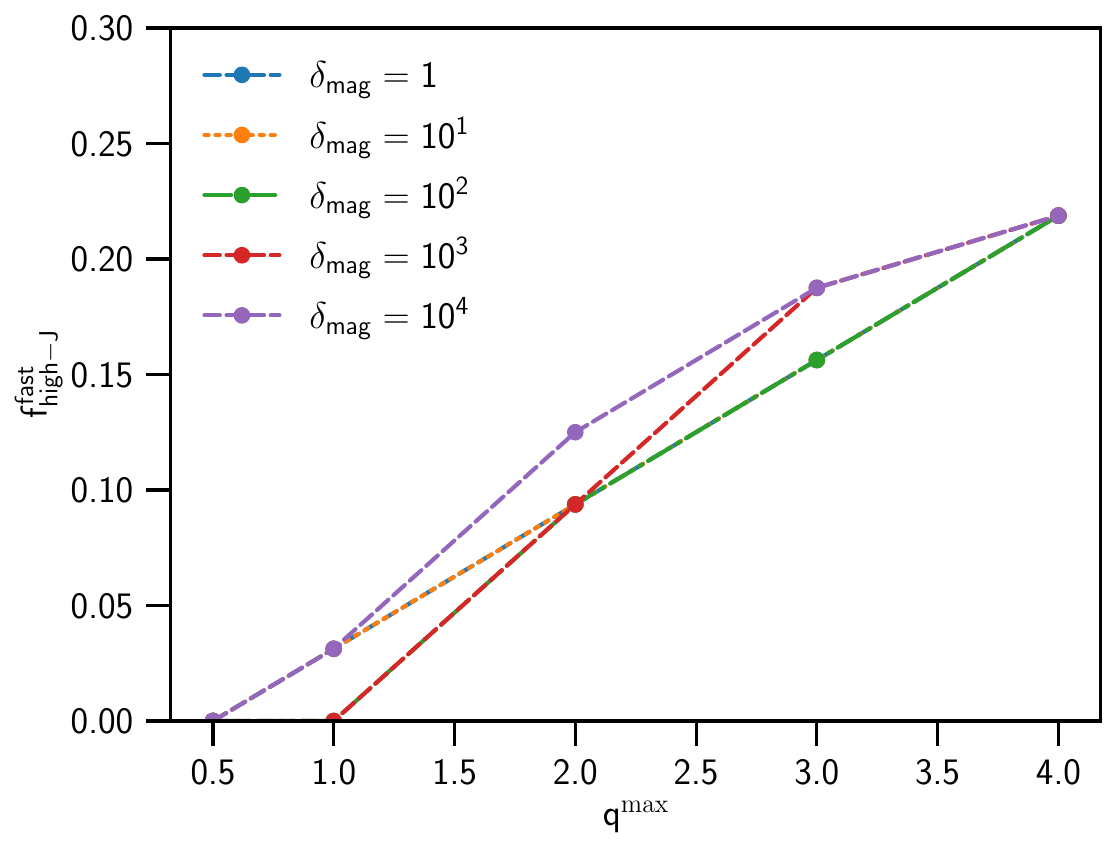}   
\put(55,15){\small \textbf{(d)} $U/(n_{1}T_{2}) = 15640$}
\put(65,27){\small RAT Alignment}
\end{overpic}	
\caption{The fraction of grains with initial random orientation that undergo fast alignment with high-J, $f_{\rm high-J}^{\rm fast}$, for different RAT parameters and magnetic relaxation, assuming $\psi=0^{\circ}$ for the different radiation fields $U/(n_{1}T_{2})$ (panels (a)-(d)). The value of $f_{\rm high-J}^{\rm fast}$ increases with $\delta_{\rm mag}$ except for $U/(n_{1}T_{2})=15640$ where magnetic relaxation becomes ineffective and alignment is dominated by RATs.}
\label{fig:fhigh_fast}
\end{figure*}

Figure \ref{fig:fhigh_fast} shows the fraction of grains that have fast alignment at high-J attractors, $f_{\rm high-J}^{\rm fast}$, for the different RAT models described by $q^{\rm max}$ and magnetic relaxation $\delta_{\rm mag}$ for different radiation strengths $U=1, 10, 10^{2}$ and $10^{4}$ (panels (a)-(c)). The maximum degree for PM grains of $\delta_{\rm mag}=1$ is $f_{\rm high-J}^{\rm fast}\sim 0.22$, whereas the alignment degree for SPM grains is higher for larger $\delta_{\rm mag}$, and it can reach $f_{\rm high-J}^{\rm fast}\sim 0.45$ for $\delta_{\rm mag}=10^{3}$. However, the fraction of $f_{\rm high-J}$ decreases with increasing $U/(n_{1}T_{2})$. For $U/(n_{1}T_{2})=15640$, $f_{\rm high-J}^{\rm fast}$ depends weakly on $\delta_{\rm mag}$ due to the reduced efficiency of magnetic relaxation, and reach $f_{\rm high-J}^{\rm fast}\sim 0.22$.

The timescale for perfect alignment is shown in Table \ref{tab:models_qmax2}. For the CD regime of $U/(n_{1}T_{2})<1$, the perfect alignment is achieved by slow transport after only $T_{\rm PA}\sim 10\tau_{\rm damp}\sim 10\tau_{\rm gas}$ for SPM grains, but it requires $T_{\rm PA}\sim 50\tau_{\rm gas}$ for PM grains (see also Figures \ref{fig:CDR_RQJ_qmax1_nH1e1_aef02_deltam1} and \ref{fig:CDR_RQJ_qmax1_nH1e1_aef02}). For the RD regime of $U/(n_{1}T_{2})>1$, grains are trapped at low-J rotations by strong radiative torques, and the perfect slow alignment is not achieved due to the RAT trapping.

\section{Discussion}\label{sec:discuss}
We generalize the MRAT theory for general astrophysical conditions characterized by the relative local radiation strength to the gas density $U/(n_{1}T_{2})$ and use numerical simulations to quantify the efficiency of grain alignment by the MRAT mechanism. We confirmed the previous results found in \cite{HoangLaz.2008,HoangLaz.2016} for the collision-dominate regime ($U/(n_{1}T_{2})\le 1$) and discovered new effects for radiation-dominated (RD) regime and extreme radiation fields of with $U/(n_{1}T_{2})>1$. Below, we discuss in detail our results and implications for a variety of astrophysical environments, from the standard ISM to cold and dense star-forming regions to hot circumstellar environments. 

\subsection{Collision-Dominated (CD) Regime: Collisional and Magnetic Excitations Transporting Grains from low-J to High-J Attractors and Time-Dependent Alignment}
In collision-dominated (CR) regimes, $U/(n_{1}T_{2})<1$, the grain angular momentum experience strong fluctuations with respect to the ambient magnetic field during the slow rotation period due to gas random collisions and magnetic fluctuations (see Figure \ref{fig:CDR_map_qmax2_nH1e1_aef02}). Such fluctuations have two important consequences depending on the MRAT properties. For grain alignment without a high-J attractor, fluctuations cause the randomization of $\bJ$ and produce negligible alignment. However, when grain alignment has high-J attractors, fluctuations can slowly scatter grains from low-J to high-J attractors, resulting in the perfect alignment. These crucially essential effects were first discovered in \cite{HoangLaz.2008} and further established in \cite{HoangLaz.2016} for the diffuse ISM. In this paper, we conducted a comprehensive study of the MRAT alignment in general conditions and confirmed the crucial role of collisional and magnetic excitations leading to the perfect alignment of grains with high-J attractors.

We also found that the alignment degrees increase slowly with time and achieve the perfect alignment after $T_{\rm PA}\sim 10\tau_{\rm gas}$ for SPM grains, but it requires $50\tau_{\rm gas}$ for PM grains due to the contribution of magnetic excitations (see Figures \ref{fig:CDR_RQJ_qmax1_nH1e1_aef02}); Table \ref{tab:models_qmax2}). Therefore, iron inclusions not only enhance grain alignment by high-J attractors but also increase the rate of transport of grains from low-J to high-J attractors. This time-dependent alignment will induce time-dependent polarization. Therefore, we suggest observations of time-dependent polarization to constrain the magnetic property of dust. 

In particular, our numerical simulations show that grain alignment is negligible if the high-J attractor points are absent in the collisional-dominated regime of $U/(n_{1}T_{2})<1$. The detection of high polarization levels from the ISM \citep{Planck.2015}, molecular clouds and filaments (\citealt{Ngoc.2023},\citealt{Ngoc.2024apj}) to protostellar environments \citep{Giangetal.2024} strongly suggests that grains contain iron inclusions to induce universal high-J attractors, which cannot be produced by RATs only. Moreover, these observational results are strong evidence for the effect of collisional and magnetic excitations in transporting grains from low-J to high-J attractors and enabling highly efficient alignment of interstellar grains.

\subsection{Radiation-Dominated (RD) Regime: Reduced Efficiency of Magnetic Relaxation and Radiative Torque Trapping}
We identified three important effects for grain alignment in strong radiation fields in which radiation is dominant, $U/(n_{1}T_{2})>1$. First, the effect of magnetic relaxation in enhancing the RAT alignment at high-J attractors is reduced at higher radiation fields due to faster driving grains to low-J attractors by RATs. Therefore, to reproduce the same high-J attractors, the critical magnetic relaxation must be increased in stronger radiation fields (see Figure \ref{fig:deltacri_lowJ_highJ}). 

Second, we find the upper limit of the radiation field or dust temperature below which the magnetic relaxation is still effective in enhancing the RAT alignment.  For the RAT model that does not produce high-J attractors (see Figure \ref{fig:deltacri_lowJ_highJ}), magnetic relaxation is only effective for the radiation field of strength $U<U^{\rm MRAT}_{\rm max}\sim 10^{7}N_{cl,5}^{3/2}\hat{B}^{3}$ (see Equation\ref{eq:Umax}), or dust temperature $T_{d}<T^{\rm MRAT}_{\rm max}\sim 220N_{cl,5}^{1/4}\hat{B}^{1/2}\K$ (see Equation \ref{eq:Td_max}). Above these thresholds, magnetic relaxation does not affect grain alignment efficiency, and grains are entirely aligned by RATs. In this regime, the fraction of grain shapes with high-J attractors is typically $f_{\rm high-J}^{\rm ens}<0.5$ (see \citealt{Herranen:2019kj}). {Moreover, the fraction of grain orientations that can be fast aligned at high-J attractors is $f_{\rm high-J}^{\rm fast}<0.25$ for the considered range of $q^{\rm max}$ (see Figure \ref{fig:fhigh_fast}), which results in a net fraction of grains at high-J attractors of $f_{\rm high-J}=f_{\rm high-J}^{\rm ens}f_{\rm high-J}^{\rm fast}< 0.5\times 0.25=0.125$ or $12.5\%$.}

Third, we found a new effect termed RAT trapping (RATT) in which strong radiative torques can trap grains at low-J rotation even in the presence of gas collisions and magnetic fluctuations when the radiation field is strong enough of $U/(n_{1}T_{2})>1$. Gas random collisions and magnetic relaxation can scatter weakly grain orientation out of the low-J attractor (see Figures \ref{fig:RDR_map_qmax2_nH1e1_aef02},\ref{fig:extreme_map_qmax2}). Therefore, for the RAT model without high-J attractors, the RATT effect can help increase the grain alignment efficiency by stabilizing grains at low-J attractors against random gas collisions. However, the net alignment degree is less that $20\%$ due to low internal alignment due to slow grain rotation.

Finally, one important consequence of the RATT is that grains trapped at low-J rotation by intense radiation would have fast radiative precession and experience the alignment along the radiation direction (${\bk}$), aka. k-RAT alignment \citep{LazHoang.2007,LazHoang.2021}). This is a key signature of the RAT theory for the RD regions. {An observational evidence for k-RAT in the vicinity of Orion Source I within OMC-1 is reported in \cite{Pattle:2021} using ALMA polarization data.}

\subsection{Radiative Torque Disruption: Fast and Slow Disruption and the Role of RAT Trapping}
Fast-rotating grains can be disrupted due to their centrifugal stress. As shown in Figures \ref{fig:RDR_map_qmax2_nH1e1_aef02} and \ref{fig:extreme_map_qmax2}, rotational disruption is related to grain alignment at high-J attractors, but some grains can be disrupted during their high-J rotation stage before being damped by RATs. 

Our numerical calculations show that collisional and magnetic excitations can slowly transport grains from low-J to high-J attractors. When the grain angular velocity exceeds $\Omega_{\rm cri}$, they are disrupted into small fragments. This {\rm slow disruption} takes a comparable time as the time required for slow perfect alignment, which is between $10-100\tau_{\rm gas}$ in CD regimes.
In the RD regime, the fast disruption can occur in less than $\tau_{\rm damp}$. However, the efficiency of fast disruption is much lower than that of slow disruption. To illustrate the fast disruption phenomenon, let us consider an ensemble of grains with different initial orientations and turn on a flash of intense radiation with a long duration. As shown in Figure \ref{fig:RDR_map_qmax1_nH1e1_aef02}, a fraction of grains driven rapidly to high-J attractors can experience {\rm fast disruption} by RATD. The grain population with fast disruption includes grains initially already aligned with the magnetic field under the interstellar radiation field. The majority of grains are trapped at low-J rotation due to strong radiative torques and can survive strong radiation fields. The RATT effect enables large grains to survive strong radiation fields, which is important for an accurate understanding of dust evolution and radiation pressure feedback.

The key difference between RATD and RATT is that RATD causes the depletion of large grains, but RATT helps large grains to survive in a strong radiation field by trapping them at low rotation. Both RATD and RATT reduce the polarization fraction, so the polarization degree alone is insufficient to distinguish the two mechanisms. Instead, we can combine the dust polarization with extinction curves and spectral index of thermal dust emission (SED) that provide constraints on maximum grain sizes. If large grains above $a_{\rm disr}$ are found in strong radiation fields, then, RATT is effective, while RATD is effective at lower efficiency. Moreover, we can also use the polarization angle and radiation direction. If large grains are dominantly trapped at low-J attractors by RATT in strong radiation fields, we expect k-RAT alignment instead of the alignment along the magnetic field (i.e., B-RAT) \citep{Hoangetal.2022}. {We note that the location of low-J attractors is determined by the zero points of the RAT parameter $\langle F\rangle$ (see Figure \ref{fig:FHmean}b). Thus, it does not depend on the radiation intensity, but the RAT model, i.e., $q^{\rm max}$.}

\subsection{Modeling Grain Alignment and Disruption Degree}
The degree of grain alignment is dominated by the fraction of grains aligned at high-J attractors because grain alignment at low-J rotation is negligible due to collisional randomization and internal thermal fluctuations. Let $f_{\rm high-J}^{\rm ens}$ be the fraction of grain shapes that have high-J attractors. In the AMO, the value of $f_{\rm high-J}^{\rm ens}$ depends on $q^{\rm max}$. For an ensemble of random shapes with RATs calculated in \cite{Herranen.2021}, they found $f_{\rm high-J}^{\rm ens}\sim 0.2-0.6$. In the presence of iron inclusions, $f_{\rm high-J}^{\rm ens}=1$ for $\delta_{\rm mag}>\delta_{\rm cri}$ (see Figure \ref{fig:deltacri_lowJ_highJ}).

For a given grain shape, assuming that grains initially have random orientations with respect to the magnetic field, only a fraction of grains from the ensemble can be aligned at high-J attractors. Therefore, one can describe the fraction of grains at high-J attractors through both the fast and slow alignment processes as
\bea 
f_{\rm high-J} = f_{\rm high-J}^{\rm ens}\left(f_{\rm high-J}^{\rm fast} + f_{\rm high-J}^{\rm slow}\right),\label{eq:fhiJ_mod}
\ena
where $f_{\rm high-J}^{\rm fast,slow}$ is the fraction of grains with fast alignment and slow alignment, respectively. The alignment degree is then $f_{\rm align}\approx f_{\rm high-J}$.

Our numerical calculations of $f_{\rm high-J}^{\rm fast}$ for different models of RATs, magnetic relaxation, and local conditions are shown in Figure \ref{fig:fhigh_fast}. Combined with our numerical calculations for the slow alignment, we obtain for SPM grains: 
\bea
f_{\rm high-J}^{\rm slow} = 1 {\rm ~for~} U/(n_{1}T_{2})\le 1,\\
f_{\rm high-J}^{\rm slow} = 0 {\rm ~for~} U/(n_{1}T_{2})>1.\label{eq:fhiJ_slow}
\ena

Therefore, in the CD regime, $f_{\rm align}=f_{\rm high-J}^{\rm ens}$, and $f_{\rm align}=1$ for SPM grains, which is the conditions of the ISM and SFRs. For the RD regime of extreme radiation field, $f_{\rm align}= f_{\rm high-J}^{\rm ens}\times f_{\rm high-J}^{\rm fast}$. 

In strong radiation fields, RATD is important for dust destruction and evolution. The RATD efficiency is given by
\bea
f_{\rm disr}=f_{\rm high-J}^{\rm ens}\times f_{\rm high-J}^{\rm fast},\label{eq:Rdisr}
\ena
which is a factor of 2 smaller than the alignment efficiency $f_{\rm align}$. Note that if initially grains are already aligned at high-J attractors and the radiation field is increased suddenly, then, those grains will be lifted further and will experience rotational disruption when $\Omega>\Omega_{\rm disr}$. In this scenario, the RATD can reach the maximum efficiency of $f_{\rm disr}=1$.

\subsection{Alignment and Disruption of Interstellar Dust in the ISM and Star-Forming Regions}
\subsubsection{The ISM, MCs to filaments and prestellar cores}
In the ISM and SFRs, including filaments and prestellar cores, grains are subject to increasing density and decreasing radiation intensity. Along this star formation process, dust grains are present in the collision-dominated regions with $U/(n_{1}T_{2})<1$ \citep{Hoang.2021}. Therefore, grains have enough time to be transported to high-J attractors, and slow alignment is efficient. Those thermally rotating grains will be aligned at higher-J by slow transport due to collisional and magnetic excitations, leading to perfect alignment over a timescale of 
\bea
T_{\rm PA}\sim 10-100 \tau_{\rm damp}\sim 10^{2}-10^{3}\left(\frac{10^{4}\cm^{-3}}{n_{\H}}\right) \yr,~~\label{eq:Tsat_SFR}
\ena
which is much shorter than the lifetime of the ISM and SFRs. Thus, one can achieve the perfect alignment of SPM grains by the MRAT mechanism.

Those grains aligned with B-fields at high-J attractors and rotating suprathermally will be embedded into molecular clouds, which collapse to form protostars. Smaller grains will be damped by increasing gas collisions and radiation attenuation. Therefore, grains with iron inclusions in star-forming regions can be perfectly aligned with B-fields on high-J attractors due to slow transport from low-J to high-J attractors, i.e., $f_{\rm high-J}^{\rm slow}=1$.

\begin{table}
	\centering
	\caption{The typical phases of the ISM and parameters.}
	\begin{tabular}{l l l l }
		& CNM & WIM & PDR \cr
		\hline\hline

		$n_{\rm H}(cm^{-3})$ & 30 & 0.1 & $10^{4}$\cr
		$T_{\rm gas}(\K)$ & 100 & $10^{4}$ & $10^{3}$\cr
		$U$ & 1 & 1 & $3\times 10^{4}$\cr
		$U/(n_{1}T_{2})$ & 0.33 & 1 & 0.3\cr
		\hline\hline
	\end{tabular}
	\label{tab:ISM}
\end{table}

\begin{figure*}
	\centering
	\begin{overpic}[width=0.32\linewidth]{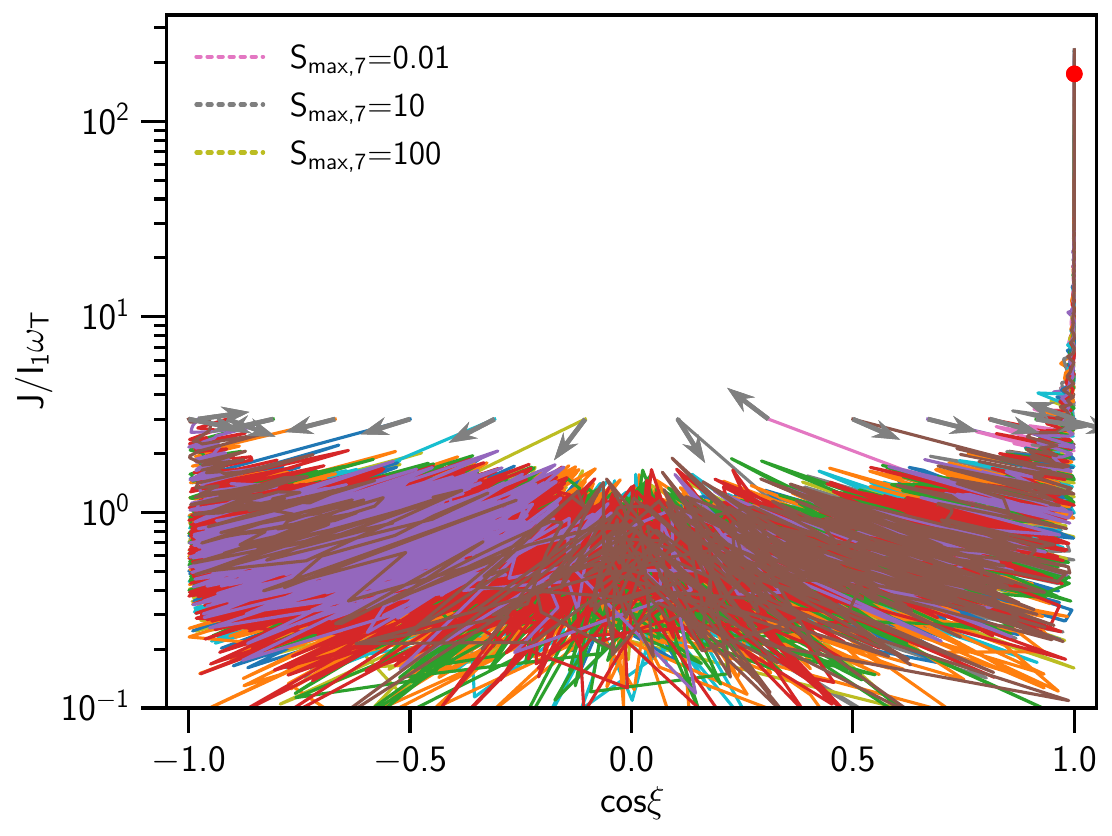}
		\put(65,75){\tiny \textbf{High-J Attractor}}		
		\put(65,60){\small \textbf{(a)} CNM}
	\end{overpic}	
	\begin{overpic}[width=0.32\linewidth]{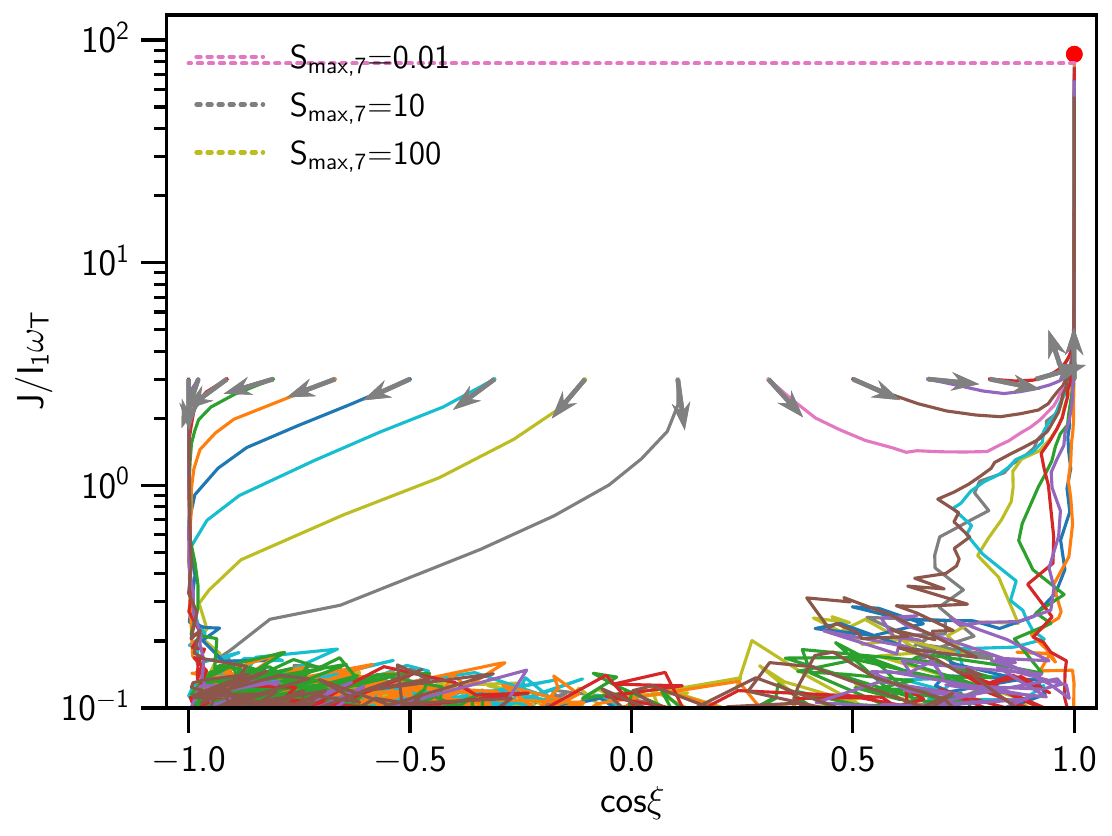}
	\put(65,75){\tiny \textbf{High-J Attractor}}		
	\put(65,60){\small \textbf{(b)} WIM}
	\end{overpic}	
	\begin{overpic}[width=0.32\linewidth]{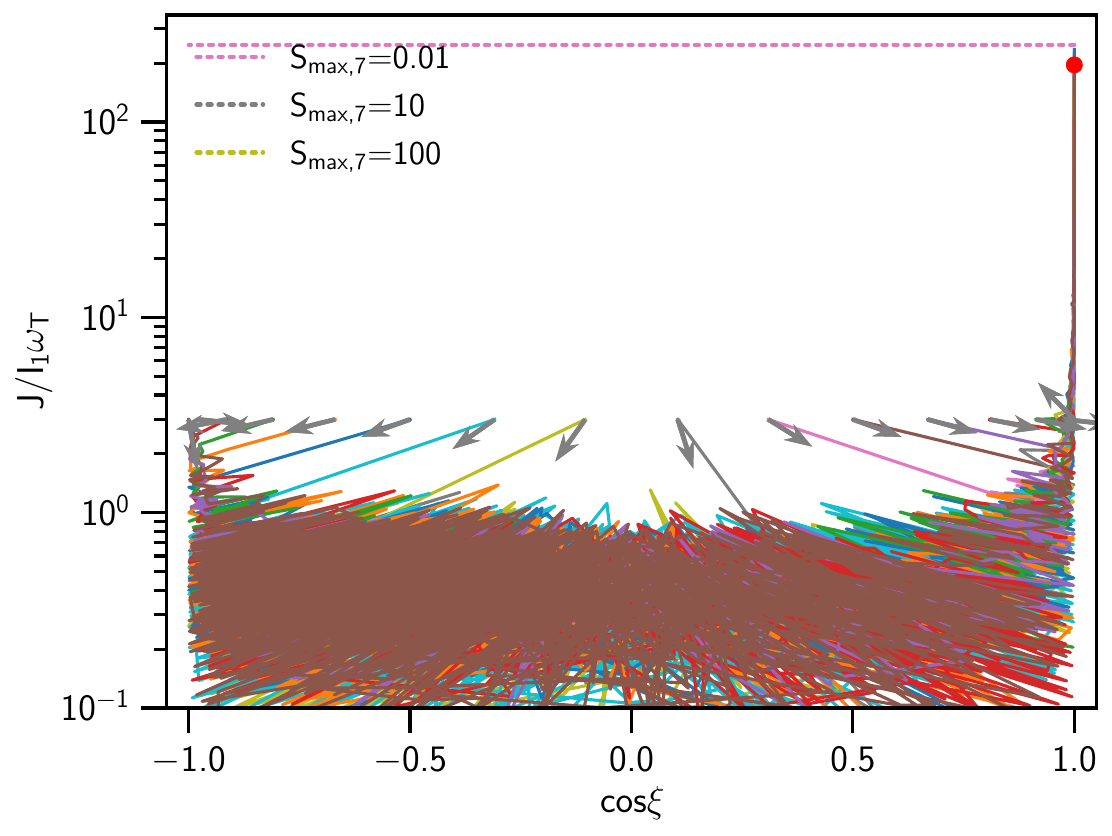}
	\put(65,75){\tiny \textbf{High-J Attractor}}		
	\put(65,60){\small \textbf{(c)} PDR}
\end{overpic}
\caption{Phase trajectory maps for grain alignment in the CNM, WIM, and PDR, assuming the RAT model with $q^{\rm max}=1$ and SPM grains of $\delta_{\rm mag}=10^{3}$. All grains align at the high-J attractor due to slow transport by gas collisions. Grains with low tensile strength can be disrupted in WIM and PDR.}
\label{fig:phasemap_ISM}
\end{figure*}

\begin{figure*}
	\centering
	\begin{overpic}[width=0.32\linewidth]{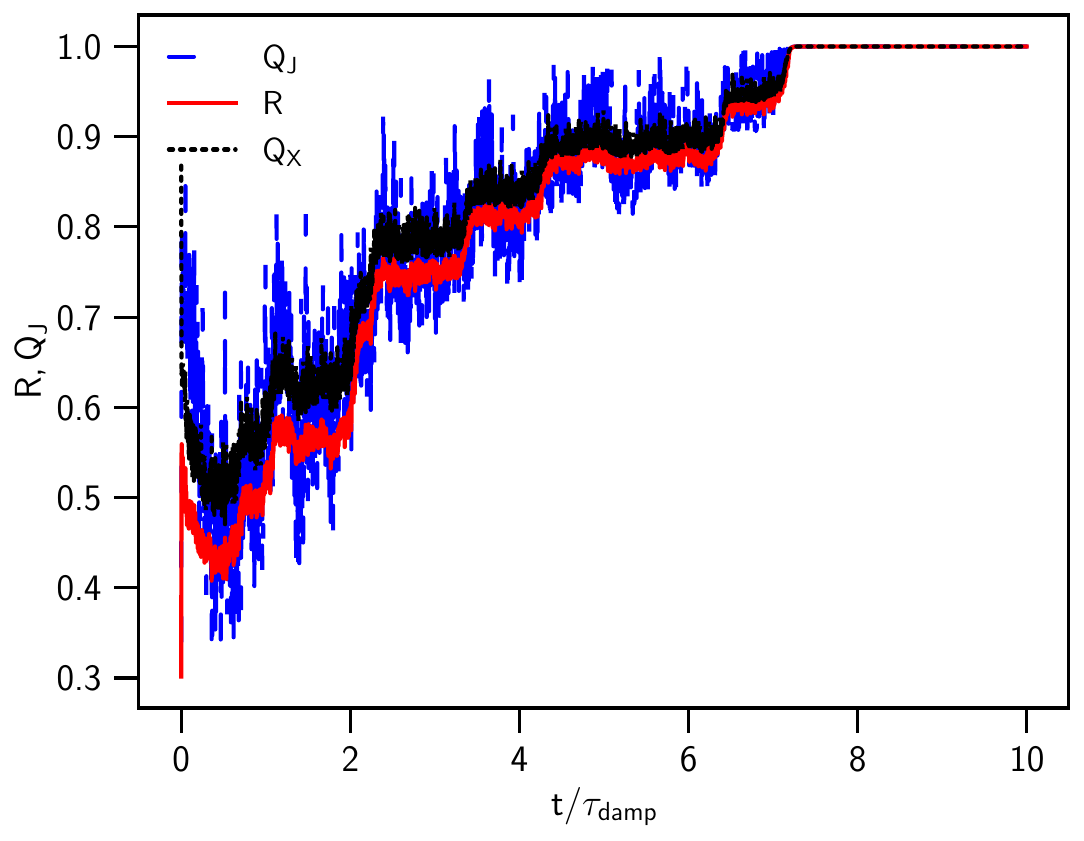}
		\put(65,20){\small \textbf{(a)} CNM}
	\end{overpic}	
	\begin{overpic}[width=0.32\linewidth]{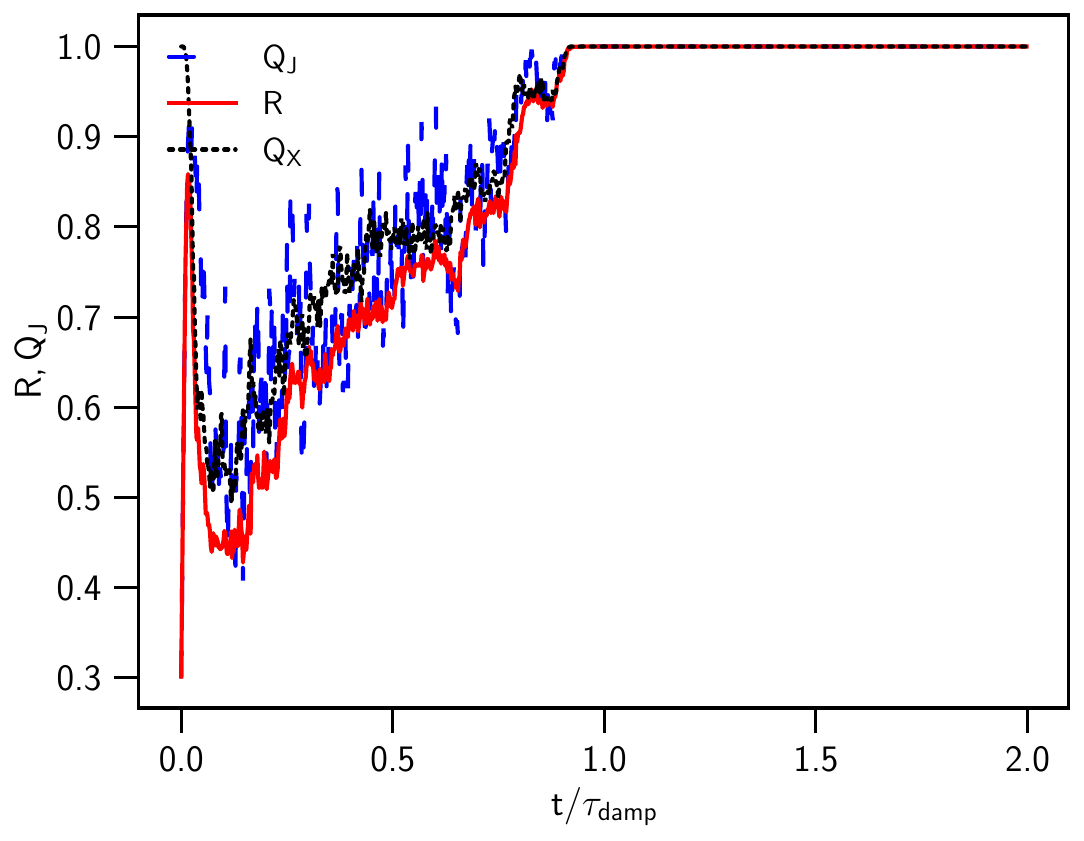}
		\put(65,20){\small \textbf{(b)} WIM}
	\end{overpic}	
	\begin{overpic}[width=0.32\linewidth]{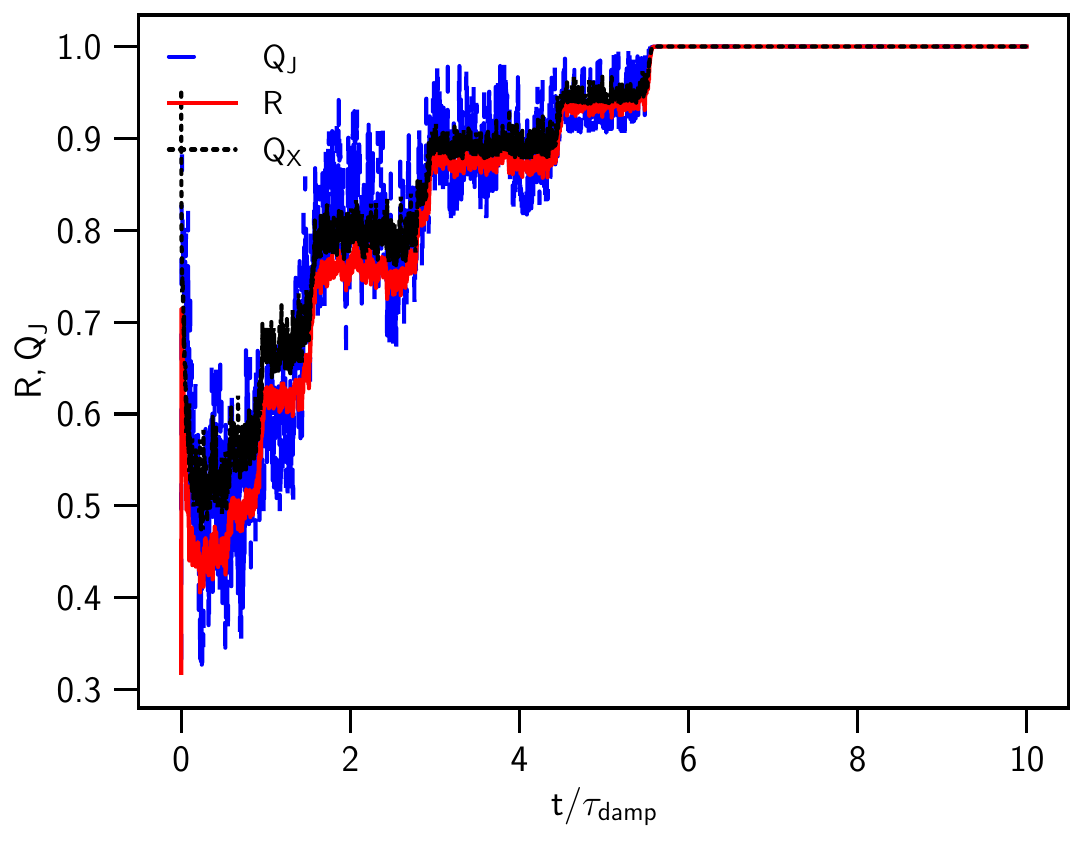}
		\put(65,20){\small \textbf{(c)} PDR}
	\end{overpic}	
\caption{Time-dependent alignment degrees of grains in the CNM, WIM, and PDR. The alignment degrees increase with time due to slow transport by collisional excitations, and achieve perfect after $10\tau_{\rm damp}$.}
\label{fig:alignment_ISM}
\end{figure*}

To illustrate the effect of local conditions, we run simulations for the cold neutral medium (CNM), warm-ionized medium (WIM), and photodissociation regions (PDR). Note that the WIM has much lower gas density but higher temperature than the CNM, whereas the PDR has much higher gas density and radiation fields. However, these three phases fall into the CD regime. Their physical parameters are shown in Table \ref{tab:ISM}. Figures \ref{fig:phasemap_ISM} and \ref{fig:alignment_ISM} show the phase trajectory maps and time-dependent alignment degrees obtained for these phases. As expected, grains in this CD regime can be perfectly aligned by slow transport of grains from low-J to high-J attractors, i.e., $f_{\rm high-J}^{\rm slow}=1$. Interestingly, in both WIM and PDRs, which have very different gas properties and radiation fields from the CNM but a similar quantity $U/(n_{1}T_{2})$ (the same CD regime), collisional and magnetic excitations can efficiently transport grains from low-J to high-J attractors, leading to the perfect alignment within $10\tau_{\rm damp}$.

\subsubsection{Protostellar environments}
Prestellar cores have a low radiation field and high density with $U/(n_{1}T_{2})<1$, and they belong to the CD regime. Therefore, large grains of micron sizes with iron inclusions in prestellar cores and protostellar envelopes are already aligned by attenuated interstellar radiation by MRATs and achieve perfect alignment due to slow collisional and magnetic excitations \citep{Hoangetal.2022}. As the central luminosity increases due to protostellar accretion and its blackbody emission, those grains at high-J attractors are further lifted to higher angular momentum. 

Let us estimate the timescales for grain alignment in protostellar cores. The gas density and radiation strength at a distance $r$ in the envelope from the protostar are given by \citep{Hoang.2021}:
\bea
n_{\rm H}&&\sim 10^{6}\left(\frac{100\au}{r}\right)^{2}\cm^{-3},\\
U&&\sim 5.2\times 10^{3}\left(\frac{100\au}{r}\right)^{2}\left(\frac{L_{\star}}{L_{\odot}}\right),\label{eq:YSO}
\ena
where the inner radius of the envelope is $r_{\rm in}=100\au$, $L_{\star}$ is the bolometric luminosity, and the attenuation by the intervening dust is disregarded. The equation yields $U/(n_{1}T_{2})\sim 10^{-3}$, which is well in the collision-dominated regime.

The rotational damping time is just the gas damping (Equation \ref{eq:taugas}) due to the subdominance of IR emission, which is equal to
\bea
\tau_{\rm damp}=\tau_{\rm gas}\sim 10^{-1}\left(\frac{r}{100\au}\right)^{2}~\yr.
\ena

Therefore, grain alignment takes about $T_{\rm PA}\sim 10-100\tau_{\rm gas}\sim 1-10\yr$ to reach the perfect alignment by the MRAT (see e.g., \ref{fig:CDR_RQJ_qmax2_nH1e1_aef02}). This timescale is much shorter than the dynamical time of protostellar cores, and one can assume the stable alignment.

One interesting effect arises when the protostellar luminosity suddenly increases due to episodic accretions. Indeed, episodic accretions onto protostars can increase the luminosity by a factor of 100 \citep{Audard.2014}, which entails $U/(n_{1}T_{2})\sim 0.1$, still in the CD regime. As a result, for episodic YSOs, increased radiation by the outburst will cause both fast and slow alignment of smaller grains over time, resulting in the time-dependent increase in the polarization degree. The polarization degree will reach the maximum level after a timescale of $T_{\rm PA}\sim 10-100 \tau_{\rm damp}\sim 1-10 \yr$ due to the perfect alignment of grains by gas and magnetic excitations. This time-dependent polarization could be tested with the polarization survey for episodic YSOs, and the physics of slow alignment by MRAT could be tested.

{Finally, we note that the above discussions above are based on numerical results for a large grain in the diffuse ISM of $a=0.2\mum$. Although grain growth to micron size is expected in protostellar environments, we expect that the general MRAT theory is applicable for such large grains as long as the Larmor precession of grains is still faster than the gas randomization so that grains can be aligned by RATs ( \citealt{Hoangetal.2022,Giangetal.2024}).}

\subsubsection{Ultracompact and compact HII regions around young massive stars}
{Ultracompact (radius $R\sim 0.1$pc) and compact (radius $D\sim 0.5$ pc) HII regions are the earliest stages of an HII region produced by young massive stars, which are the subsequent stage of hot cores. The typical gas density is $n_{\H}\sim 10^{6}-10^{9}\cm^{-3}$, gas temperature $T_{\gas}\sim 100-200\K$, and the dust temperature is $T_{d}\sim 150\K$, corresponding to $U\sim 10^{6}$, heated by the central luminosity $L\sim 10^{4}-10^{6}L_{\odot}$ \citep{Churchwell:2002ju}. This results in $U/(n_{1}T_{2})<1$ for $n_{\H}>10^{7}\cm^{-3}$ and $U/(n_{1}T_{2})>1$ for $n_{\H}<10^{7}\cm^{-3}$. Therefore, grain alignment and disruption in UCHII regions follows the CD regime, and it transits into RD regime in compact HII regions when the HII region expands. The effect of RATD on radiation pressure in the hot core with embedded massive protostars is studied in \cite{Hoang.2021b}. In this study, the hot dust is assumed to survive against RATD. According to our RATT effect, such an assumption is justified because grains are trapped at low-J attractors and not disrupted by RATD.}

\subsubsection{HII bubbles around massive stars}
HII bubbles (classical HII regions) are the ionized gas and dust shell produced by central massive stars (aka. Stromgren spheres), which is the result of the expansion of ultracompact HII regions. Gas is heated mainly by photoelectric effects of atomic hydrogen, while dust originating from the ISM (ISD) is heated by both stellar radiation and gas collisions. The typical gas temperature is $T_{\gas}\sim 10^{4}\K$, and dust temperature is $\sim 100\K$ ($U\sim 4\times 10^{4}$). The gas density is $n_{\H}< 10^{3}\cm^{-3}$, which corresponds to $U/(n_{1}T_{2})>1$, so HII bubbles fall into the RD regime. 

Since dust in HII bubbles mostly originates from the ISM, large grains have already been aligned at high-J attractors by the ISRF prior to massive star formation, and would be disrupted efficiently by massive (proto)stellar radiation. As a result, HII regions are expected to contain only small grains due to the depletion of large grains by RATD. In realistic situations, HII regions may contain dust grains newly formed from gas condensation in stellar winds. For this stardust, grains can experience both RATD and RATT, but the alignment and disruption degree is lower (see Figure \ref{fig:fhigh_fast}).
	
For the local HII bubble, \cite{Medan.2019} found the polarization efficiency $p/A_{V}\sim 0.1\%/{\rm mag}$ at $A_{V}=1$, which is much lower than the maximum polarization efficiency observed in the ISM of $p/A_{V}\simeq 3\%/{\rm mag}$. Such a lower polarization efficiency could be explained by both RATD and RATT. Further observational polarization observations and data analysis are required to distinguish the RATD and RATT in HII bubbles.

\subsection{Dust in active galactic nuclei (AGN) torus}
Dust in the torus ($\sim 1-10$ pc) around supermassive black holes is subject to strong radiation fields from accretion disks. The gas density at a radial distance $r$ in the torus is $n_{\H}\sim n_{0}r_{\pc}^{-1}\cm^{-3}$ where $r_{pc}=r/1{\pc}$ and $n_{0}$ is the hydrogen density at $r=1\pc$. The radiation strength is approximated as $U(r)\sim 10^{8}r_{\pc}^{-2+\eta}$ where $\eta>0$ accounts for the reddening by intervening dust \citep{Gianghoang.2021}. Therefore, one estimates $U/(n_{1}T_{2})\sim 10^{4}(n_{0}/10^{5}\cm^{-3})r_{\pc}^{-\eta}$ where $T_{\rm gas}\sim T_{d}\sim 20U^{1/6}$ is the gas temperature. Numerical calculations in \cite{Gianghoang.2021} shows $\eta\sim 3$ for $n_{0}=10^{5}\cm^{-3}$, which yields $U/(n_{1}T_{2})>1$ for $r<10\pc$, and grain alignment and disruption follow the RD regime. Therefore, large grains of interstellar origin in the AGN torus may be rotating at high-J attractors and can be perfectly disrupted. However, the newly formed dust population experiences only the fast alignment and disruption due to RATT in the RD regime.

\subsection{Alignment and Disruption of Circumstellar Dust (Stardust) in the Envelope of Evolved Stars}
Circumstellar dust is the dust newly formed around stars from the condensation of dense gas expelled from stars (aka stardust). As soon as they are formed, stardust grains are thermally rotating due to gas collisions and subject to strong radiation. This is radically different from ISD where grains are initially subject to the average interstellar radiation and can be perfectly aligned at high-J attractors due to the slow transport process. Below, we describe briefly the typical parameters and discuss the alignment and disruption of stardust using our new general theory.

The gas density and radiation strength in the envelopes of AGB stars can be given by \citep{TramHoang.2020}
\bea
n_{\rm H}&&\simeq 10^{6}\left(\frac{10^{15}\cm}{r}\right)^{2}\left(\frac{\dot{M}}{10^{-5}M_{\odot}yr^{-1}}\right)~\cm^{-3},\label{eq:AGB_n}\\
U&&\simeq 1.2\times 10^{8}\left(\frac{10^{15}\cm}{r}\right)^{2}\left(\frac{L_{\star}}{10^{4}L_{\odot}}\right),\label{eq:AGB_U}
\ena
where $\dot{M}$ is the mass loss rate and $L_{\star}$ is the bolometric luminosity of the evolved star.

For the gas temperature $T_{\rm gas}\sim 200 (r/10^{15}\cm)^{-2}$ \citep{TramHoang.2020}, Equations (\ref{eq:AGB_n}) and (\ref{eq:AGB_U}) yield the parameter $U/(n_{1}T_{2})\sim 500$, implying that the AGB envelopes fall into the radiation-dominated regime. Therefore, in the inner region of $r<10^{15}\cm$, only a fraction of grains can be fast aligned at high-J attractors, and the rest of the grains are trapped at low-J rotations due to the RATT (see Figure \ref{fig:RDR_map_qmax2_nH1e1_aef02}). However, collisional and magnetic excitations are inefficient in transporting them to high-J attractors due to RATT, leading to the alignment $f_{\rm align}=f_{\rm high-J}^{\rm fast}<0.5$ (see Figure \ref{fig:fhigh_fast}), which is a factor of 2 lower than in the collision-dominated regime of the ISM and SFRs where grain alignment can be perfect by the MRAT mechanism. The maximum degree of fast alignment is $f_{\rm align}=f_{\rm high-J}^{\rm ens}\times f_{\rm high-J}^{\rm fast}>0.5\times 0.25\sim 0.125$, assuming $f_{\rm high-J}^{\rm ens}=0.5$ and $f_{\rm high-J}^{\rm fast}=0.25$ for RAT alignment (see Figure \ref{fig:fhigh_fast}). When the grain temperature decreases in the outer regions, the magnetic relaxation becomes effective and can increase the fraction of grains in the ensembles with high-J attractor to $f_{\rm high-J}^{\rm ens}=1$. As a result, the alignment degree increases to $f_{\rm align}=f_{\rm high-J}^{\rm fast}\sim 0.25$, implying the increase in the alignment degree from the inner to outer region in the the envelope of O-rich AGB stars. For carbonaceous grains in C-rich AGB stars, the alignment is purely determined by RATs due to the nonmagnetic relaxation, therefore, the alignment degree for an ensemble of grains is $f_{\rm align}\sim f_{\rm high-J}^{\rm ens}\times f_{\rm high-J}^{\rm fast}>0.25\times 0.25\sim 0.125$. Detailed numerical calculations for grain alignment in the envelopes of AGB stars are required to accurately understand the spatial variation of grain alignment.

\subsection{Alignment and Disruption of Dust in the Solar and Exoplanetary Systems}
Dust in planetary systems (solar and exoplanetary systems) is mainly produced from collisions of larger bodies and released from comets. Dust in planetary systems is warm and hot due to strong stellar radiation. 

The radiation strength at distance $r$ from a star of luminosity $L_{\star}$ is given by
\bea
U\simeq 5.3\times 10^{7}\left(\frac{L_{\star}}{L_{\odot}}\right)\left(\frac{1 \au}{R}\right)^{2},\label{eq:U_solar}
\ena
where $L_{\star}=L_{\odot}$ for our solar system.

For the typical density of the local interplanetary medium is $n_{\rm H}\sim 0.1\cm^{-3}$ and $T_{\rm gas}\sim 10^{5}\K$, one obtains 
\bea
\frac{U}{(n_{1}T_{2})}\sim 5.3\times 10^{6}\left(\frac{(L_{\star}/L_{\odot})}{n_{1}}\right)\left(\frac{1\au}{R}\right)^{2}.\label{eq:Un1_solar}
\ena

Our results show that for this extreme radiation field of $U>U^{\rm MRAT}_{\rm max}$ (see Equation \ref{eq:Umax}), grain alignment is entirely determined by RATs due to the suppression of magnetic relaxation. The maximum alignment degree is $R_{\rm max}\sim 0.25$ as shown in Figure \ref{fig:fhigh_fast}. Moreover, only grains that are fast aligned toward high-J attractors can be disrupted, and slow transport is ineffective due to the strong radiation field. The disruption efficiency is $f_{\rm disr}\sim f_{\rm high-J}^{\rm ens}\times f_{\rm high-J}^{\rm fast}\sim 0.1$ for the range of $q^{\rm max}$ as shown in Figure \ref{fig:fhigh_fast}. This can better constrain the evolution of Zodiacal dust by the RATD effect \citep{Hoang_Fcorona.2021,Ng.2025} and help resolve the tension of low $\beta$ meteoroids than predicted by the perfect RATD in the Zodiacal disk \citep{Silsbee.2025}.

Moreover, the degree of grain alignment of dust in the Zodiacal/ exozodiacal clouds is estimated to be $f_{\rm align}\sim f_{\rm high-J}^{\rm ens}\times f_{\rm high-J}^{\rm fast}\sim 0.1$, resulting in a low degree of thermal dust polarization. Furthermore, the k-RAT alignment might be dominant due to the RATT effect. We suggest high-sensitivity observations toward debris disks to test this prediction. Interestingly, ALMA has reported a marginal detection of dust polarization at $\sim 0.51\%$ \citep{Hull.2022}. Combining the total emission intensity (SED) and polarization data of hot/warm dust from exoplanetary systems would be crucial to test the RATD and RATT effects.

Table \ref{tab:GRAT_summary} summarizes the main properties of RAT alignment, disruption, and trapping of interstellar dust and stardust in different environments.

\begin{table*}
	\centering
	\caption{Summary of RAT Alignment, Disruption and Trapping in different environments}
	\begin{tabular}{l l l l l l }
	Environments & $U/(n_{1}T_{2})$ & Dust & RAT Alignment & RAT Disruption & RAT Trapping \cr
	\hline\hline\cr
	ISM and SFRs & $<1$ & ISD & Fast and Slow & Fast and Slow & N\cr
	Evolved Star's Envelopes & $> 1$ & Stardust & Fast & Fast & Y \cr
	Planetary Systems & $\gg 1$ & Stardust & Fast & Fast & Y\cr
	Supernova Ejecta &  $\gg 1$ & Stardust & Fast & Fast & Y\cr
	AGN torus & $> 1$ & ISD and Stardust & Fast and Slow & Fast and Slow & Y\cr
	\hline\hline
	\end{tabular}
	\label{tab:GRAT_summary}
\end{table*}

\section{Summary}\label{sec:summary}
We generalized the MRAT theory for arbitrary radiation fields and gas density, and used numerical simulations of the MRAT alignment to quantify the efficiency of grain alignment and disruption. Our main results are summarized as follows:
\begin{itemize}
    
\item We derive the critical magnetic relaxation strength for which grains of different shapes have high-J attractors due to the MRAT mechanism in general local gas density and radiation fields described by the parameter $U/(n_{1}T_{2})$. The critical relaxation must be increased in the environments with higher $U/(n_{1}T_{2})$ to overcome the disalignment effect of RATs.

\item  We also derive the critical radiation field or dust temperature below which SPM relaxation is effective so that grain alignment is described by the MRAT. For grains with embedded iron clusters, we found $T^{\rm MRAT}_{\rm max}\sim 220N_{cl,5}^{1/4}\hat{B}^{1/2}\K$. Above these thresholds, grain alignment is purely determined by RATs.

\item
We find that, for the collision-dominated (CD) regime of $U/(n_{1}T_{2})\leq 1$, collisional excitations can slowly transport large grains of $a_{\eff}=0.2\mum$ to high-J attractor so that these grains can achieve perfect {\it slow alignment} within $\sim 10-100\tau_{\rm gas}$ due to MRAT. 

\item For the radiation-dominated (RD) regime with $U/(n_{1}T_{2})>1$, only a fraction of grains can be rapidly aligned at high-J attractors by MRATs, and the majority of grains are driven to low-J rotation and trapped there due to strong radiation, which is termed {\it radiative torque (RAT) trapping}. In this RD regime, grains only have the {\it fast alignment} by MRATs with the degree less than $50\%$. 

\item For extreme radiation fields of $U/(n_{1}T_{2})>10^{4}$, we find that the efficiency of superparamagnetic relaxation in enhancing the RAT alignment is suppressed, and grains can have fast alignment and disruption purely determined by RATs with degrees below $\sim 23\%$. 

\item We found that magnetic relaxation and excitations become ineffective at extreme radiation fields of $U/(n_{1}T_{2})>\delta_{\rm mag}$ due to the dominance of RATs, and grain alignment is determined by RATs only. Rotational disruption only occurs with a small fraction of grains with fast alignment to high-J attractor $f_{\rm high-J}^{\rm fast}$. We found that for strong radiation fields, grains can be rotationally disrupted during the alignment process of grains before being damped by RATs.

\item We quantified the values of $f_{\rm high-J}^{\rm fast}$ for different RAT models, magnetic relaxation, and $U/(n_{1}T_{2})$. We find that the maximum $f_{\rm high-J}^{\rm fast}$ can reach $45\%$ for SPM grains by the MRAT and $22\%$ for the RAT mechanisms. 

\item We introduce a new parametric model for grain alignment efficiency based on the dimensionless parameter $U/(n_{1}T_{2})$. Our general RAT theory can be applied to model grain alignment and disruption for different astrophysical environments, from the interstellar dust in the ISM and SFRs to stardust newly formed in the envelope of evolved stars and supernova ejecta.

\item The newly discovered the RAT trapping effect helps explain the existence of hot/warm large dust grains in stellar systems, AGN tori, and the survival of newly formed dust in the envelopes of evolved stars and supernova ejecta against RATD.

\end{itemize}

{We are grateful to the referee for constructive comments that has improved the presentation of the paper.} We thank Nguyen Chau Giang, Nguyen Bich Ngoc, and Nguyen Tat Thang for reading the first draft and for their useful comments. T.H. acknowledges the support from the major research project (No. 2025186902) from Korea Astronomy and Space Science Institute (KASI).

\appendix

\section{Results for different models of RATs, Magnetic Properties and Radiation Fields}
Here we show the results for the different models of RATs with $q^{\rm max}=1,2,3$ and SPM grains with $\delta_{\rm mag}=10^{3}$ for both CD and RD regimes. For this SPM grains, high-J attractors are always present regardless of the value of $q^{\rm max}$ (see Figure \ref{fig:deltacri_lowJ_highJ}). Therefore, we expect the results are similar to the case of $q^{\rm max}=2$, but shows for completeness.

\subsection{Results for different RAT Properties, $q^{\rm max}$}
\subsubsection{SPM grains in the collision-dominated regime}

\begin{figure*}
	\centering
\begin{overpic}[width=0.32\linewidth]{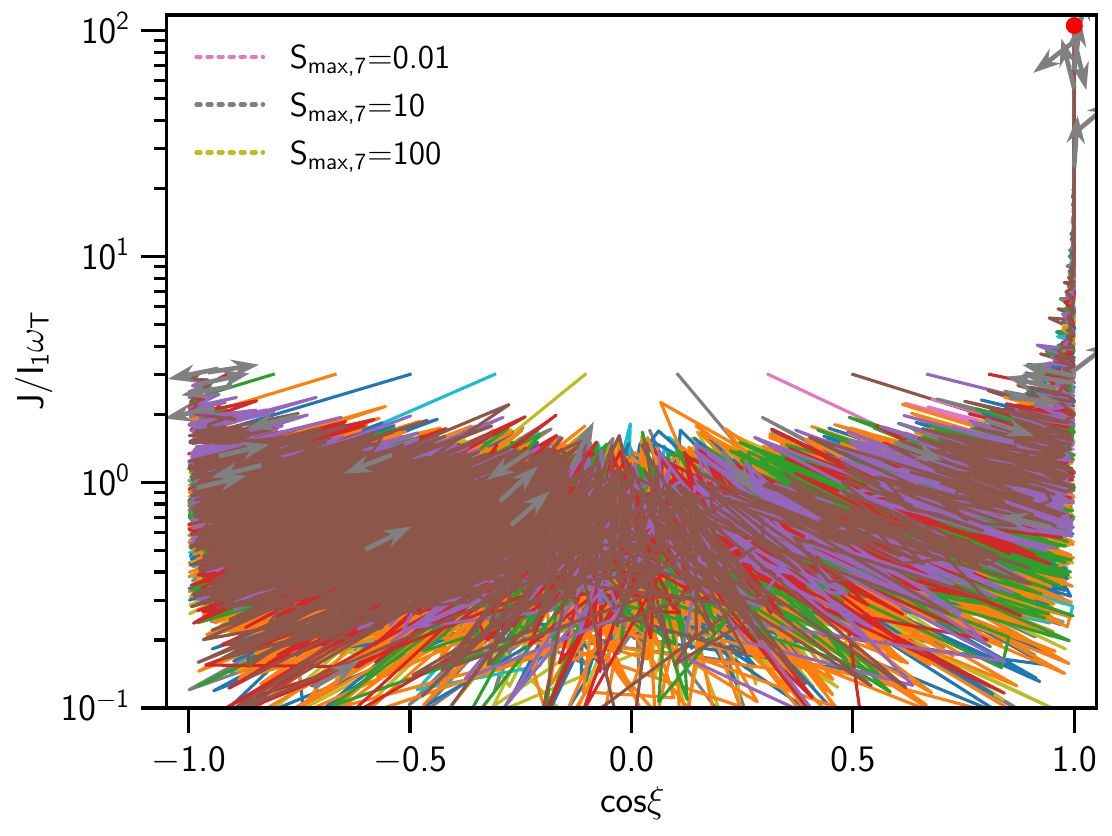}
			\put(65,75){\tiny \textbf{High-J Attractor}}
\put(45,65){\small \textbf{(a)} $U/(n_{1}T_{2}) = 0.16$}
\end{overpic}	
\begin{overpic}[width=0.32\linewidth]{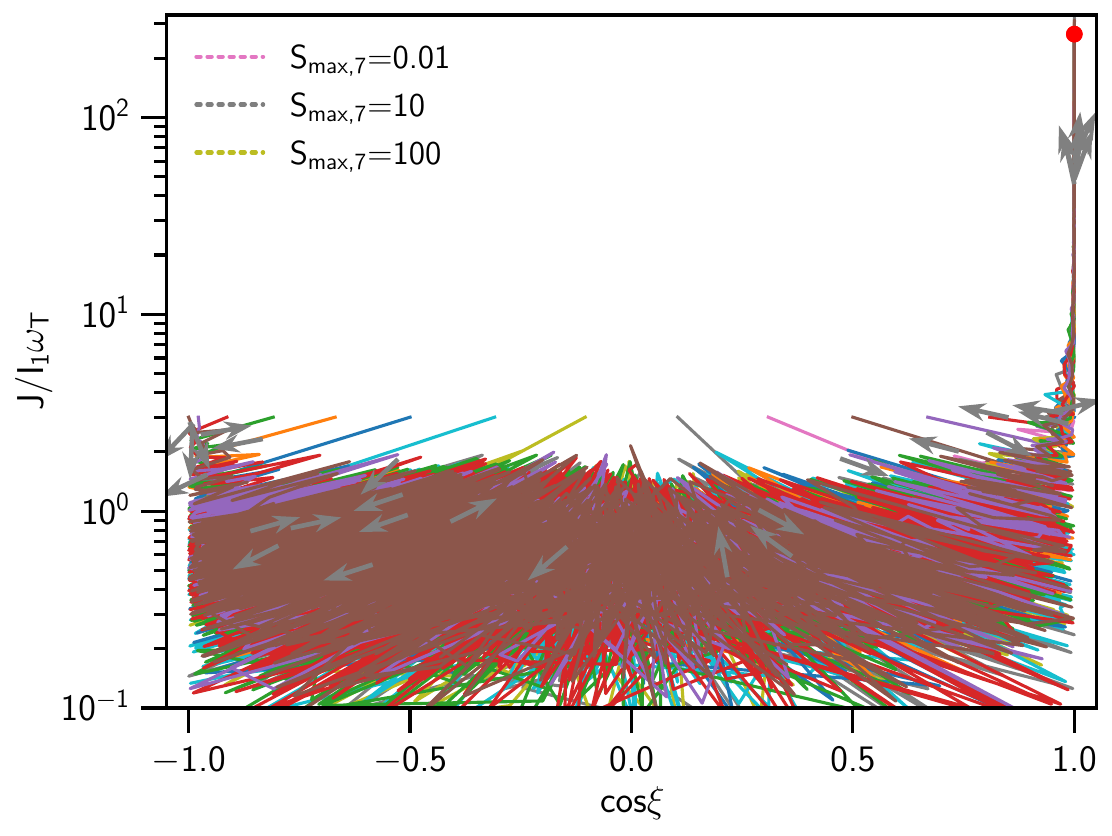} 
			\put(65,75){\tiny \textbf{High-J Attractor}}
	\put(45,65){\small \textbf{(b)} $U/(n_{1}T_{2}) = 0.6$}
	\end{overpic}	
	\begin{overpic}[width=0.32\linewidth]{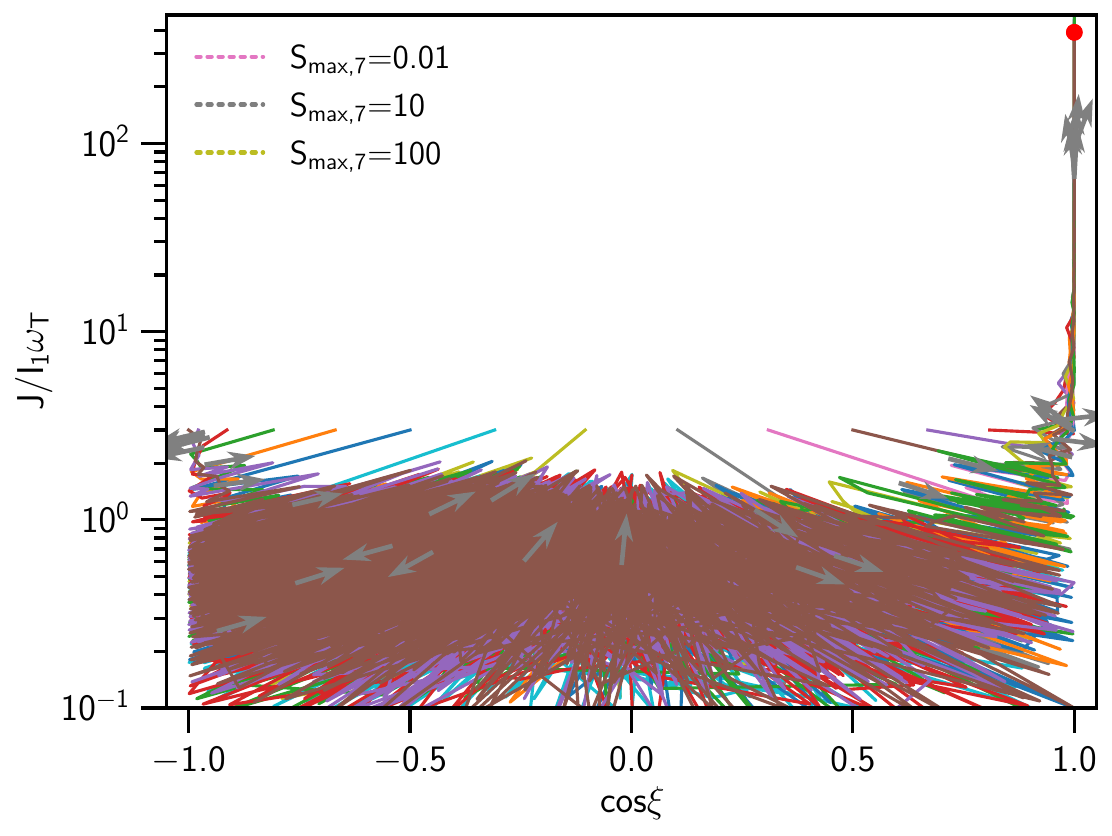}  
				\put(65,75){\tiny \textbf{High-J Attractor}}
	\put(45,65){\small \textbf{(c)} $U/(n_{1}T_{2}) = 1$}
\end{overpic}	
	\caption{Phase trajectory maps of grain alignment for the RAT model of $q^{\rm max}=1$ and SPM grains with $\delta_{\rm mag}=10^{3}$ in the CD regime and different $U/(n_{1}T_{2})$ (panels (a)-(c)). Some grains rapidly spin up to the high-J attractor, and the rest of the grains are slowly transported to high-J attractors due to gas collisions and magnetic fluctuations.}
	\label{fig:CDR_map_qmax1_nH1e1_aef02}
\end{figure*}

\begin{figure*}
\centering
\begin{overpic}[width=0.32\textwidth]{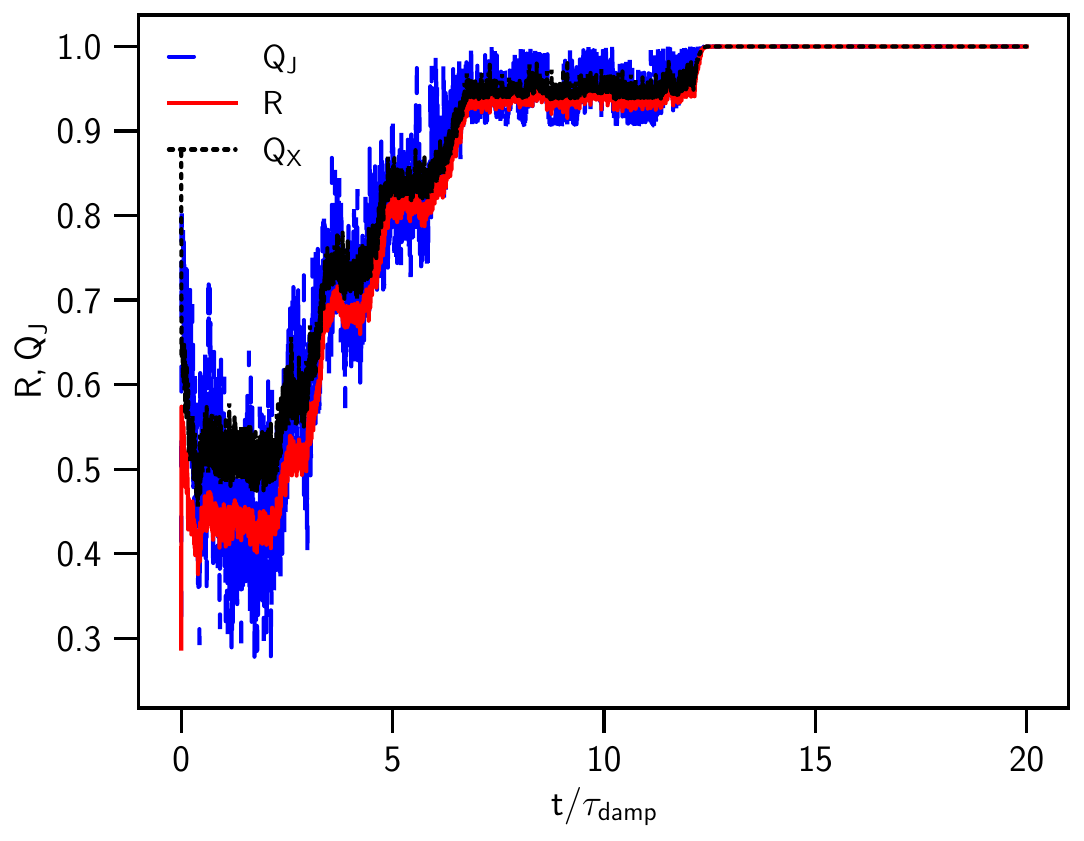}
	\put(40,20){\small \textbf{(a)} $U/(n_{1}T_{2}) = 0.16$}
	\end{overpic}	
\begin{overpic}[width=0.32\textwidth]{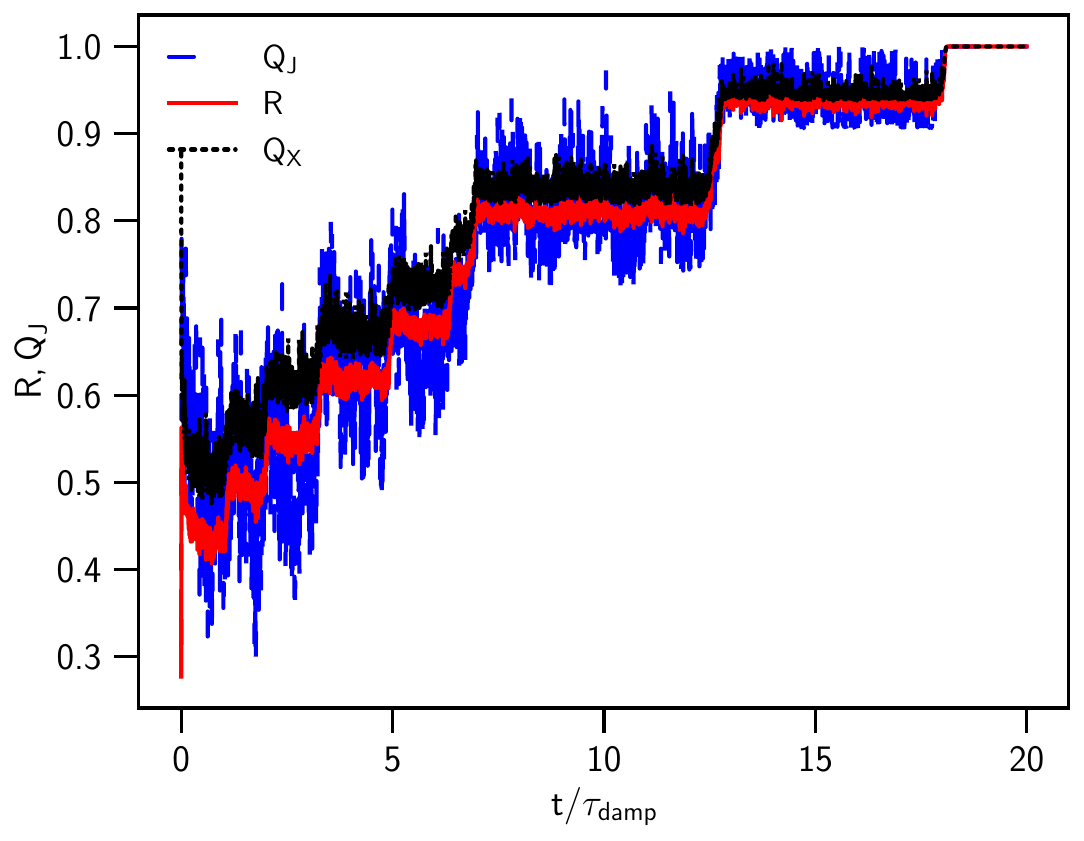}
	\put(40,20){\small \textbf{(b)} $U/(n_{1}T_{2}) = 0.6$}
	\end{overpic}	
\begin{overpic}[width=0.32\textwidth]{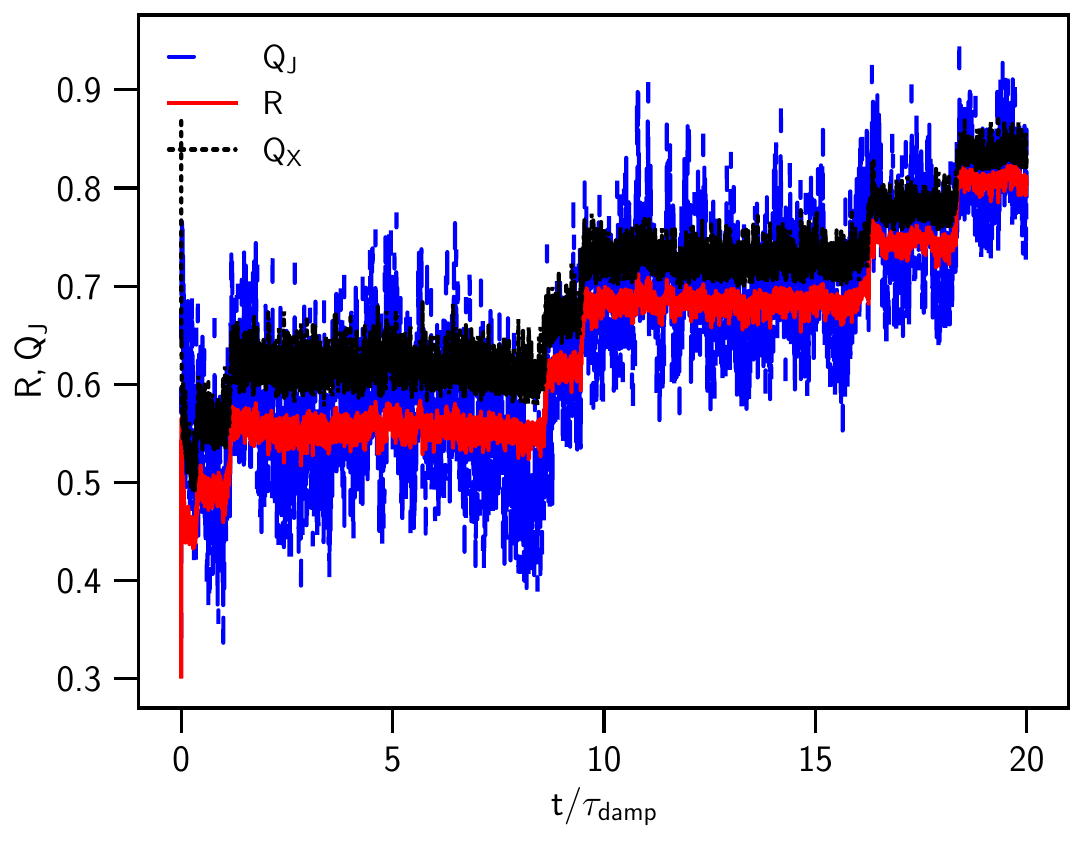}
	\put(40,20){\small \textbf{(c)} $U/(n_{1}T_{2}) = 1$}
	\end{overpic}	
	\caption{Same as Figure \ref{fig:CDR_map_qmax1_nH1e1_aef02} but for temporal evolution of the alignment degrees. Alignment degree increases with time due to the random transport of grains to high-J attractors and achieves perfect alignment at $T_{\rm PA}\sim 10-20\tau_{\rm damp}$.}
	\label{fig:CDR_RQJ_qmax1_nH1e1_aef02}	
\end{figure*}

Figures \ref{fig:CDR_map_qmax1_nH1e1_aef02} and \ref{fig:CDR_RQJ_qmax1_nH1e1_aef02} shows the phase map and time-dependent alignment degrees for the RAT model of $q^{\rm max}=1$ and SPM grains of $\delta_{\rm mag}=10^{3}$ in the CD regime. Some grains are directly driven to high-J attractors, but the rest are first driven to low-J rotation, where gas collisions and magnetic relaxation transport grains to high-J attractors. Grain alignment degrees increase with time and achieve perfect alignment due to collisional and magnetic excitations of grains from low-J to high-J attractors.

\begin{figure*}
	\centering
\begin{overpic}[width=0.32\linewidth]{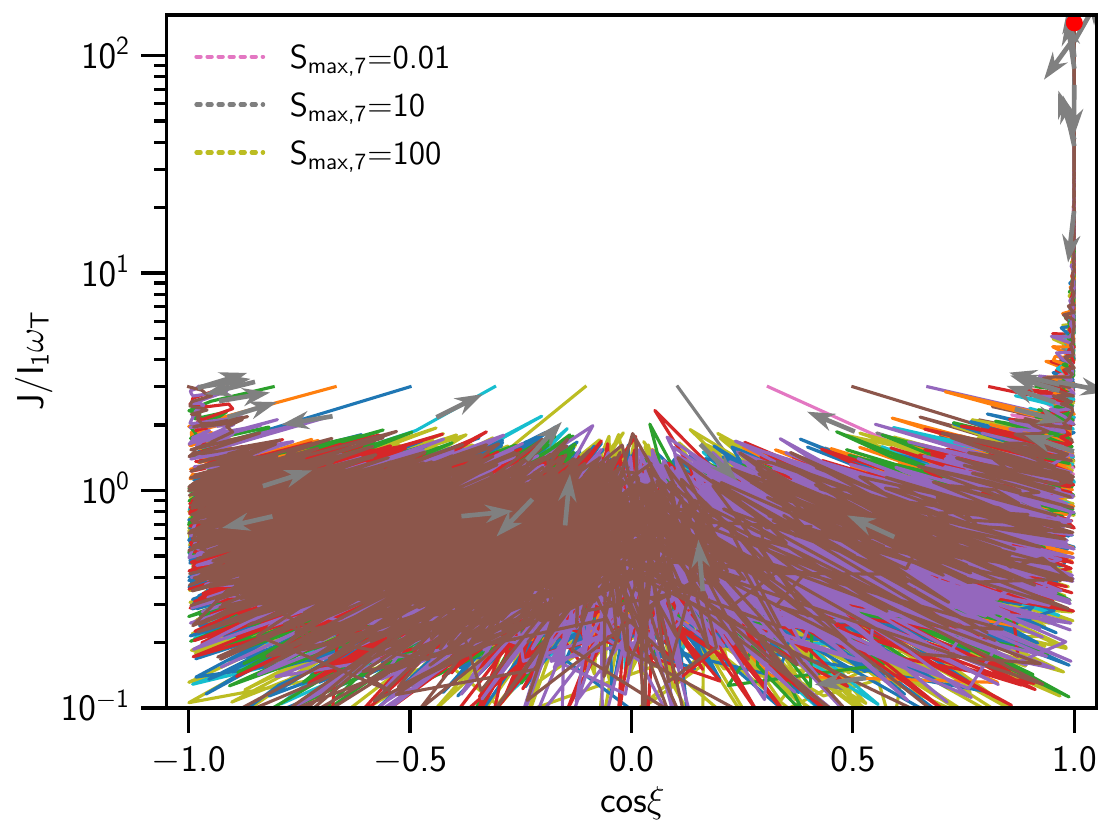}
			\put(65,75){\tiny \textbf{High-J Attractor}}
	\put(40,65){\small \textbf{(a)} $U/(n_{1}T_{2}) = 0.16$}
\end{overpic}	
\begin{overpic}[width=0.32\linewidth]{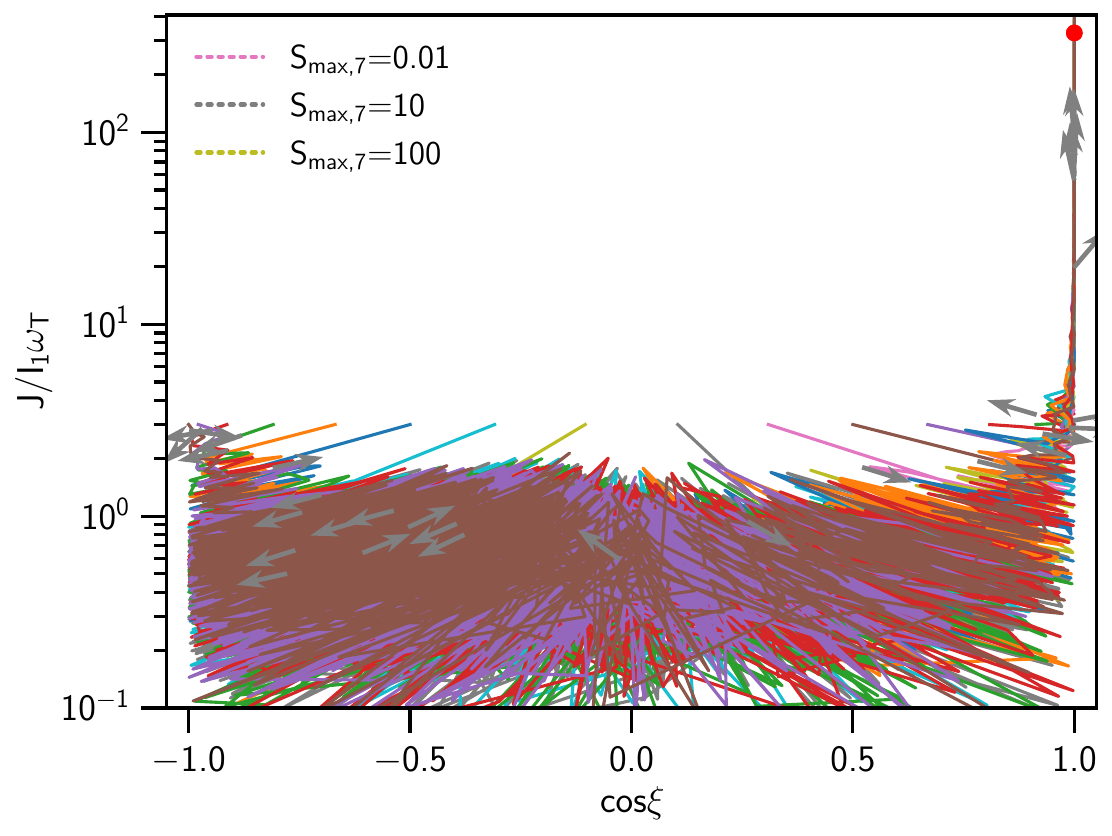}
			\put(65,75){\tiny \textbf{High-J Attractor}}
		\put(40,65){\small \textbf{(b)} $U/(n_{1}T_{2}) = 0.6$}
	\end{overpic}	
	 \begin{overpic}[width=0.32\linewidth]{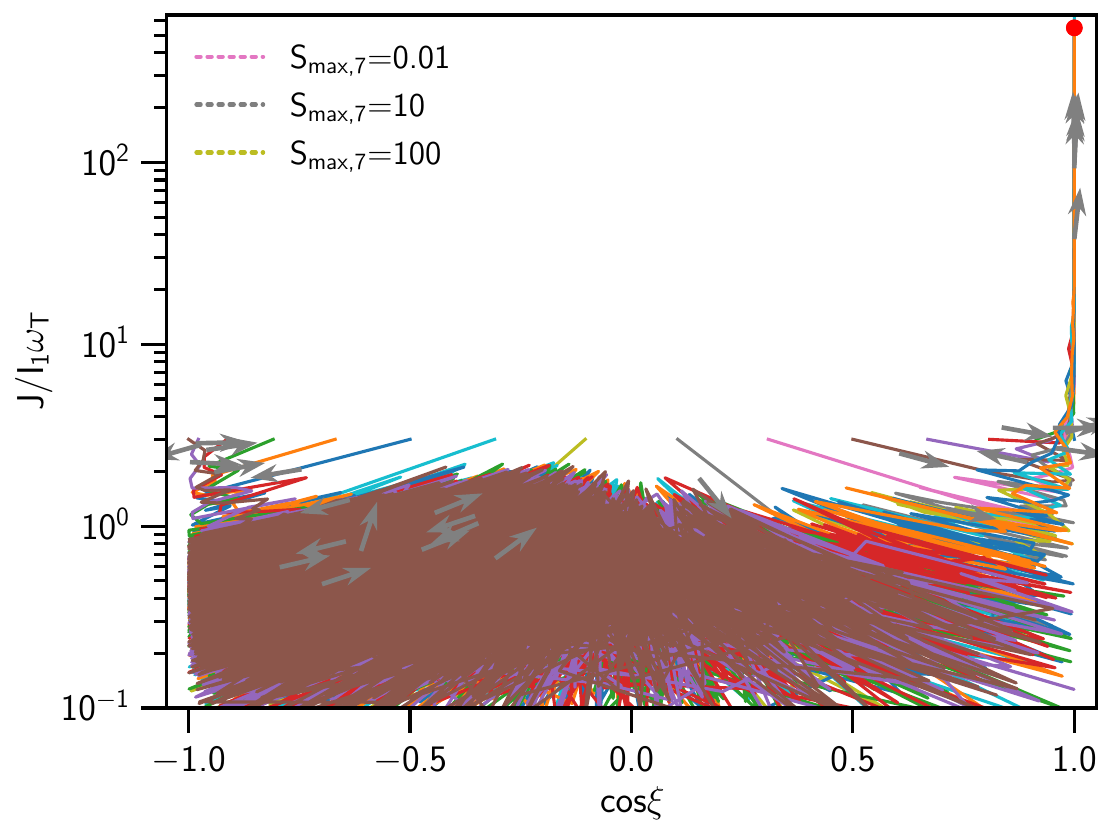}  
	 			\put(65,75){\tiny \textbf{High-J Attractor}}
	 	    	\put(40,65){\small \textbf{(c)} $U/(n_{1}T_{2}) = 1$}
	 \end{overpic}	
	\caption{Phase trajectory maps of grain alignment for the RAT model of $q^{\rm max}=3$ and SPM grains of $\delta_{\rm mag}=10^{3}$ in the CD regime of different $U/(n_{1}T_{2})$ (panels (a)-(c)). Some grains are rapidly spun up to the high-J attractor, but the rest of the grains are driven to the low-J rotation, followed by slow transport to high-J attractors by collisional and magnetic excitations.}
	\label{fig:CDR_map_qmax3_nH1e1_aef02}
\end{figure*}

\begin{figure*}	
	\centering
\begin{overpic}[width=0.32\linewidth]{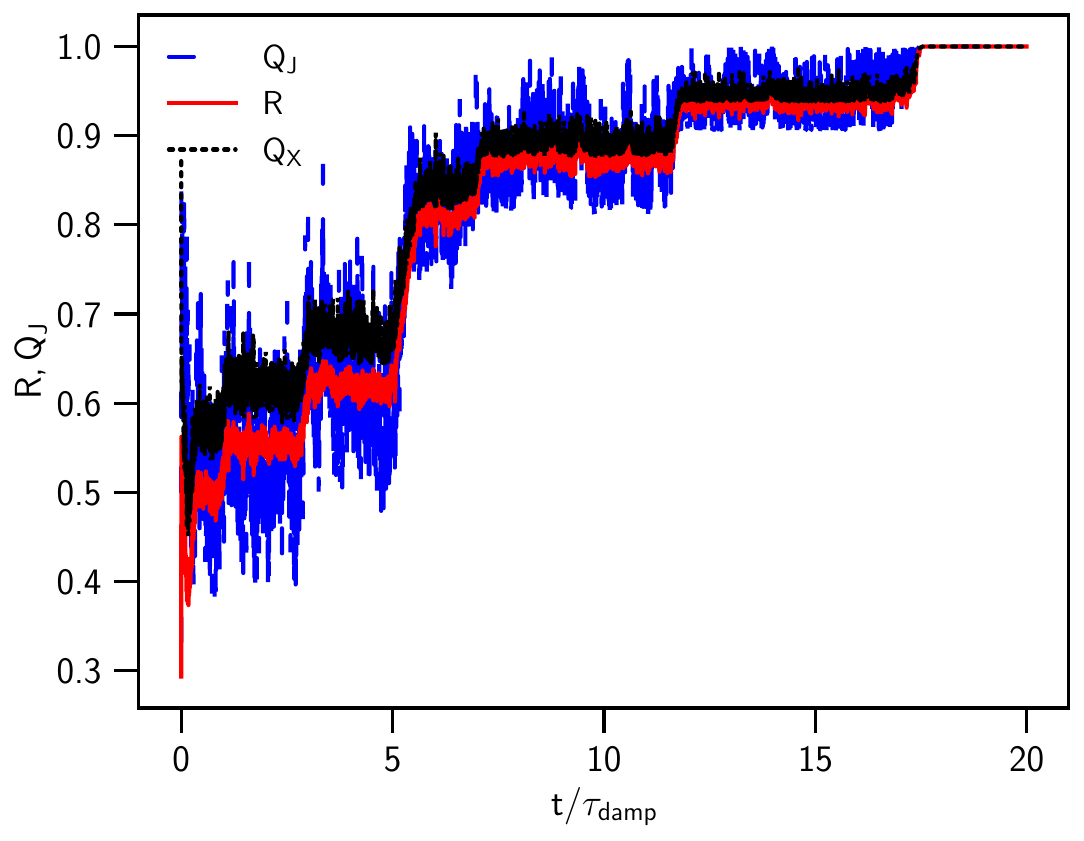}
		\put(40,20){\small \textbf{(a)} $U/(n_{1}T_{2}) = 0.16$}
	\end{overpic}	
\begin{overpic}[width=0.32\linewidth]{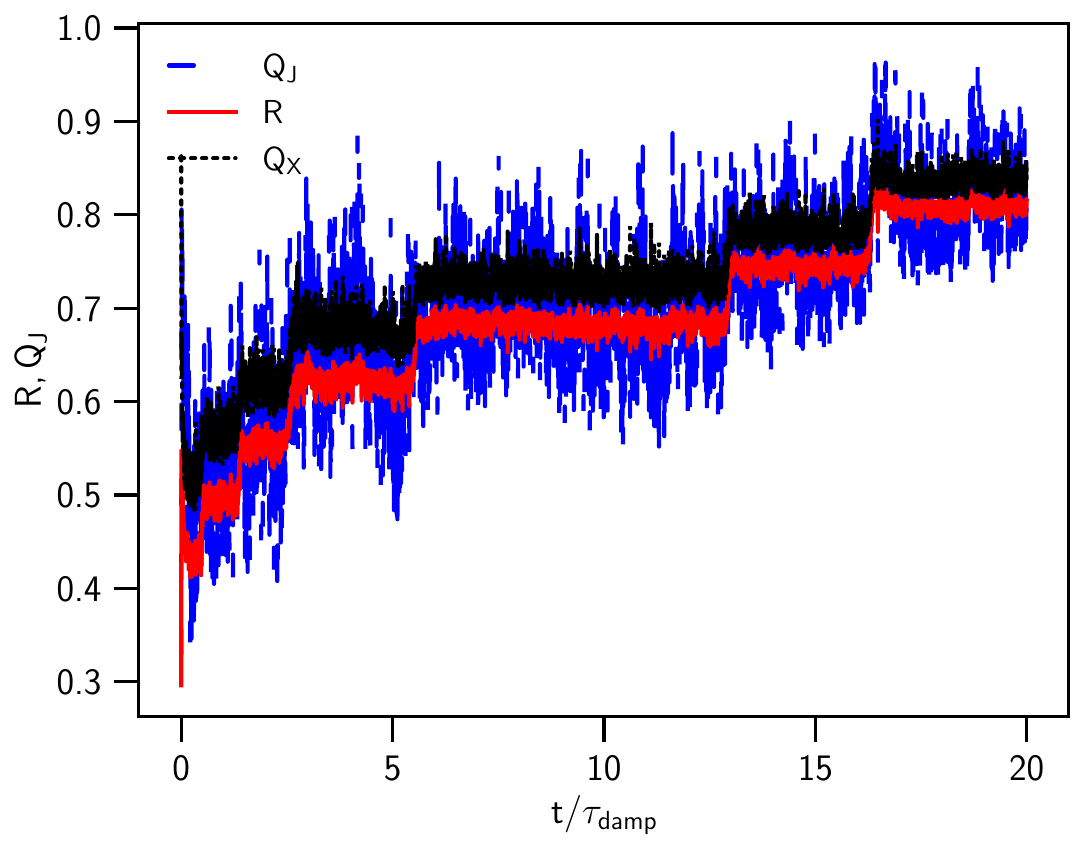}
		\put(40,20){\small \textbf{(b)} $U/(n_{1}T_{2}) = 0.6$}
	\end{overpic}	
\begin{overpic}[width=0.32\linewidth]{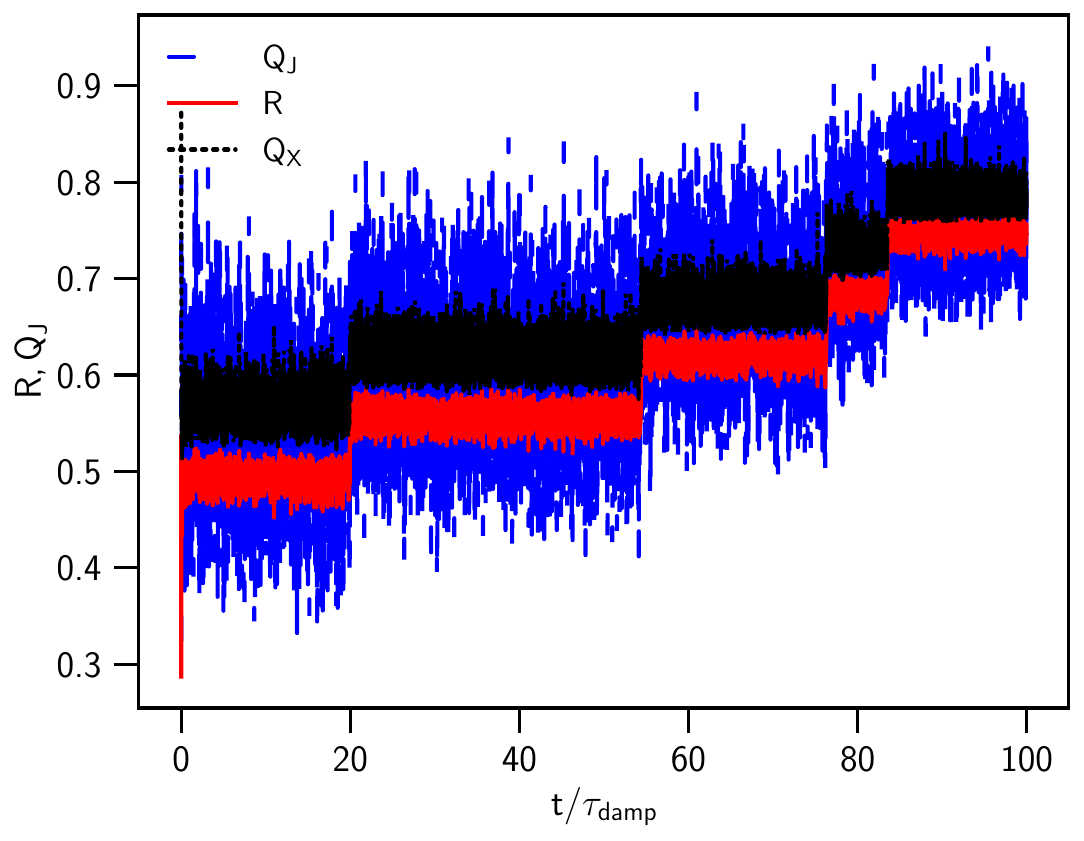}
		\put(40,20){\small \textbf{(c)} $U/(n_{1}T_{2}) = 1$}
	\end{overpic}	
	\caption{Same as Figure \ref{fig:CDR_map_qmax3_nH1e1_aef02} but for temporal evolution of the alignment degrees. Alignment degree increases with time due to slow transport of grains to high-J attractors by collisional and magnetic excitations and achieve perfect at $T_{\rm PA}\sim 20-100\tau_{\rm damp}$.}
	\label{fig:CDR_RQJ_qmax3_nH1e1_aef02}
\end{figure*}

Figures \ref{fig:CDR_map_qmax3_nH1e1_aef02} and \ref{fig:CDR_RQJ_qmax3_nH1e1_aef02} show the phase map and time-dependent alignment degrees for $q^{\rm max}=3$. The results are similar to the RAT model of $q^{\rm max}=2$ shown in \ref{fig:CDR_map_qmax2_nH1e1_aef02} and \ref{fig:CDR_RQJ_qmax2_nH1e1_aef02}. This arises from the fact that for the SPM grains of $\delta_{\rm mag}=10^{3}$, grains have high-J attractors due to the MRAT (see Figure \ref{fig:deltacri_lowJ_highJ}).

\subsubsection{SPM grains in the radiation-dominated regimes}

\begin{figure*}
	\centering
\begin{overpic}[width=0.32\linewidth]{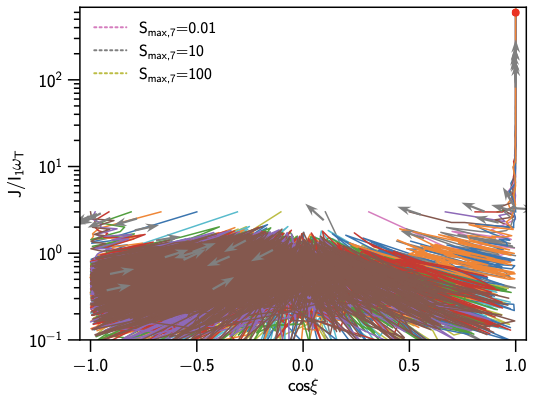}
			\put(65,75){\tiny \textbf{High-J Attractor}}
	\put(20,35){\tiny RAT Trapping}
			\put(40,65){\small \textbf{(a)} $U/(n_{1}T_{2}) = 1.9$}
	\end{overpic}	
\begin{overpic}[width=0.32\linewidth]{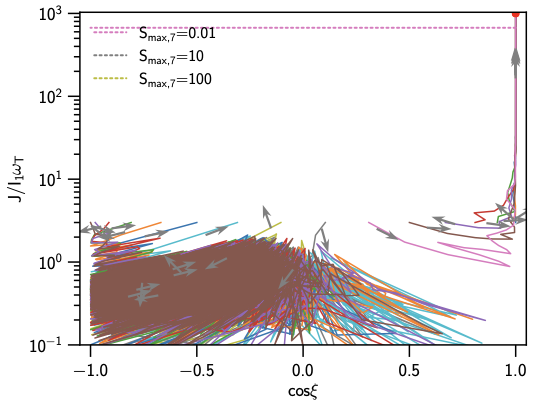} 
			\put(65,75){\tiny \textbf{High-J Attractor}}
	\put(20,35){\tiny RAT Trapping}
			\put(40,65){\small \textbf{(b)} $U/(n_{1}T_{2}) = 4.1$}
	\end{overpic}	     
\begin{overpic}[width=0.32\linewidth]{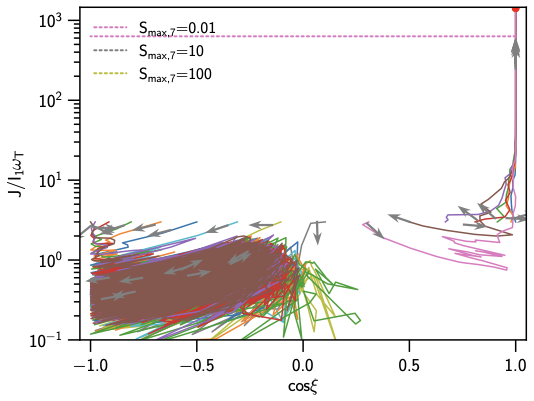}    
			\put(65,75){\tiny \textbf{High-J Attractor}}
	\put(20,35){\tiny RAT Trapping}
			\put(40,65){\small \textbf{(c)} $U/(n_{1}T_{2}) = 7.3$}
	\end{overpic}	
	\caption{Phase trajectory maps of grain alignment for the RAT model of $q^{\rm max}=1$ and SPM grains of  $\delta_{\rm mag}=10^{3}$  in the RD regime with different $U/(n_{1}T_{2})$ (panels (a)-(c)). The value of is fixed. Some grains are driven rapidly to be aligned at the high-J attractor, and the rest of the grains are driven to low-J rotation and trapped by the RAT trapping.}
	\label{fig:RDR_map_qmax1_nH1e1_aef02}
\end{figure*}

\begin{figure*}	
\begin{overpic}[width=0.32\linewidth]{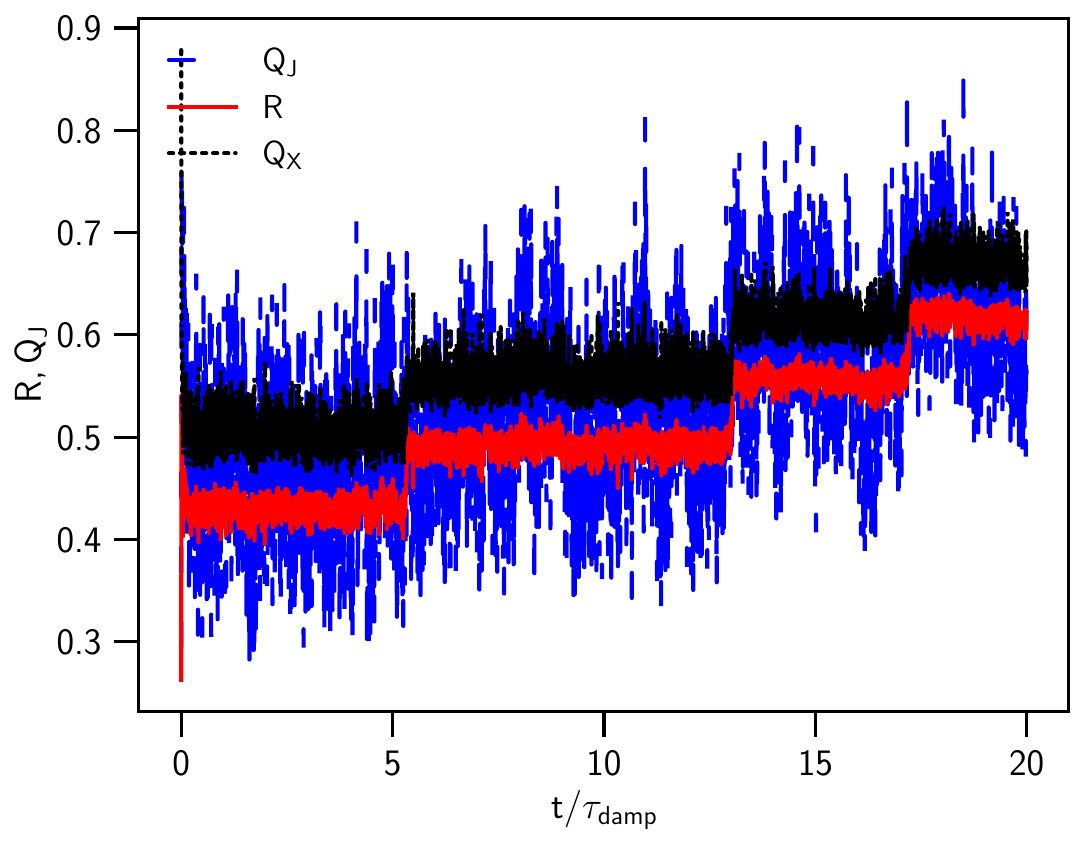}
			\put(40,67){\small \textbf{(a)} $U/(n_{1}T_{2}) = 1.9$}
	\end{overpic}	
\begin{overpic}[width=0.32\linewidth]{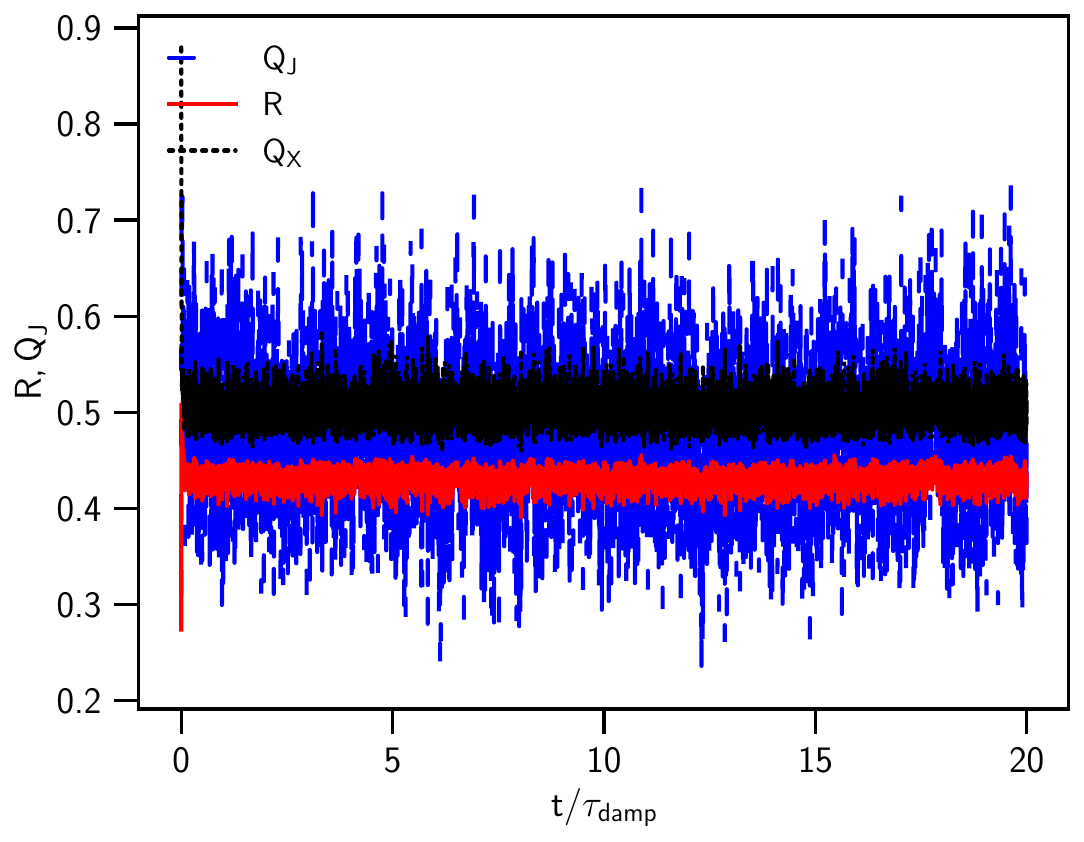}
			\put(40,67){\small \textbf{(b)} $U/(n_{1}T_{2}) = 4.1$}
	\end{overpic}	
\begin{overpic}[width=0.32\linewidth]{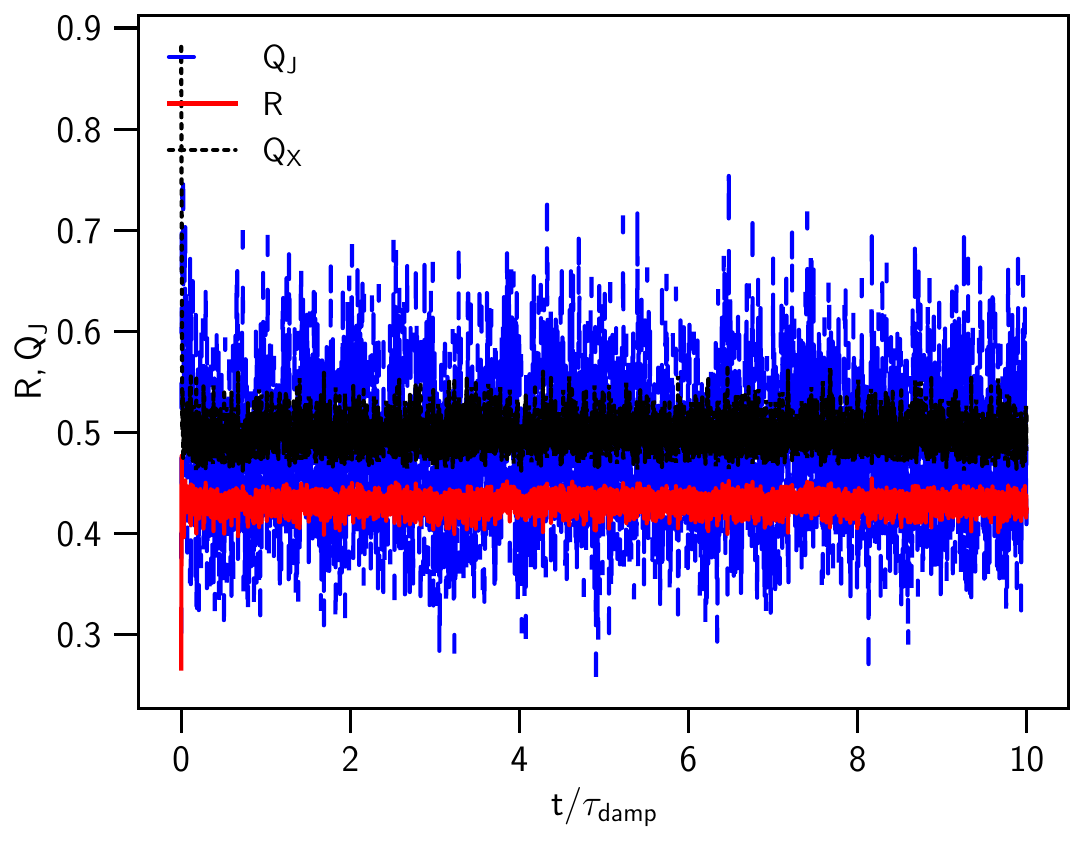}
			\put(40,67){\small \textbf{(c)} $U/(n_{1}T_{2}) = 7.3$}
	\end{overpic}	
	\caption{Same as Figure \ref{fig:RDR_map_qmax1_nH1e1_aef02} but for temporal evolution of the alignment degrees. Alignment degree increases rapidly within $0.01\tau_{\rm damp}$ and becomes saturated at $R\sim 0.4$ for intense radiation due to RAT trapping.}
	\label{fig:RDR_RQJ_qmax1_nH1e1_aef02}
\end{figure*}

Figures \ref{fig:RDR_map_qmax1_nH1e1_aef02} and \ref{fig:RDR_RQJ_qmax1_nH1e1_aef02} show the 
shows the phase map and time-dependent alignment degrees for the RAT model of $q^{\rm max}=1$ and SPM grains of $\delta_{\rm mag}=10^{3}$ in the RD regime. A fraction of grains is rapidly aligned at high-J attractor by the MRAT mechanism, but the majority of grains are driven to low-J and trapped there due to dominant radiative torques. Collisions and magnetic fluctuations can scatter grain orientation at low-J rotation, but are not strong enough to overcome RAT trapping. Grain alignment degrees increase rapidly first due to the MRAT alignment for a fraction of grains heading to high-J attractors, and then they get saturated at low alignment degrees due to RAT trapping of grains at low-J rotation.

\begin{figure*}
	\centering
\begin{overpic}[width=0.32\linewidth]{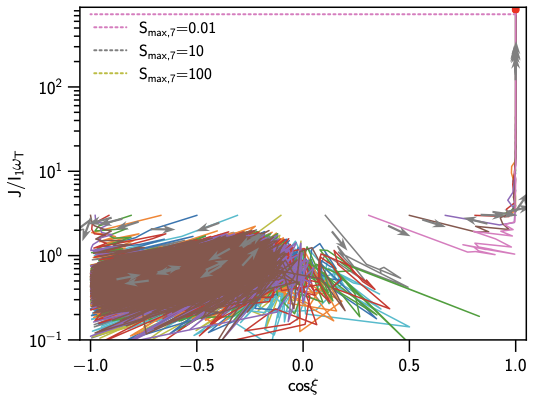}
			\put(65,75){\tiny \textbf{High-J Attractor}}
				\put(40,65){\small \textbf{(a)} $U/(n_{1}T_{2}) = 1.9$}
	\end{overpic}	
\begin{overpic}[width=0.32\linewidth]{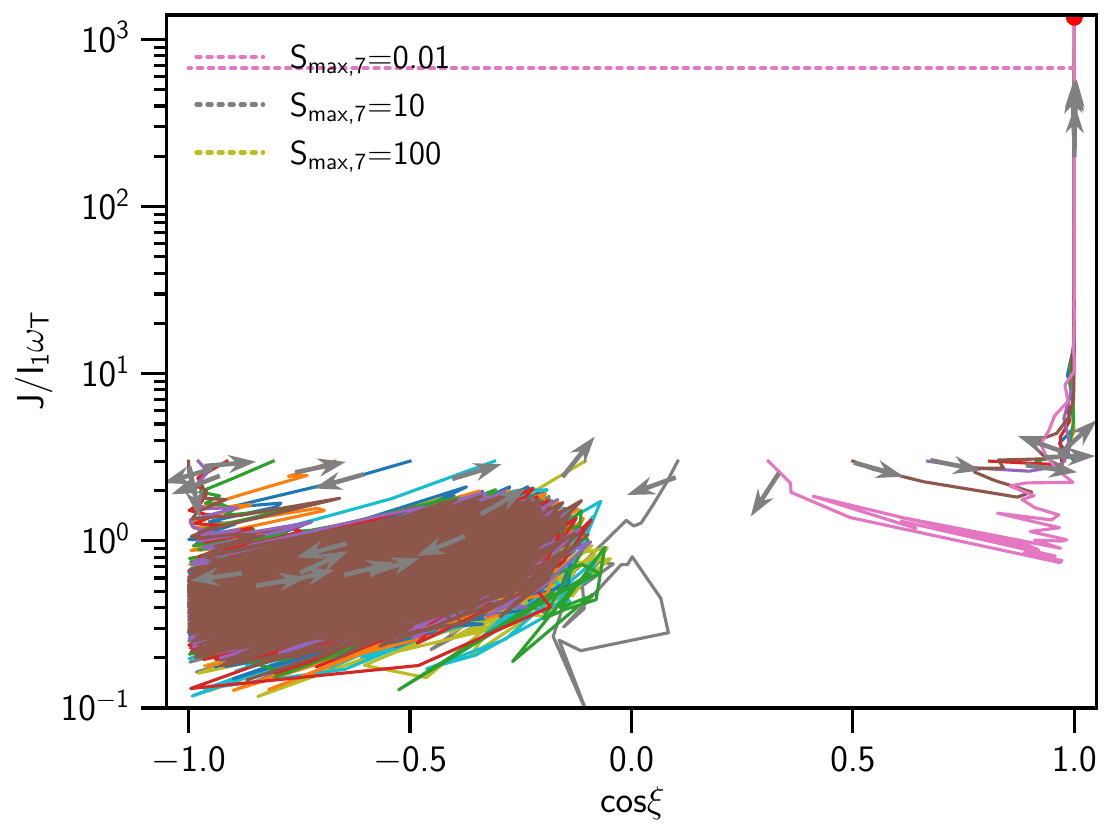}  
			\put(65,75){\tiny \textbf{High-J Attractor}}
				\put(40,65){\small \textbf{(b)} $U/(n_{1}T_{2}) = 4.1$}
	\end{overpic}
\begin{overpic}[width=0.32\linewidth]{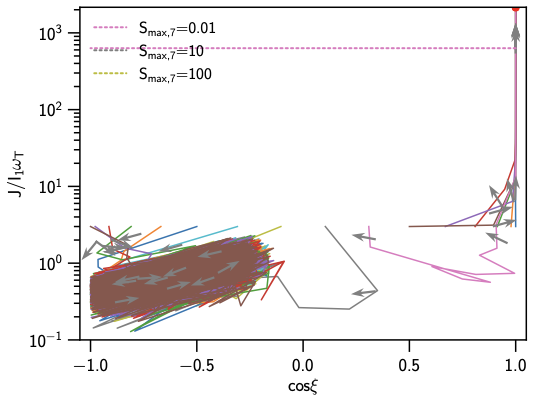} 
			\put(65,75){\tiny \textbf{High-J Attractor}}
				\put(40,65){\small \textbf{(c)} $U/(n_{1}T_{2}) = 7.3$}
	\end{overpic}
	\caption{Phase trajectory maps of grain alignment for the RAT model of $q^{\rm max}=3$ and SPM grains of $\delta_{\rm mag}=10^{3}$ in the RD regime of different $U/(n_{1}T_{2})$ (panels (a)-(c)). Some grains are driven rapidly to be aligned at the high-J attractor, and the rest of the grains are driven to low-J rotation and trapped by the RAT trapping.}
	\label{fig:RDR_map_qmax3_nH1e1_aef02}
\end{figure*}

\begin{figure*}	
\centering
\begin{overpic}[width=0.32\linewidth]{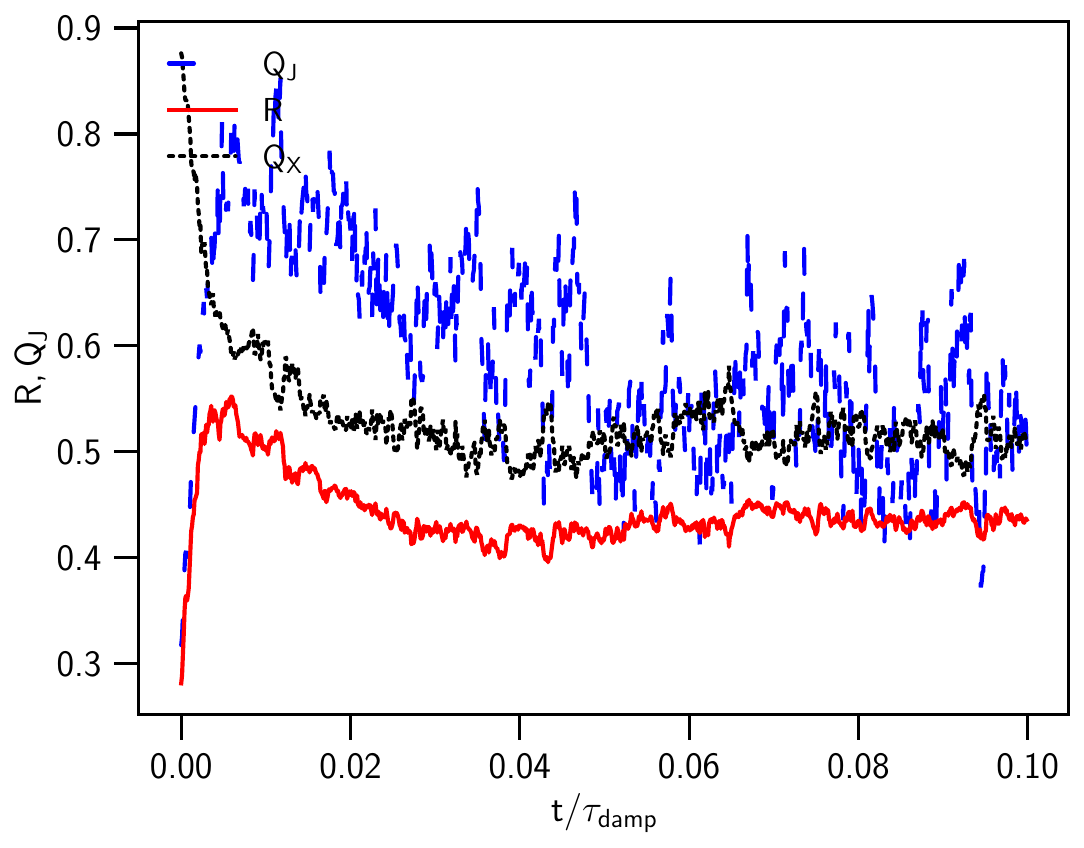}
				\put(40,65){\small \textbf{(a)} $U/(n_{1}T_{2}) = 1.9$}
	\end{overpic}	
\begin{overpic}[width=0.32\linewidth]{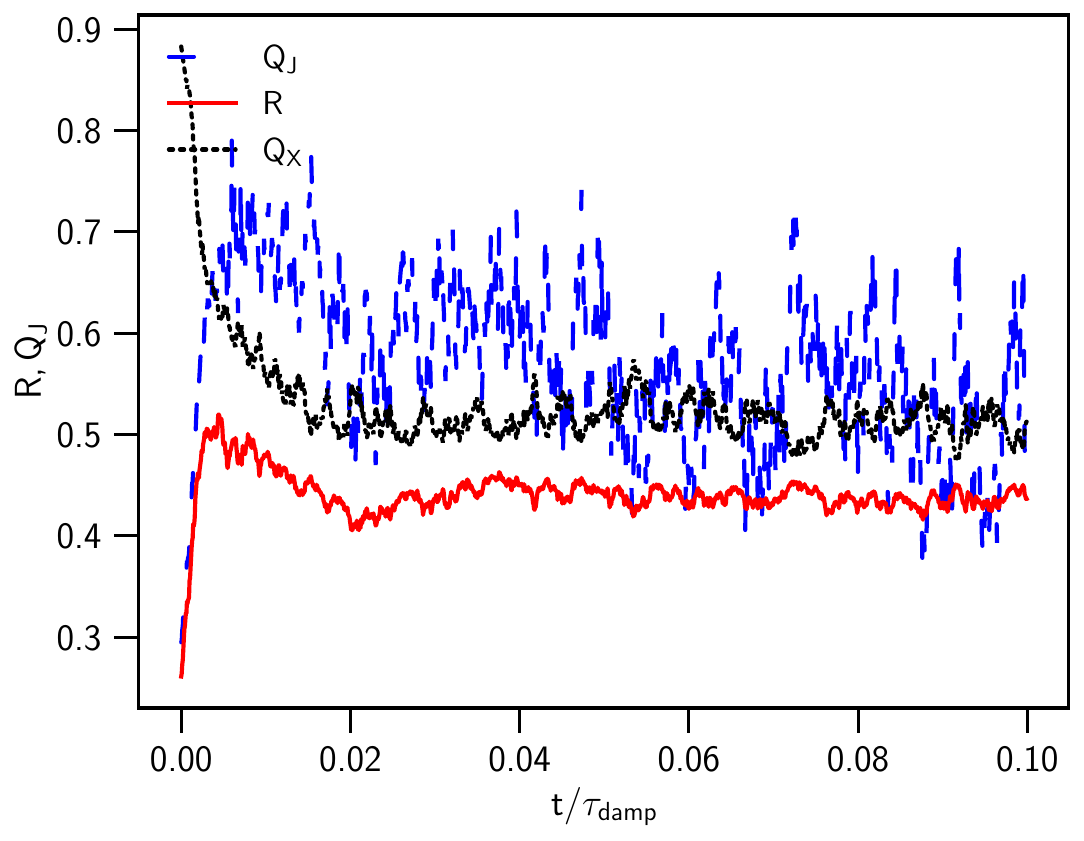}
				\put(40,65){\small \textbf{(b)} $U/(n_{1}T_{2}) = 4.1$}
	\end{overpic}	
\begin{overpic}[width=0.32\linewidth]{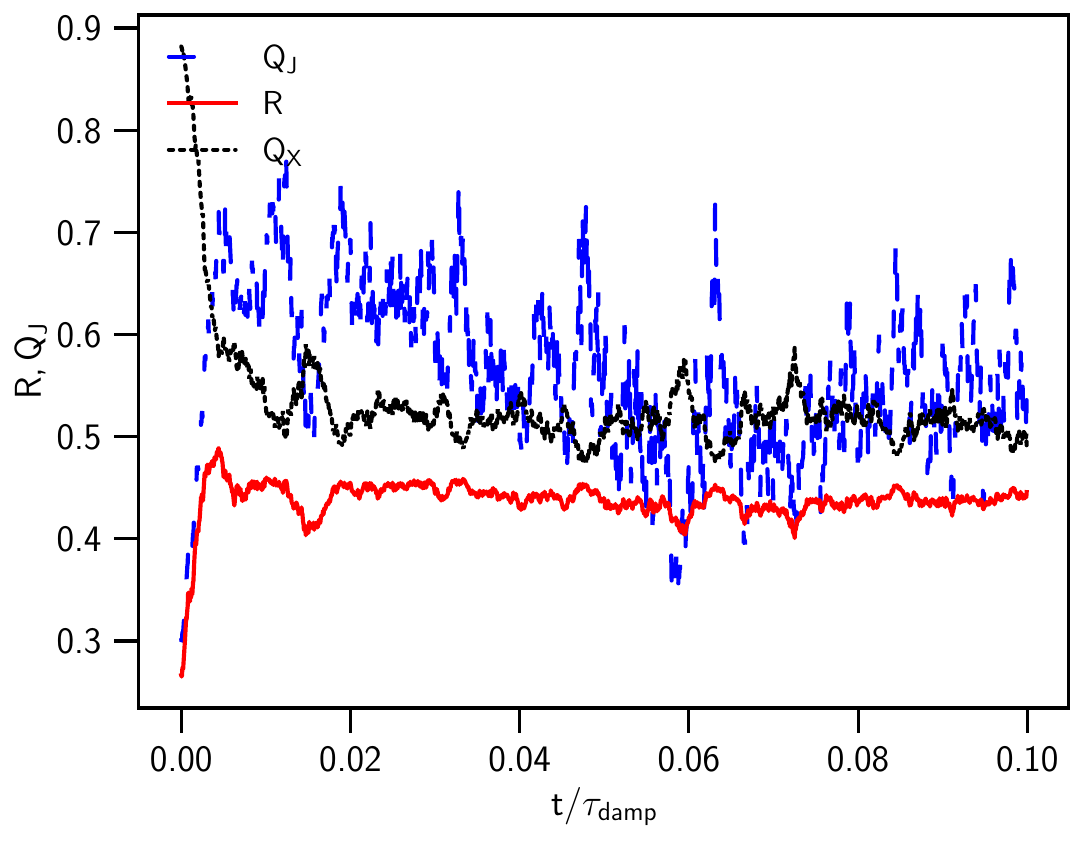}
				\put(40,65){\small \textbf{(c)} $U/(n_{1}T_{2}) = 7.3$}
	\end{overpic}	
	\caption{Same as Figure \ref{fig:RDR_map_qmax3_nH1e1_aef02} but for temporal evolution of the alignment degrees for $\delta_{\rm mag}=10^{3}$. Alignment degrees increase rapidly within $0.01\tau_{\rm damp}$ and become saturated at $R\sim 0.45$ for intense radiation due to RAT trapping.}
	\label{fig:RDR_RQJ_qmax3_nH1e1_aef02}
\end{figure*}

Figures \ref{fig:RDR_map_qmax3_nH1e1_aef02} and \ref{fig:RDR_RQJ_qmax3_nH1e1_aef02} show the phase-map and time-dependence alignment degrees for the RAT model of $q^{\rm max}=3$ in the RD regime. The similar features are observed as in $q^{\rm max}=1,2$ because the high-J attractor is produced by SPM relaxation with $\delta_{\rm mag}=10^{3}$.

\subsection{PM Grains in the CD and RD regimes}

\begin{figure*}
\centering
\begin{overpic}[width=0.32\linewidth]{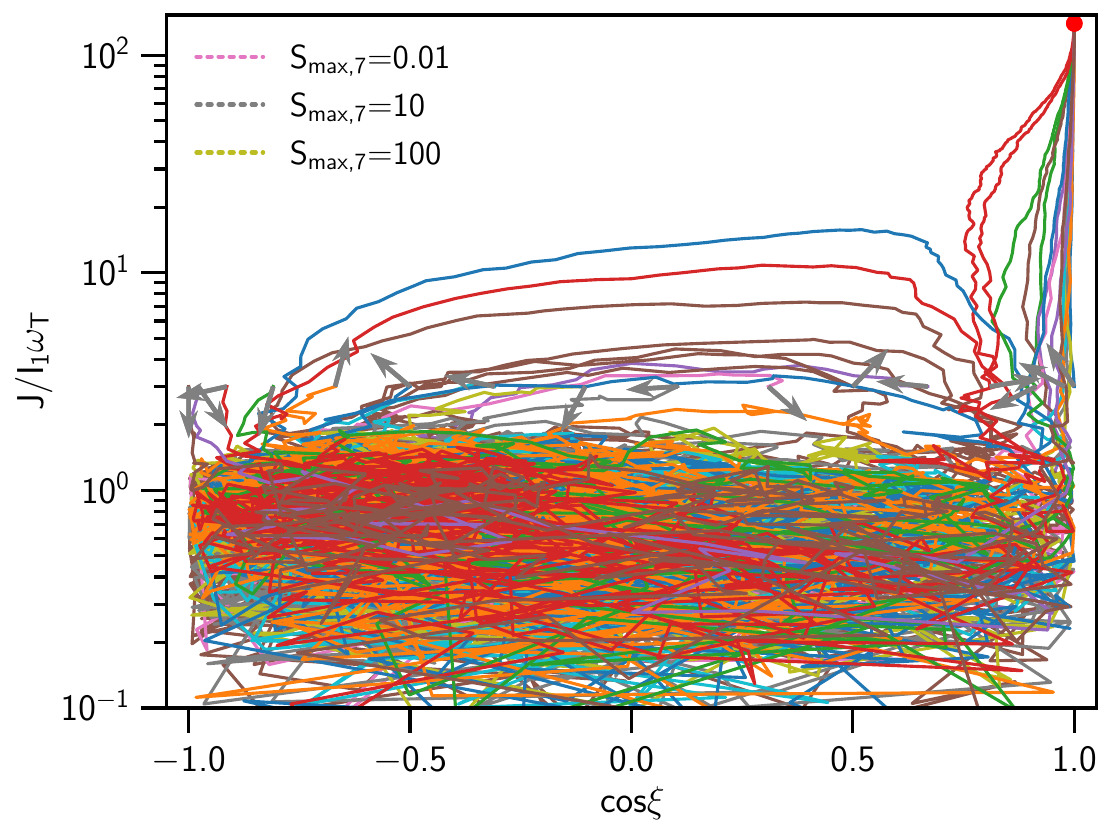}
					\put(40,65){\small \textbf{(a)} $U/(n_{1}T_{2}) = 0.16$}
	\end{overpic}	
\begin{overpic}[width=0.32\linewidth]{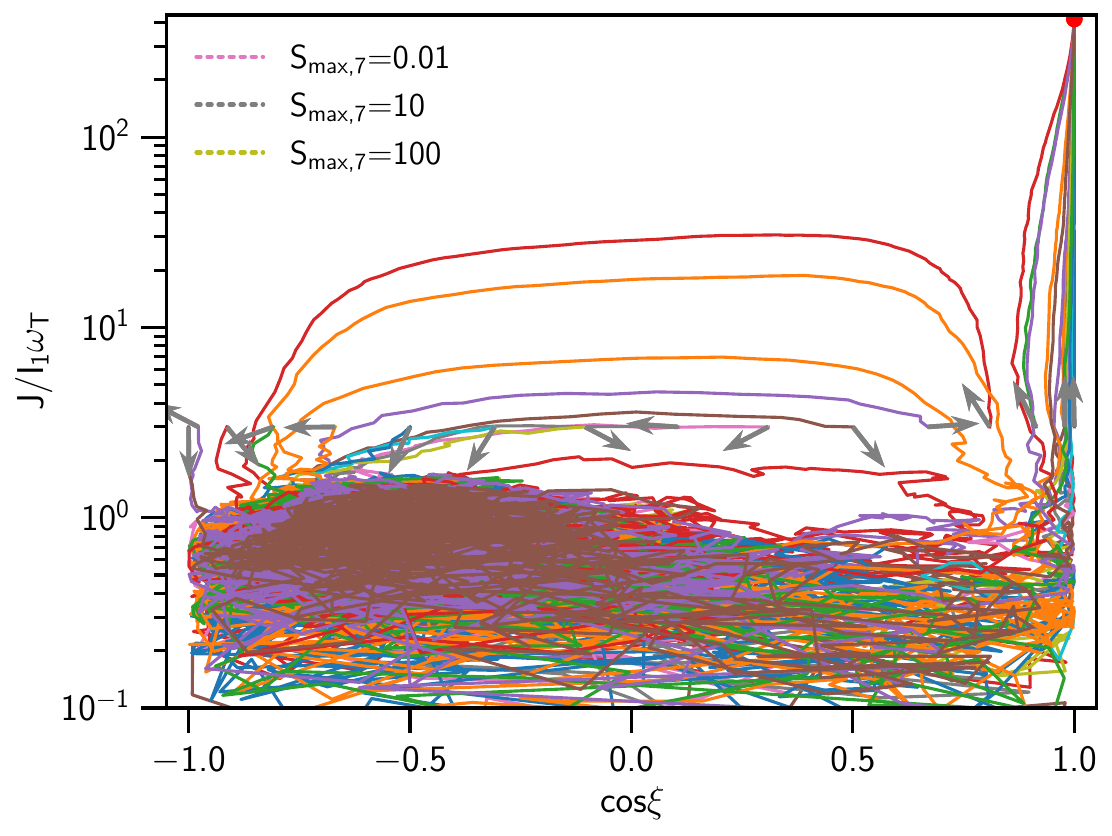}
					\put(40,65){\small \textbf{(b)} $U/(n_{1}T_{2}) = 0.6$}
	\end{overpic}	  
\begin{overpic}[width=0.32\linewidth]{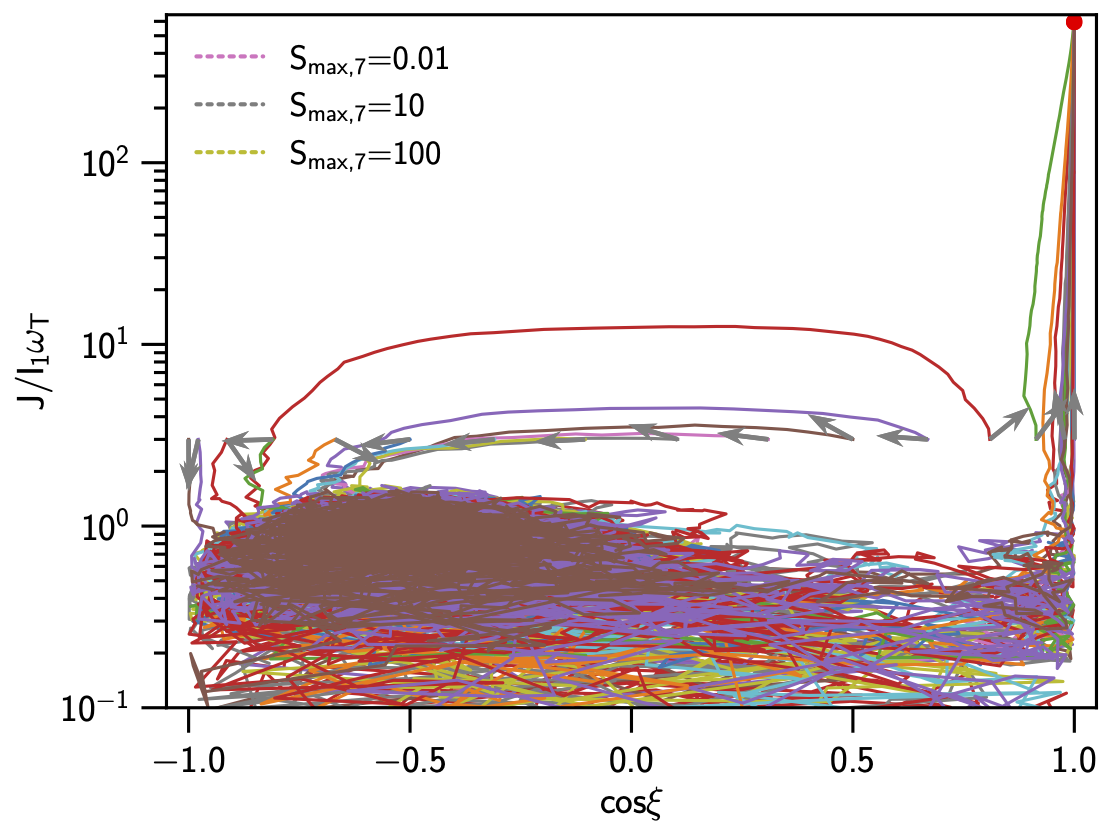}      
					\put(40,65){\small \textbf{(c)} $U/(n_{1}T_{2}) = 1$}
	\end{overpic}	        
	\caption{Phase trajectory maps for PM grains of $\delta_{\rm mag}=1$ and the RAT model of $q^{\rm max}=3$ in CD regime of different $U/(n_{1}T_{2})$ (panels (a)-(c)). Grains are eventually aligned at high-J attractors by RATs due to collisional and magnetic excitations.}
	\label{fig:CDR_map_qmax3_nH1e1_aef02_deltam1}
\end{figure*}

\begin{figure*}	
\begin{overpic}[width=0.32\linewidth]{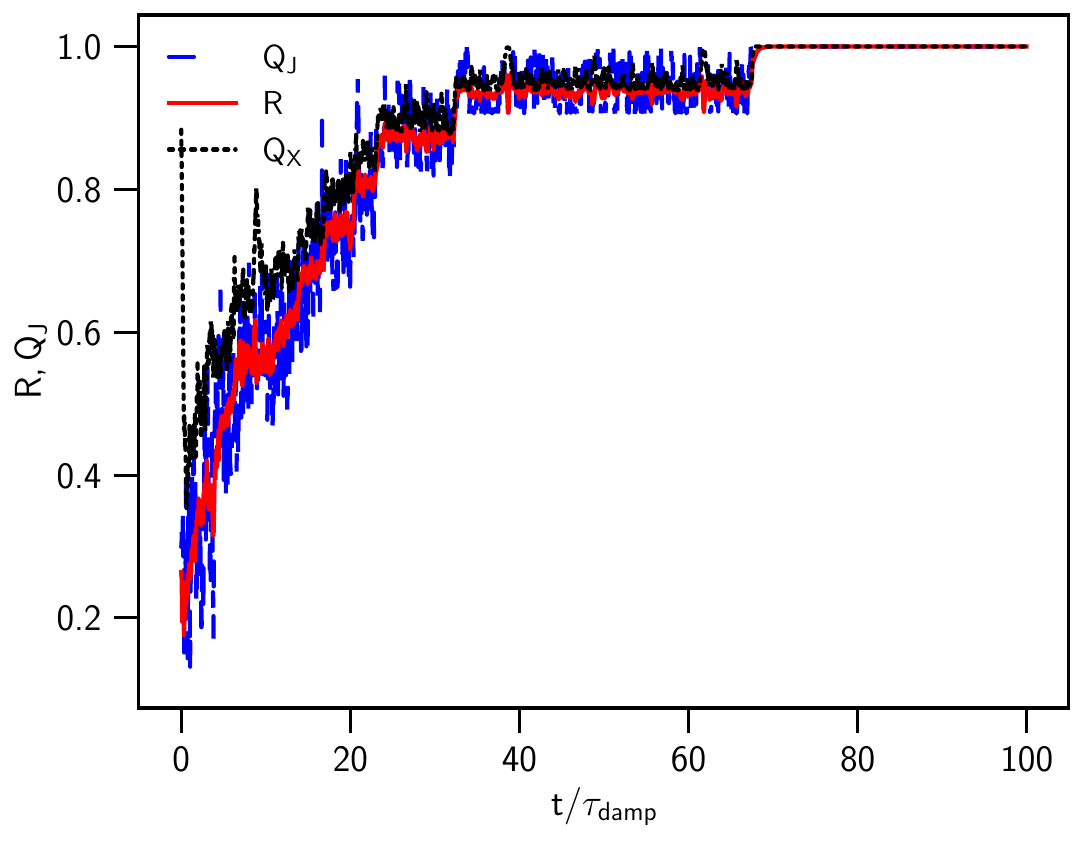}
					\put(40,20){\small \textbf{(a)} $U/(n_{1}T_{2}) = 0.16$}
	\end{overpic}	
\begin{overpic}[width=0.32\linewidth]{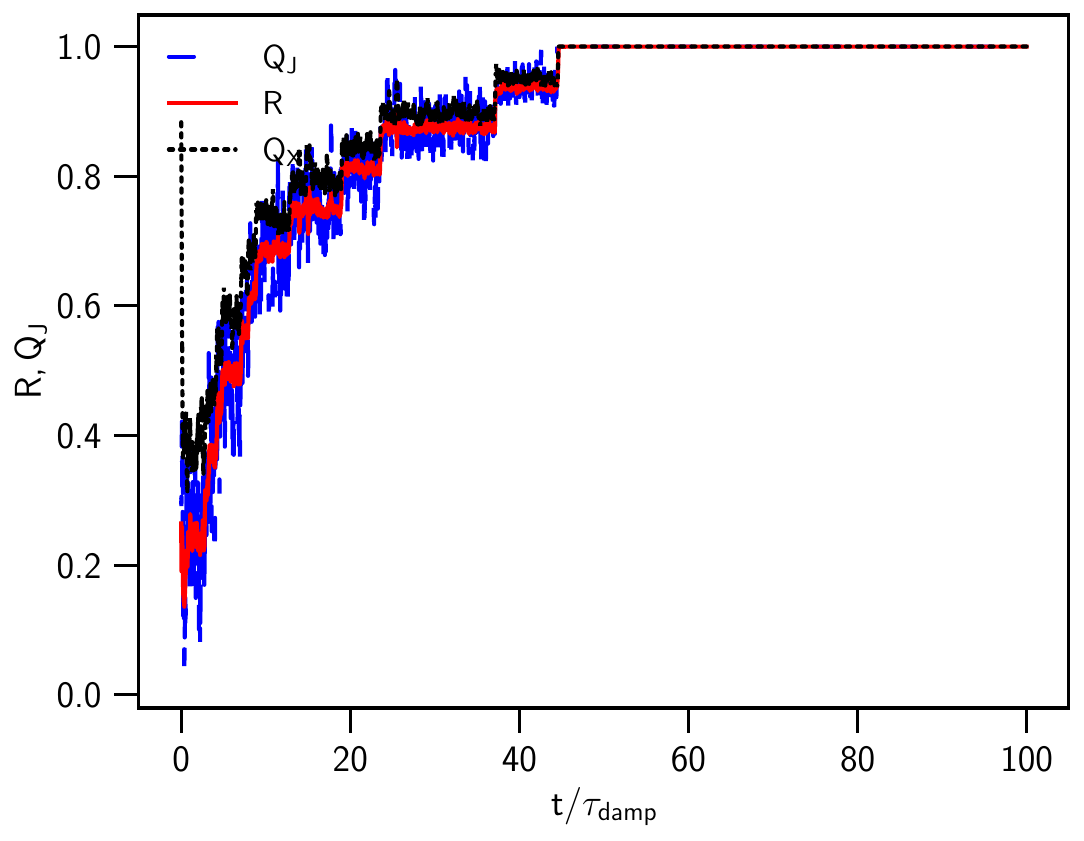}
					\put(40,20){\small \textbf{(b)} $U/(n_{1}T_{2}) = 0.6$}
	\end{overpic}	
\begin{overpic}[width=0.32\linewidth]{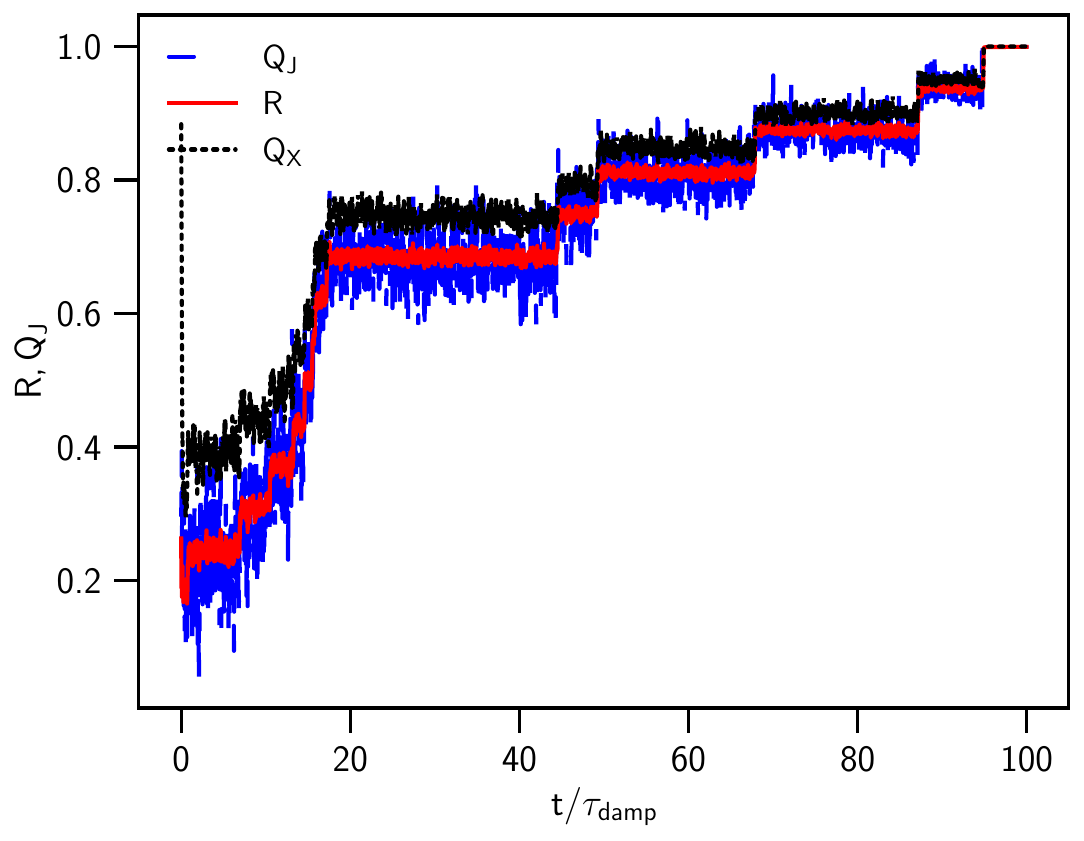}
					\put(40,20){\small \textbf{(c)} $U/(n_{1}T_{2}) = 1$}
	\end{overpic}	
	\caption{Same as Figure \ref{fig:CDR_map_qmax3_nH1e1_aef02_deltam1} but for the time-dependent alignment. Grain alignment degrees slowly increase with time and reach perfect alignment after $50-100\tau_{\rm damp}$.}
	\label{fig:CDR_RQJ_qmax3_nH1e1_aef02_deltam1}
\end{figure*}

Figures \ref{fig:CDR_map_qmax3_nH1e1_aef02_deltam1} and \ref{fig:CDR_RQJ_qmax3_nH1e1_aef02_deltam1} shows the phase trajectory alignment maps and time-dependent alignment degrees for the RAT model of $q^{\rm max}=3$ and PM grains of $\delta_{\rm mag}=1$ in the CD regime. A fraction of grains is rapidly aligned at high-J attractor, but the majority is first driven to low-J rotations, during which collisional and magnetic excitations slowly transport them to high-J attractor. The alignment degrees increase with time and reach perfect slow alignment after $50-100\tau_{\rm damp}$. Compared to Figure \ref{fig:CDR_RQJ_qmax3_nH1e1_aef02} for SPM grains, one can see that the timescale for reaching the perfect alignment is longer by a factor of 5 for the PM grains. This can be understood due to the effect of magnetic excitations that facilitate the scattering of grains out of the low-J rotation to high-J attractors.

\begin{figure*}
	 \centering
\begin{overpic}[width=0.32\linewidth]{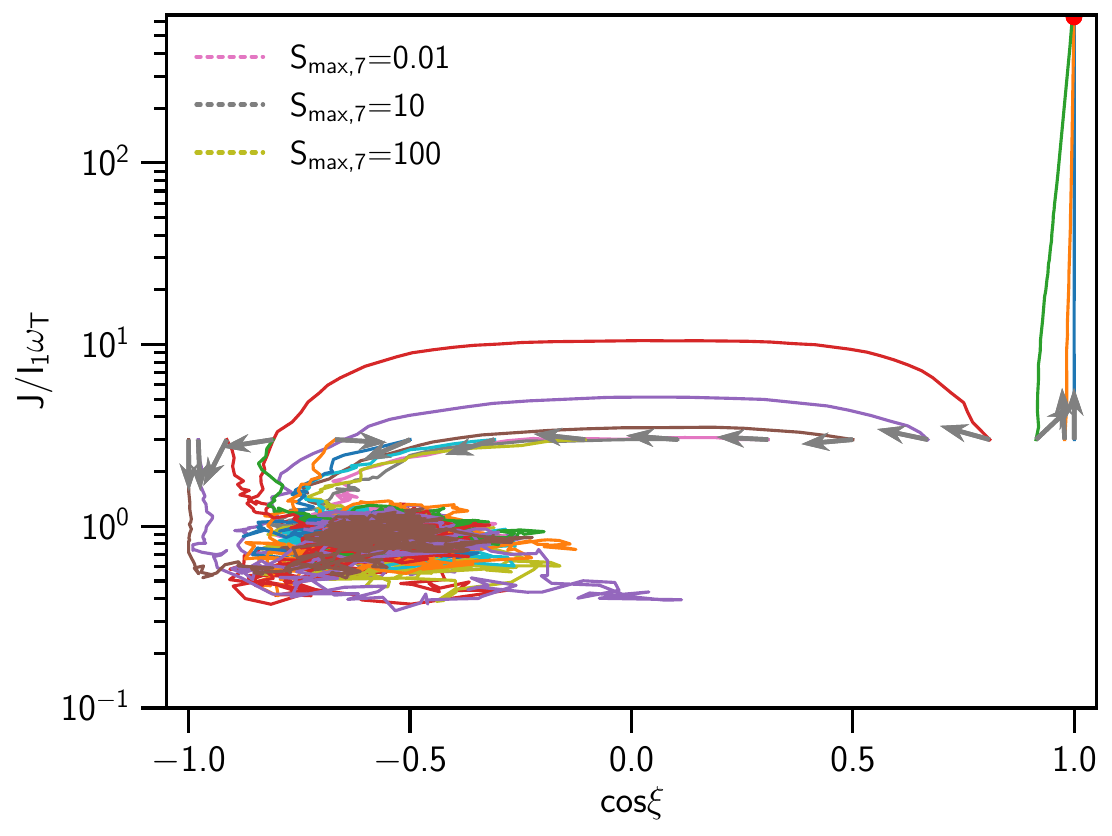}
			\put(65,75){\tiny \textbf{High-J Attractor}}
	\put(20,15){\tiny RAT Trapping}
					\put(40,65){\small \textbf{(a)} $U/(n_{1}T_{2}) = 1.9$}
	\end{overpic}	
\begin{overpic}[width=0.32\linewidth]{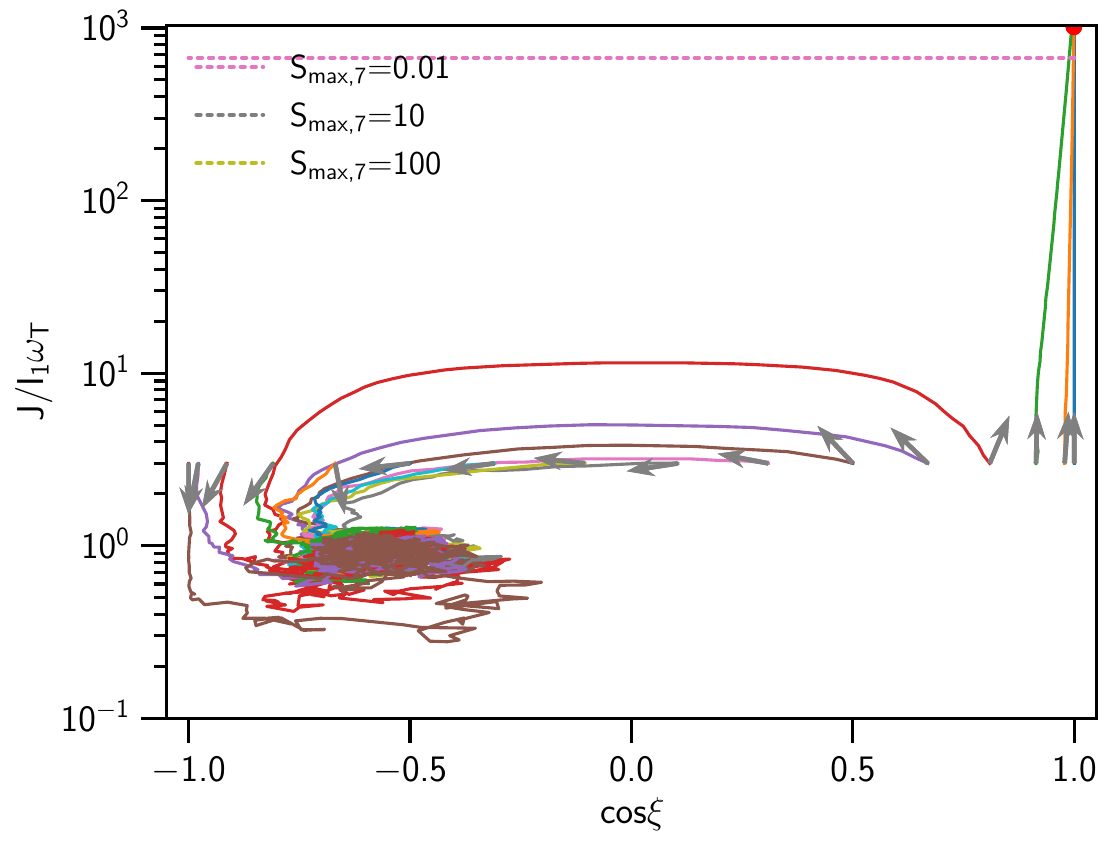}  
			\put(65,75){\tiny \textbf{High-J Attractor}}
	\put(20,15){\tiny RAT Trapping}
					\put(40,65){\small \textbf{(b)} $U/(n_{1}T_{2}) = 4.1$}
	\end{overpic}	
\begin{overpic}[width=0.32\linewidth]{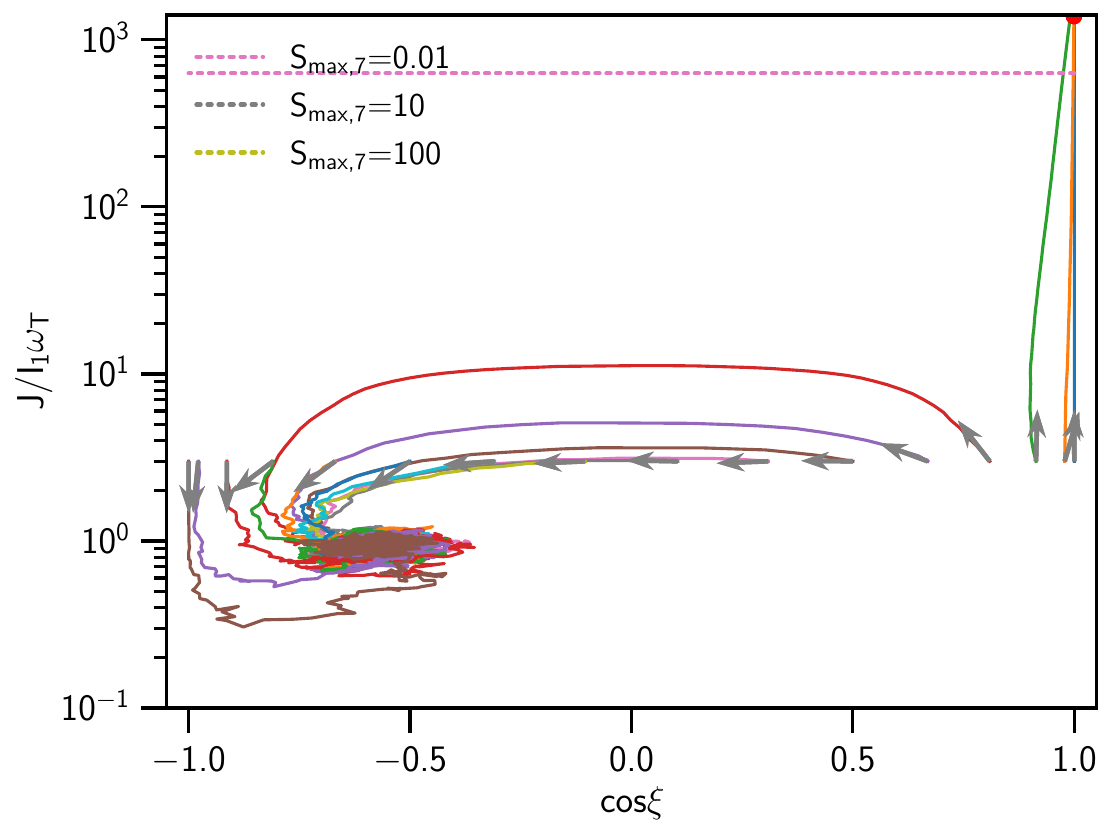}  
			\put(65,75){\tiny \textbf{High-J Attractor}}
	          				\put(40,65){\small \textbf{(c)} $U/(n_{1}T_{2}) = 7.3$}
	          				\put(20,15){\tiny RAT Trapping}
	\end{overpic}	
	
	\caption{The phase trajectory map for the RAT model of $q^{\rm max}=3$ and PM grains of $\delta_{\rm mag}=1$ in the RD regime of different $U/(n_{1}T_{2})$ (panels (a)-(c)). Some grains are rapidly spun up to the high-J attractor, but the majority of grains are driven to low-J attractors and trapped at low-J rotation. Collisional and magnetic excitations are insufficient to significantly scatter grain orientations to transport them to high-J attractors compared to the CD regime, and grains are trapped by RATs.}
	\label{fig:RDR_map_qmax3_nH1e1_aef02_deltam1}
\end{figure*}

\begin{figure*}	
	\centering
\begin{overpic}[width=0.32\linewidth]{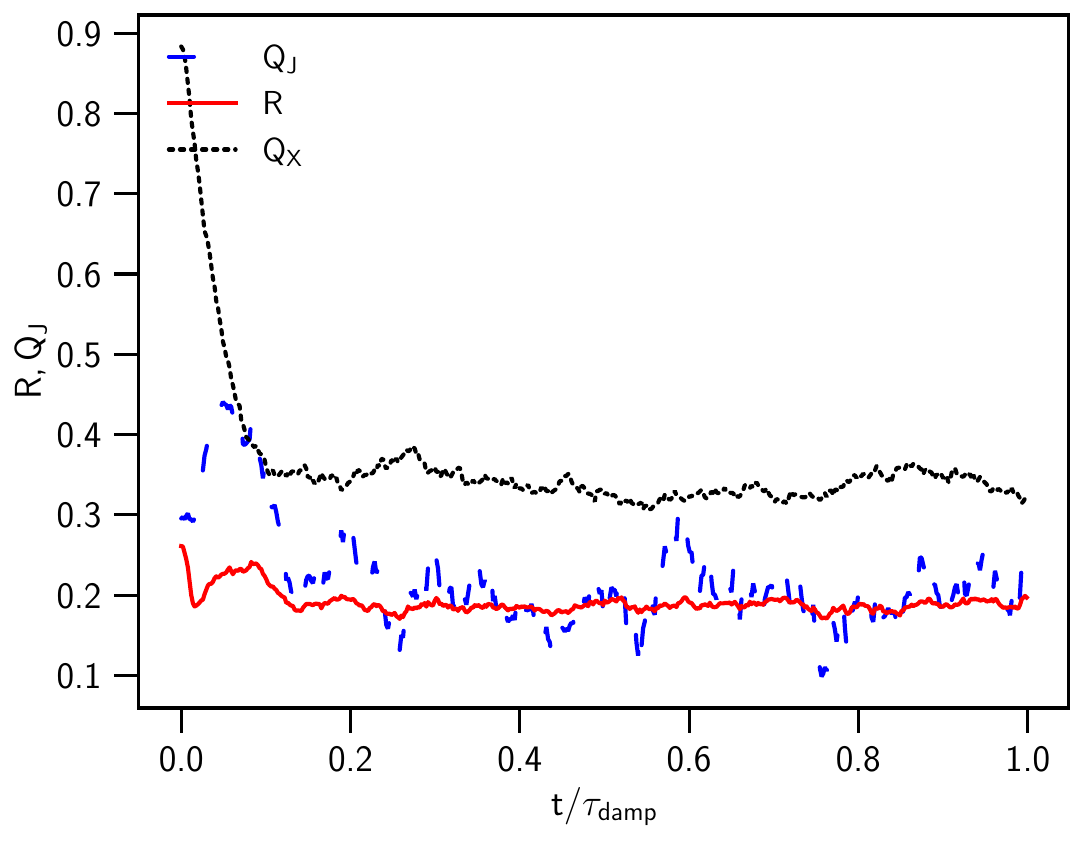}
					\put(40,65){\small \textbf{(a)} $U/(n_{1}T_{2}) = 1.9$}
	\end{overpic}	
\begin{overpic}[width=0.32\linewidth]{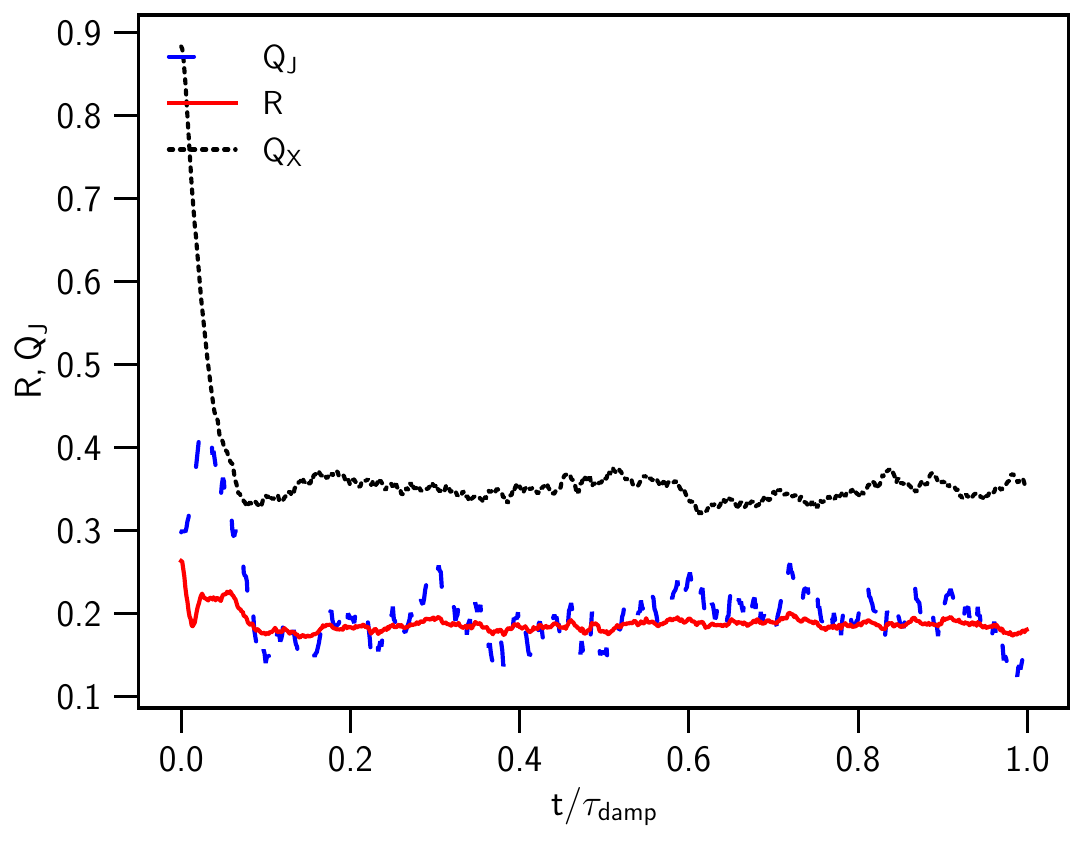}
					\put(40,65){\small \textbf{(b)} $U/(n_{1}T_{2}) = 4.1$}
	\end{overpic}	
\begin{overpic}[width=0.32\linewidth]{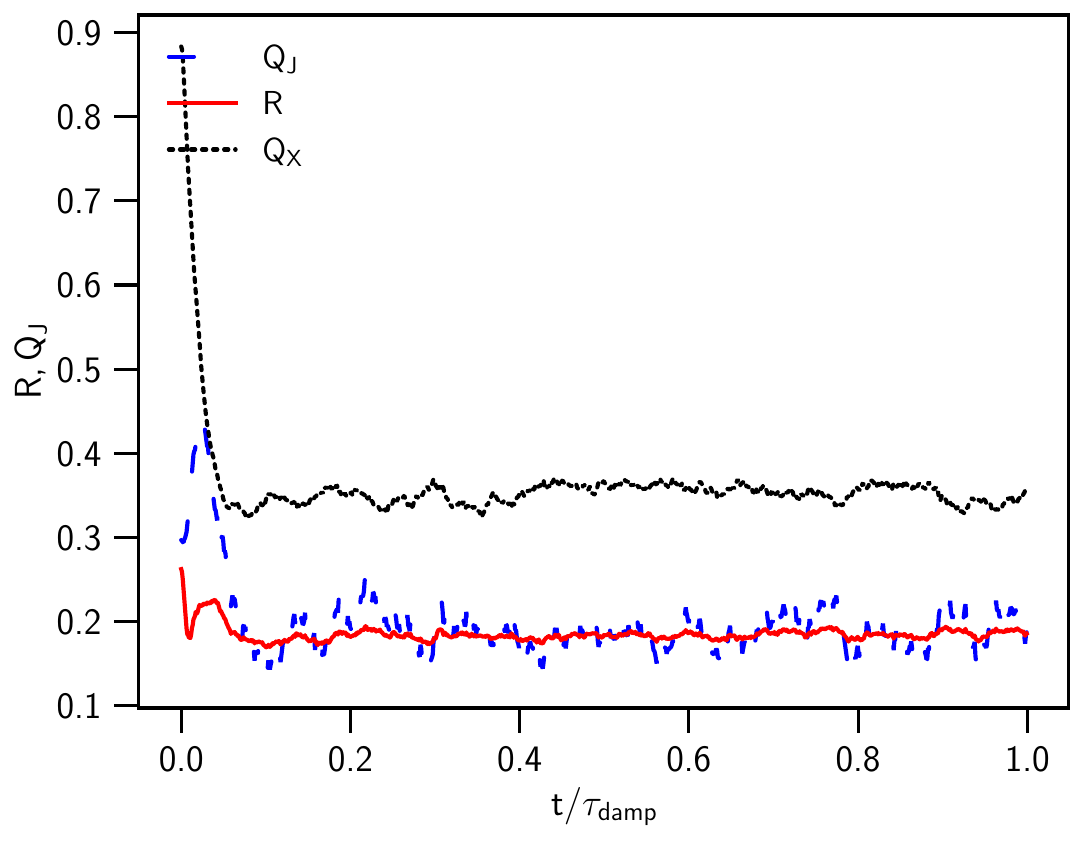}
					\put(40,65){\small \textbf{(c)} $U/(n_{1}T_{2}) = 7.3$}
	\end{overpic}	
	\caption{Same as Figure \ref{fig:RDR_map_qmax3_nH1e1_aef02_deltam1} but for time-dependent alignment. Alignment degree increases rapidly within $0.01\tau_{\rm damp}$ and becomes saturated at $R\sim 0.2$ for intense radiation due to RAT trapping.}
	\label{fig:RDR_RQJ_qmax3_nH1e1_aef02_deltam1}
\end{figure*}

Figure \ref{fig:RDR_map_qmax3_nH1e1_aef02_deltam1} shows the phase trajectory map for the RAT model of $q^{\rm max}=3$ and PM grains of $\delta_{\rm mag}=1$ for the RD regime. Collisional and magnetic excitations are insufficient to significantly scatter grain orientations to transport them to high-J attractors compared to the CD regime. Figure \ref{fig:RDR_RQJ_qmax3_nH1e1_aef02_deltam1} shows the time-dependent alignment degrees. Grain alignment is caused by grains with fast alignment at high-J attractors and is saturated due to RAT trapping of grains at low-J rotation.

\begin{figure*}
	    \centering
\begin{overpic}[width=0.32\linewidth]{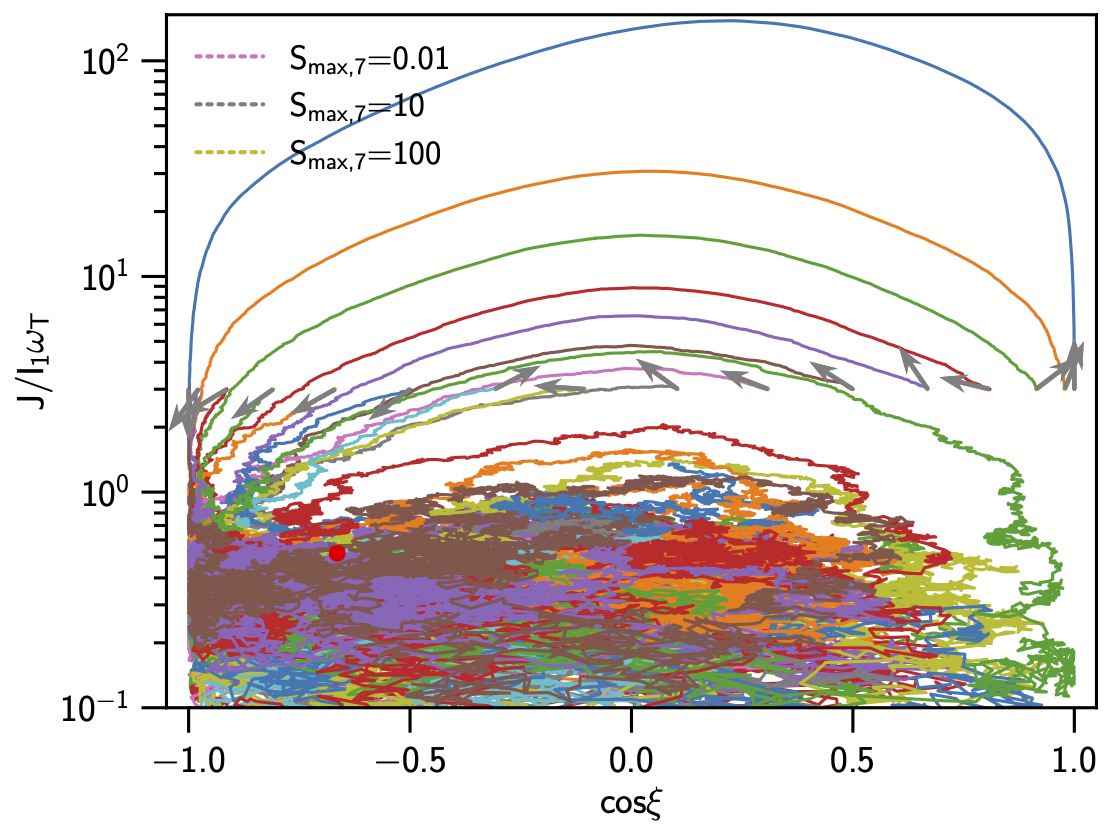}
					\put(40,65){\small \textbf{(a)} $U/(n_{1}T_{2}) = 1.9$}
	\end{overpic}	
\begin{overpic}[width=0.32\linewidth]{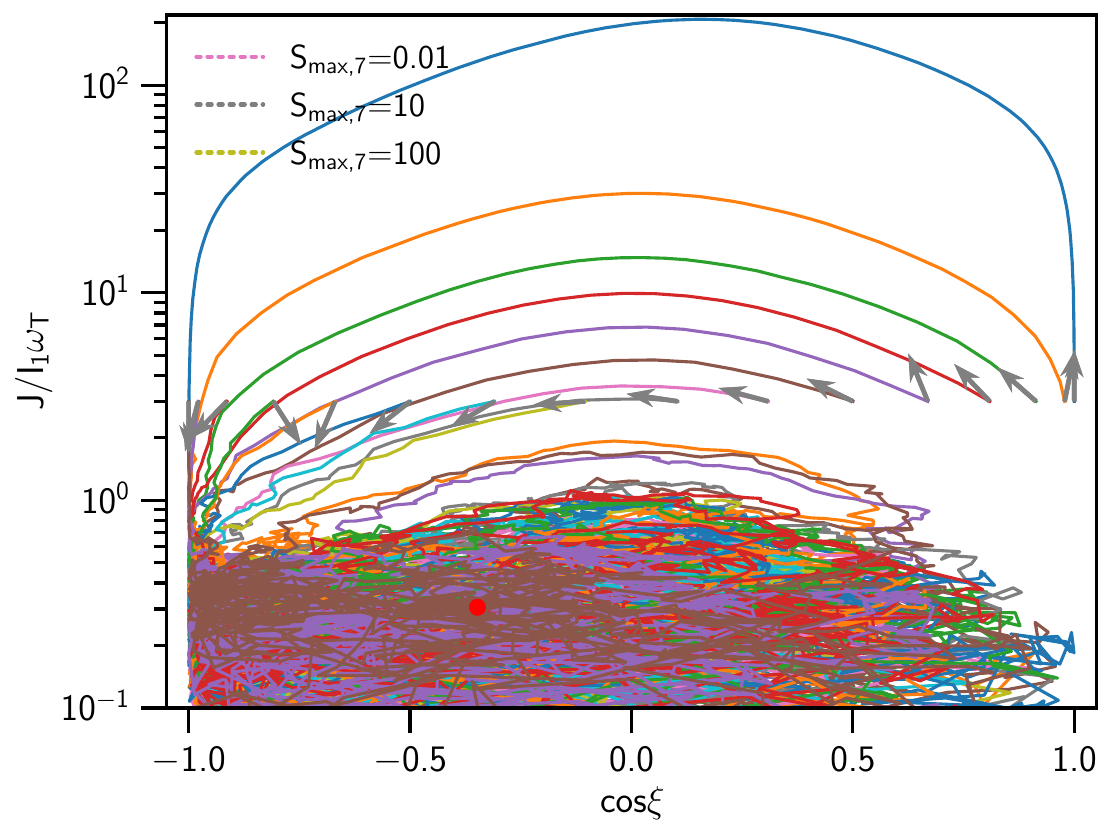}        
				\put(40,65){\small \textbf{(b)} $U/(n_{1}T_{2}) = 4.1$}
\end{overpic}	
\begin{overpic}[width=0.32\linewidth]{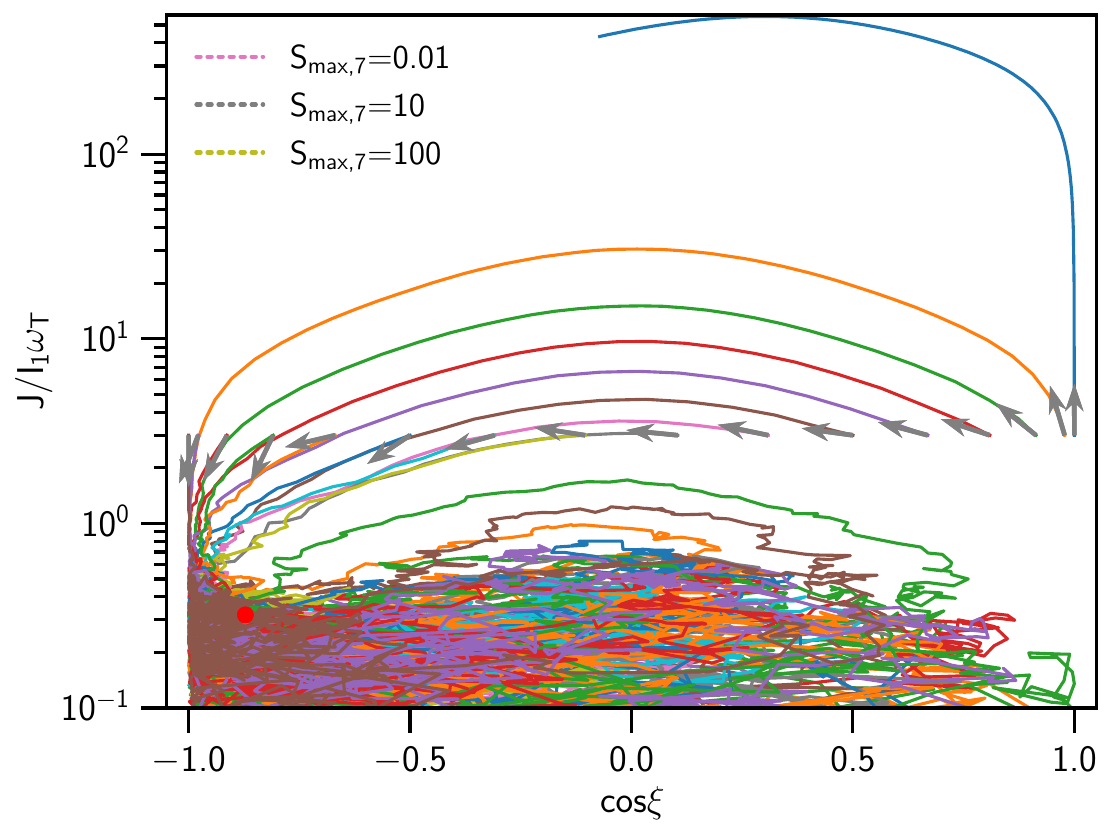}  
					\put(40,65){\small \textbf{(c)} $U/(n_{1}T_{2}) = 7.3$}
	\end{overpic}	
	\caption{Same as Figure \ref{fig:RDR_map_qmax3_nH1e1_aef02_deltam1} but for $q^{\rm max}=1$. Grains have no high-J attractors and undergo a random trajectory at thermal rotation and deterministic trajectories at suprathermal rotation of $J/I_{1}\omega_{T}>3$.}
	\label{fig:RDR_map_qmax1_nH1e1_aef02_deltam1}
\end{figure*}

\begin{figure*}	
	\centering
\begin{overpic}[width=0.32\linewidth]{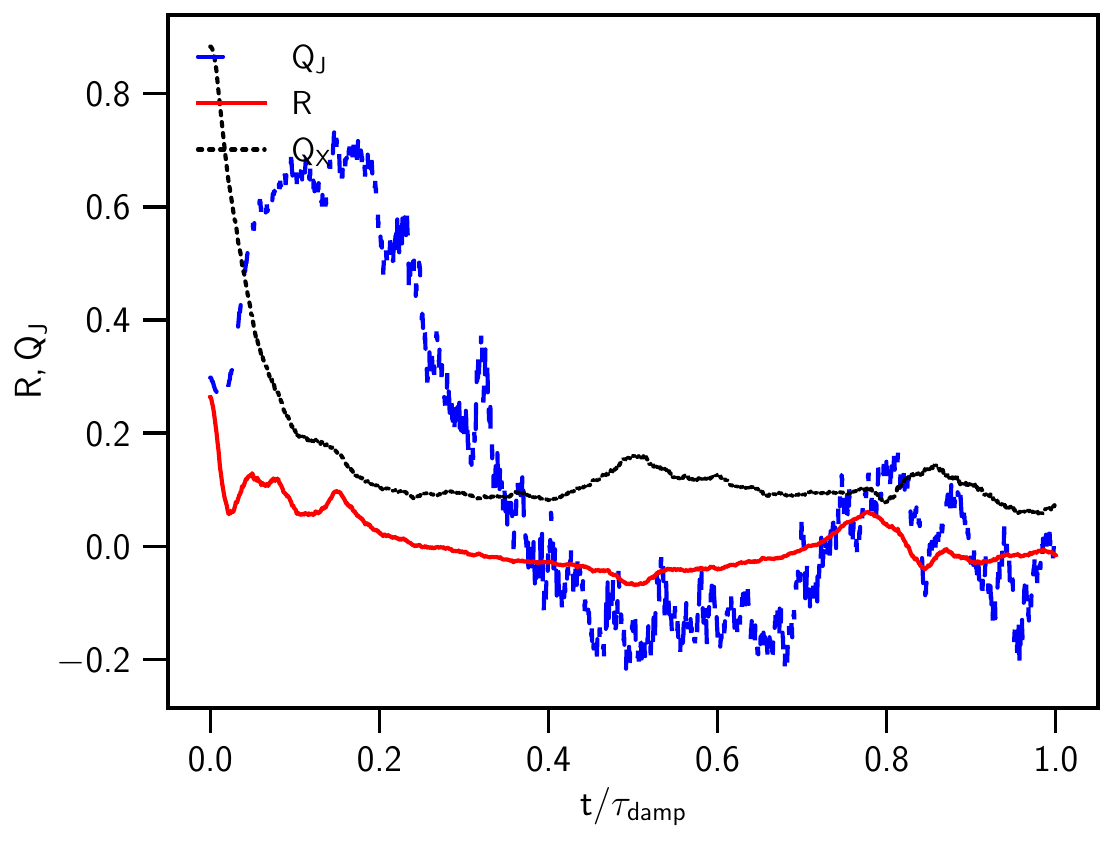}
					\put(40,65){\small \textbf{(a)} $U/(n_{1}T_{2}) = 1.9$}
	\end{overpic}	
\begin{overpic}[width=0.32\linewidth]{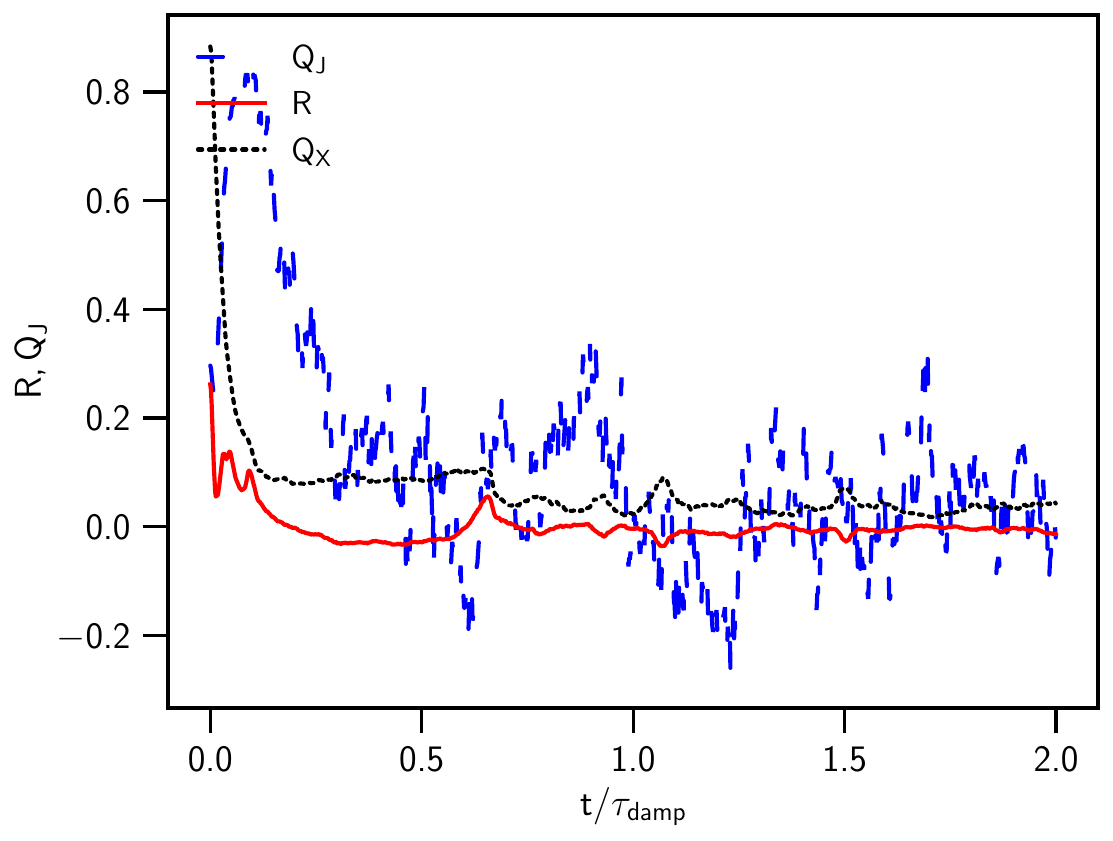}
					\put(40,65){\small \textbf{(b)} $U/(n_{1}T_{2}) = 4.1$}
	\end{overpic}	
\begin{overpic}[width=0.32\linewidth]{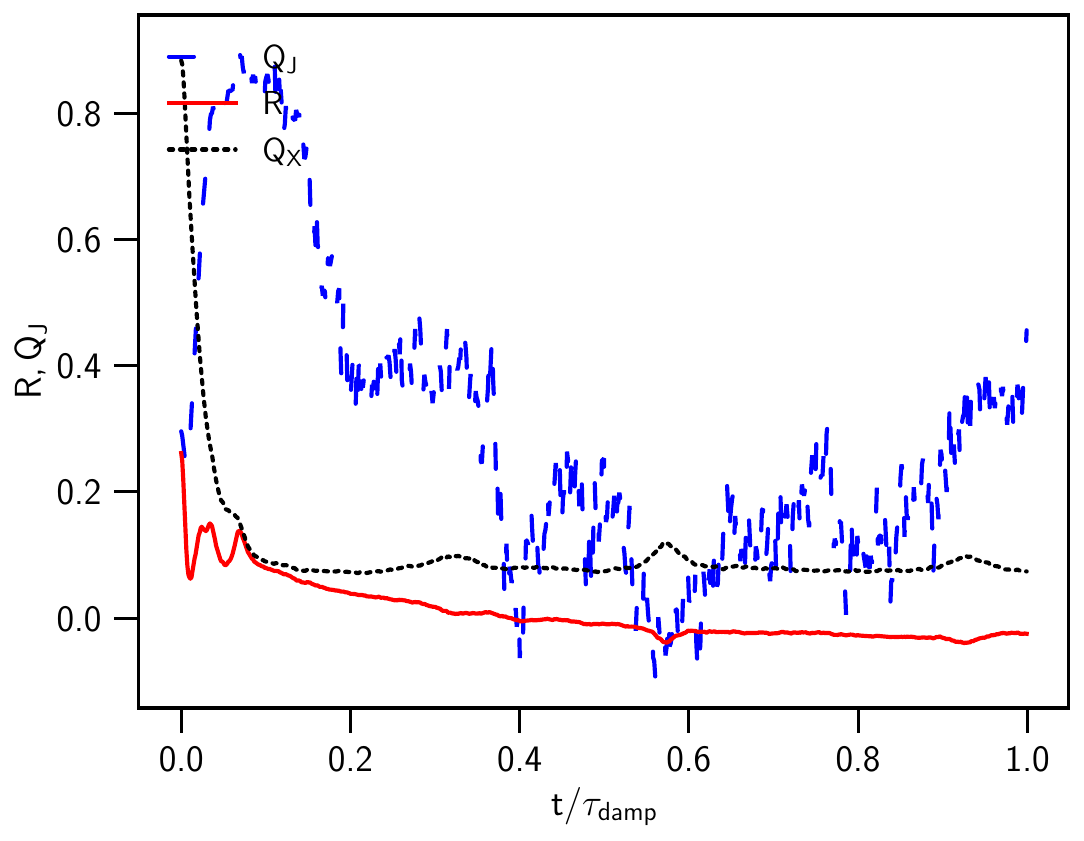}
					\put(40,65){\small \textbf{(c)} $U/(n_{1}T_{2}) = 7.3$}
	\end{overpic}	
	\caption{Same as Figure \ref{fig:RDR_map_qmax1_nH1e1_aef02_deltam1} but for time-dependent alignment. Grains have negligible alignment due to the lack of a high-J attractor.}
	\label{fig:RDR_RQJ_qmax1_nH1e1_aef02_deltam1}
\end{figure*}

Figures \ref{fig:RDR_map_qmax1_nH1e1_aef02_deltam1} and \ref{fig:RDR_RQJ_qmax1_nH1e1_aef02_deltam1} show the phase alignment maps and time-dependent alignment degrees for the RAT model of $q^{\rm max}=1$ and PM grains in the RD regime. Grain alignment is negligible due to the lack of high-J attractors.

\subsection{SPM Grains in Extreme Radiation Fields}
\begin{figure*}
	    \centering
\begin{overpic}[width=0.42\linewidth]{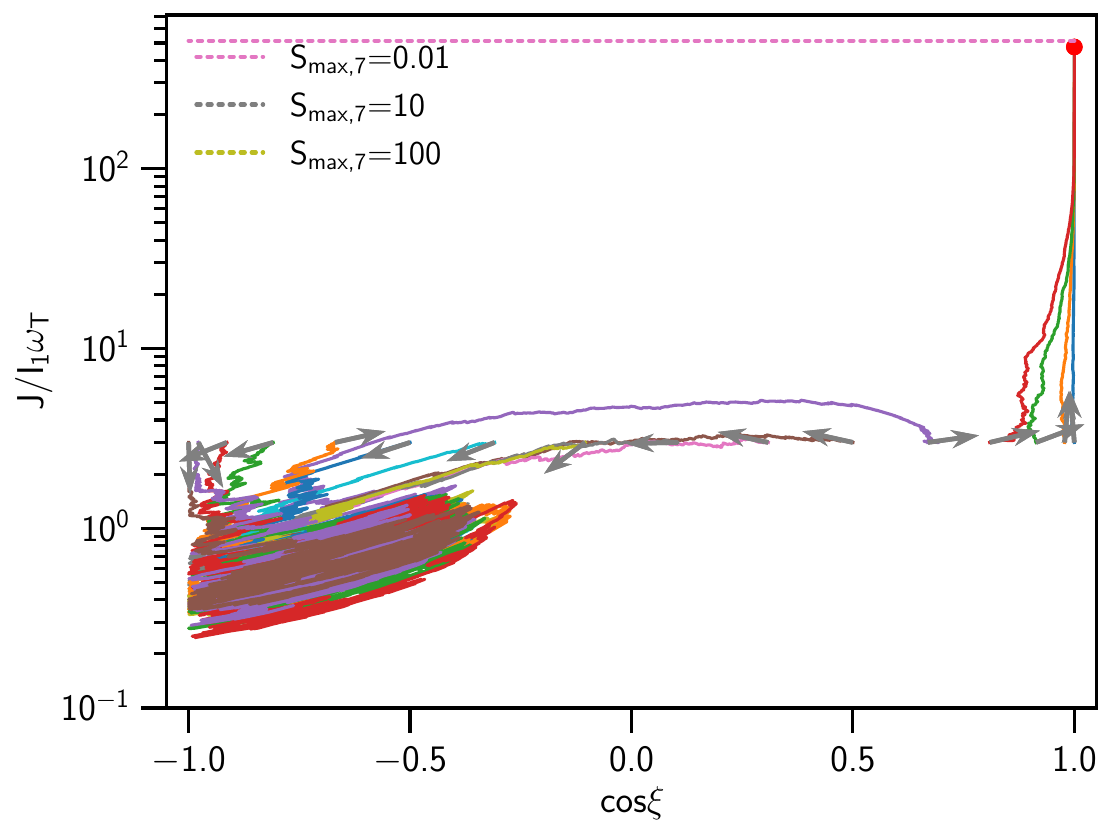}
			\put(72,75){\tiny \textbf{High-J Attractor}}
	\put(20,14){\tiny \textbf{RAT Trapping}}
					\put(40,20){\small \textbf{(a)} $U/(n_{1}T_{2}) = 49.5$}
	\end{overpic}	
\begin{overpic}[width=0.42\linewidth]{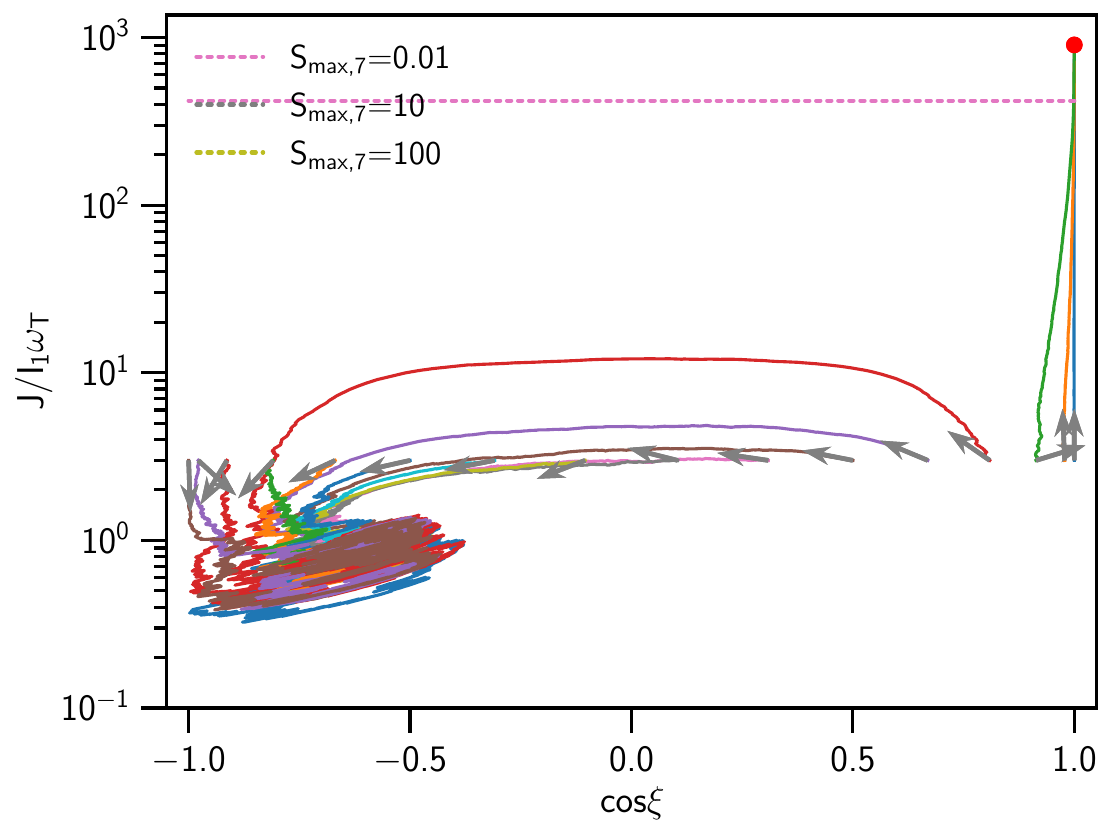}
			\put(72,75){\tiny \textbf{High-J Attractor}}
	\put(20,14){\tiny \textbf{RAT Trapping}}
					\put(40,20){\small \textbf{(b)} $U/(n_{1}T_{2}) = 337$}
	\end{overpic}	
\begin{overpic}[width=0.42\linewidth]{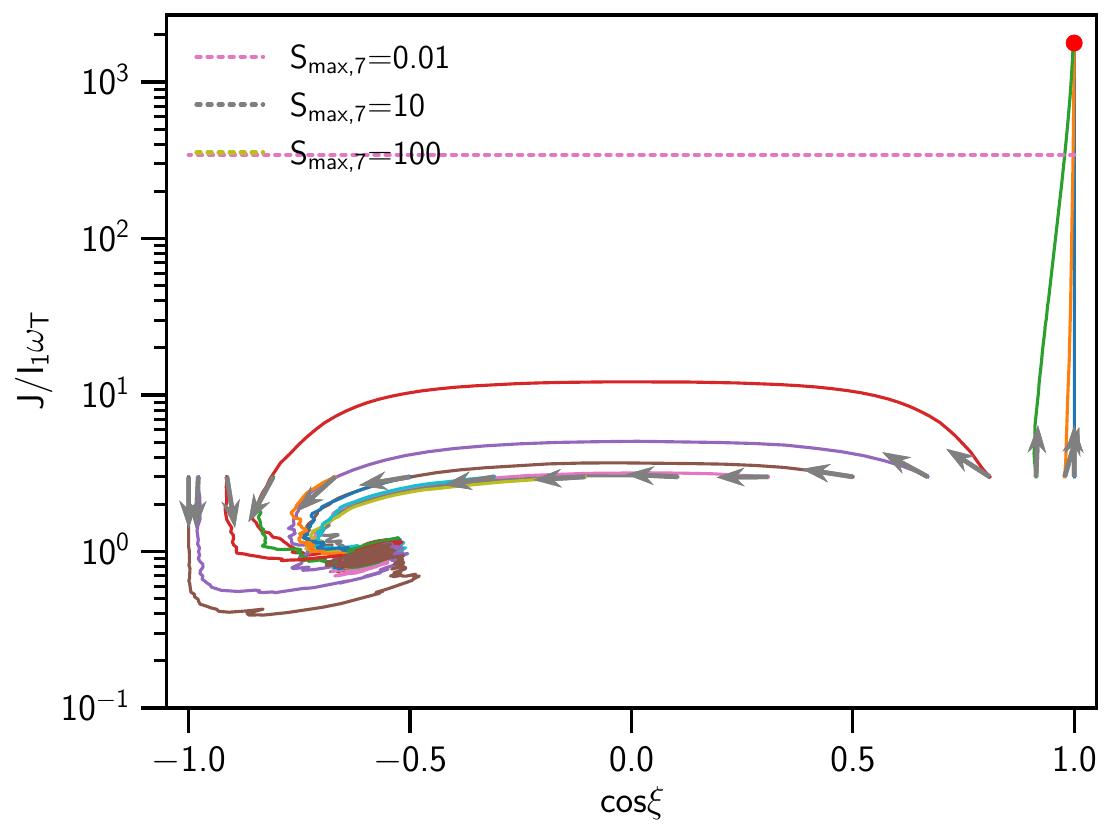}    
			\put(72,75){\tiny \textbf{High-J Attractor}}
	  \put(20,14){\tiny \textbf{RAT Trapping}}
				\put(40,20){\small \textbf{(c)} $U/(n_{1}T_{2}) = 2296$}
\end{overpic}	
\begin{overpic}[width=0.42\linewidth]{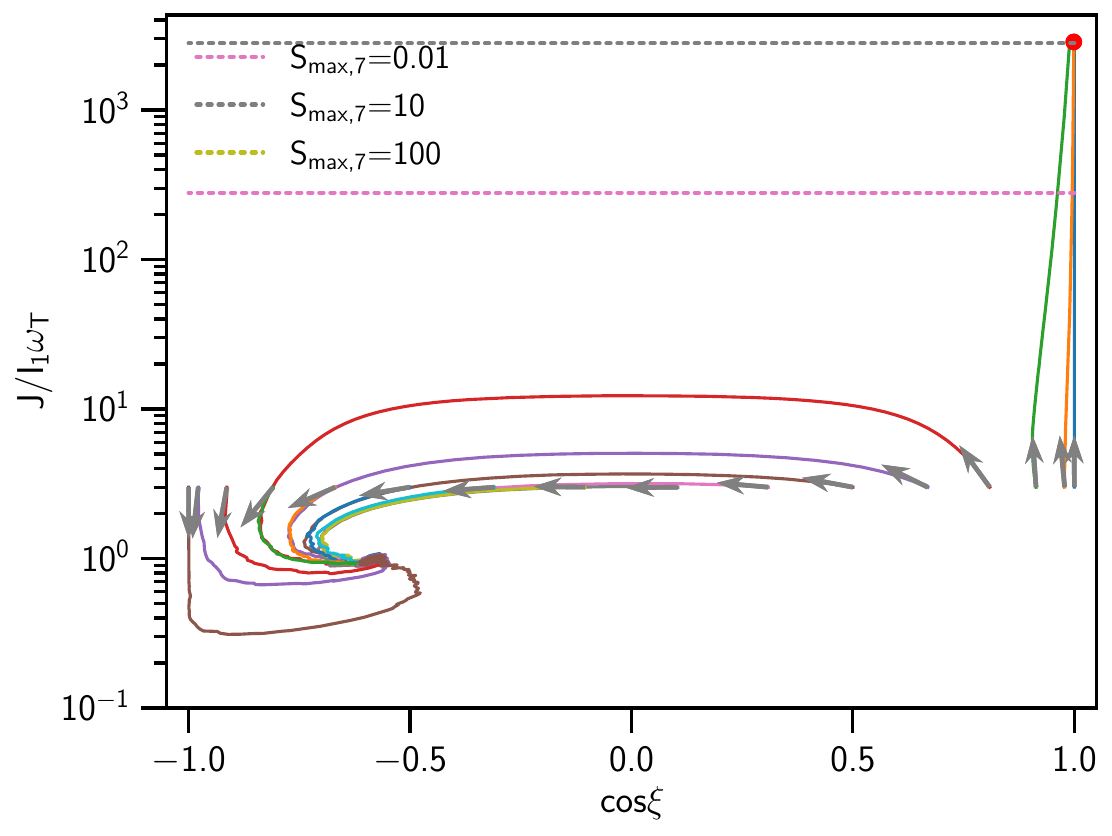}  
			\put(72,75){\tiny \textbf{High-J Attractor}}
	\put(20,14){\tiny \textbf{RAT Trapping}}
					\put(40,20){\small \textbf{(d)} $U/(n_{1}T_{2}) = 15640$}
	\end{overpic}	
	
	\caption{Phase trajectory maps for the RAT model of $q^{\rm max}=3$ and SPM grains of $\delta_{\rm mag}=10^{3}$ in extreme radiation fields (panels (a)-(d)). The fraction of grains that have fast alignment at high-J attractor decreases with increasing $U/(n_{1}T_{2})$ due to reduced efficiency of magnetic relaxation.}
	\label{fig:extreme_map_qmax3}
\end{figure*}

\begin{figure*}	
	\centering
\begin{overpic}[width=0.42\linewidth]{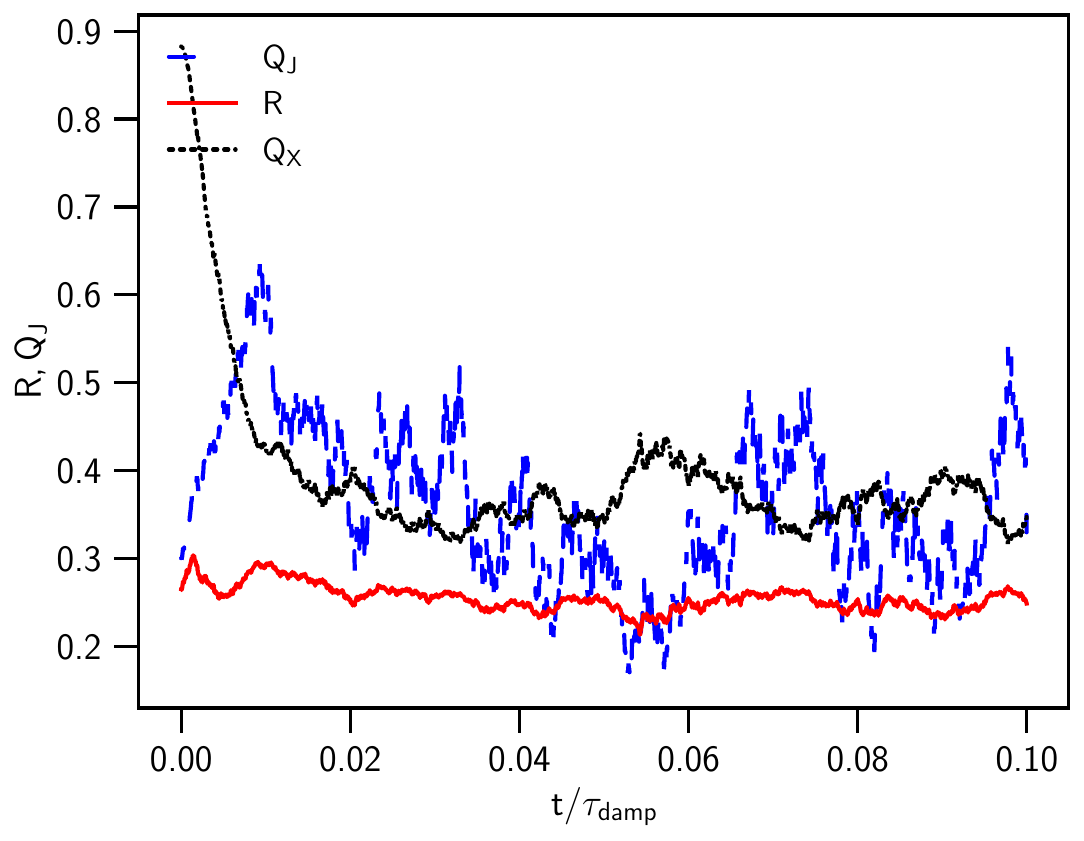}
					\put(45,65){\small \textbf{(a)} $U/(n_{1}T_{2}) =  49.5$}
	\end{overpic}	
\begin{overpic}[width=0.42\linewidth]{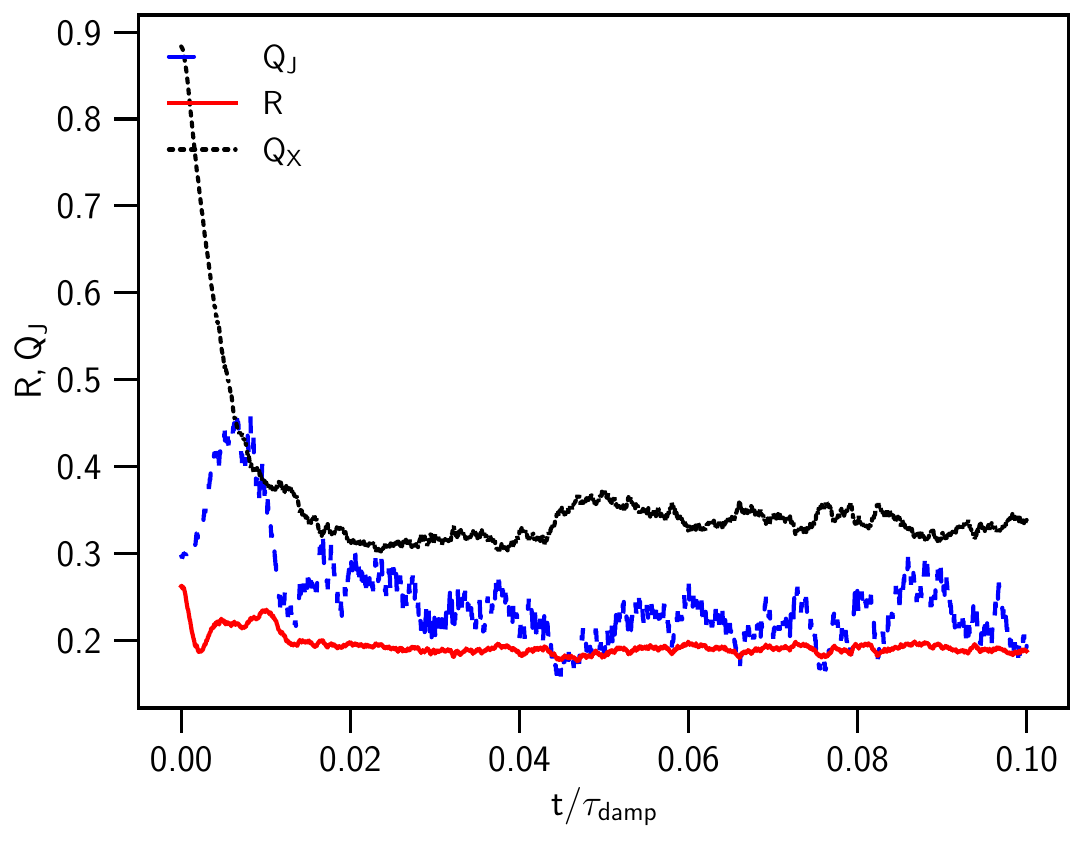}
				\put(45,65){\small \textbf{(b)} $U/(n_{1}T_{2}) =  337$}
\end{overpic}	
\begin{overpic}[width=0.42\linewidth]{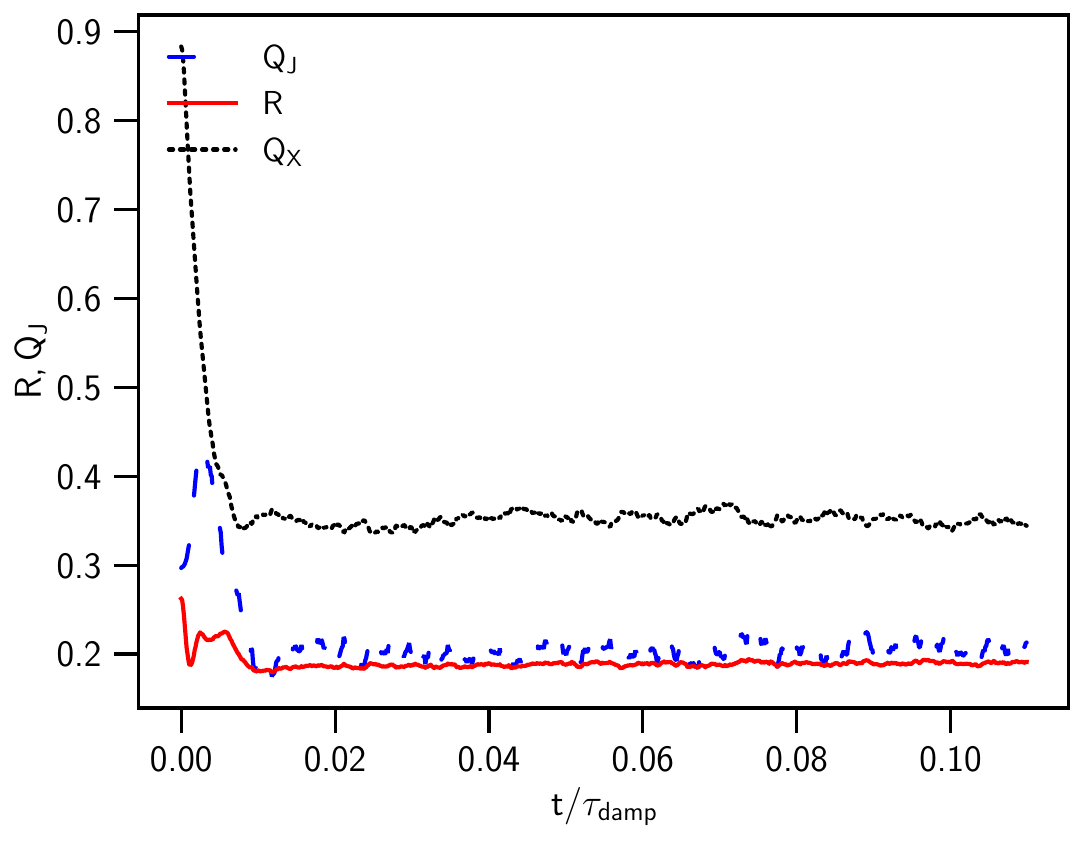}
					\put(45,65){\small \textbf{(c)} $U/(n_{1}T_{2}) =  2296$}
	\end{overpic}	
\begin{overpic}[width=0.42\linewidth]{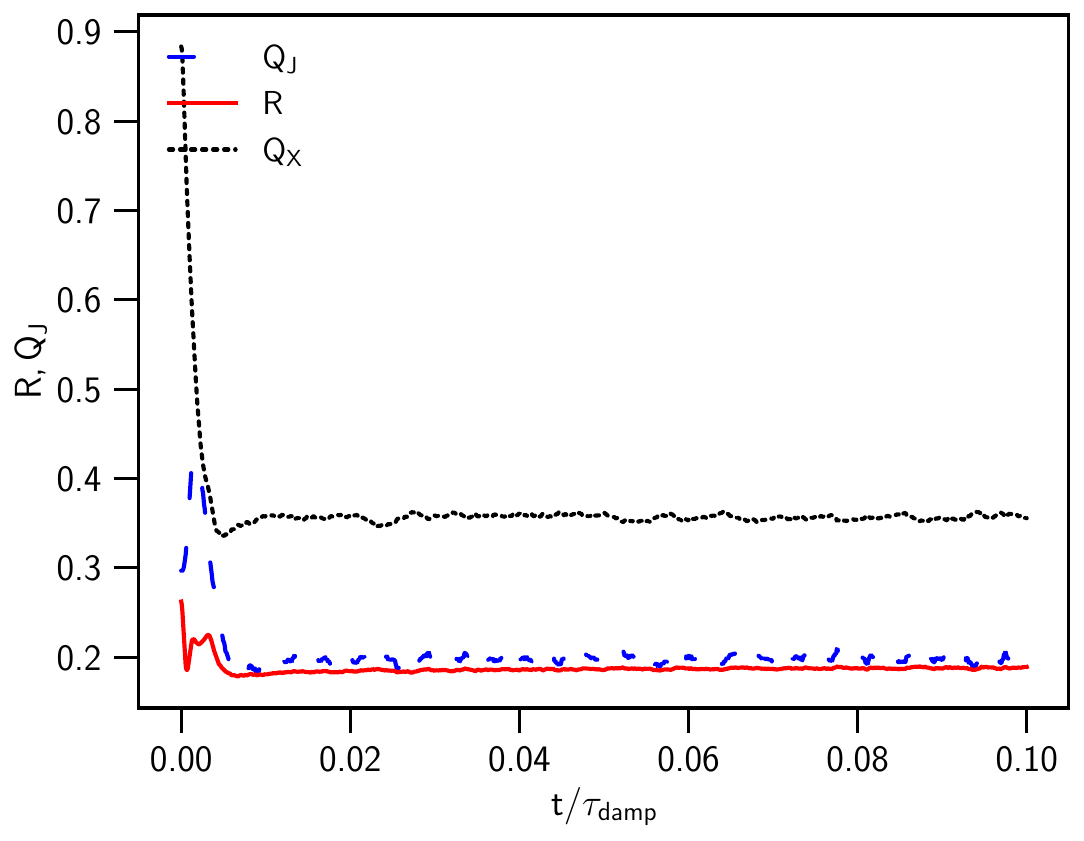}
					\put(45,65){\small \textbf{(d)} $U/(n_{1}T_{2}) =  15640$}
	\end{overpic}	
	\caption{Same as Figure \ref{fig:extreme_map_qmax3} but for the time-dependent grain alignment degrees. The net grain alignment degree ($R$) decreases with increasing $U/(n_{1}T_{2})$ due to reduced efficiency of magnetic relaxation.}
	\label{fig:extreme_RQJ_qmax3}
\end{figure*}

In Figure \ref{fig:extreme_map_qmax3}, we show the results for SPM grains using the RAT model of  $q^{\rm max}=3$ in extreme radiation fields.
A fraction of grains is fast aligned at high-J attractors, whereas the rest is driven to low-J rotation and trapped there due to RAT trapping. The degrees of alignment shown in Figure \ref{fig:extreme_RQJ_qmax3}) are less than $30\%$ and become more stable for stronger radiation fields after a short initial period due to radiative trapping.

\section{Radiative Torque Disruption Sizes and Timescales}\label{apdx:RATD}

If the grain is being spun-up to the high-J attractor point by the radiation source with stable luminosity, the radiative torque $\Gamma_{\rm RAT}$ is constant, and the grain velocity is steadily increased over time. The maximum angular velocity of grains spun-up by RATs is given by
\bea
\Omega_{\rm RAT}=\frac{\Gamma_{\rm RAT}\tau_{\rm damp}}{I_{1}},~~~~~\label{eq:omega_RAT0}
\ena
where $\Gamma_{\rm RAT}$ is given by Equation (\ref{eq:GammaRAT}). The situation is different for grains being moved to the low-J attractor point. The angular momentum of such grains decreases. 

For a general radiation field, the maximum rotation rate induced by RATs is given by
\bea
\Omega_{\rm RAT}&\simeq & 9.22\times 10^{7}\gamma_{-1} a_{\eff,-5}^{0.7}\bar{\lambda}_{0.5}^{-1.7}
\left(\frac{U}{n_{1}T_{2}^{1/2}}\right) \left(\frac{1}{1+F_{\rm IR}}\right)\rad\s^{-1},\label{eq:omega_RAT}
\ena
for grains with $a_{\eff}\lesssim a_{\rm trans}$, and
\bea
\omega_{\rm RAT}\simeq 1.42\times 10^{9}\gamma_{-1}a_{\eff,-5}^{-2}\bar{\lambda}_{0.5}
\left(\frac{U}{n_{1}T_{2}^{1/2}}\right) \left(\frac{1}{1+F_{\rm IR}}\right)\rad\s^{-1},
\ena
for grains with $a_{\rm eff}> a_{\rm trans}$, where $\gamma_{-1}=\gamma/0.1$.

The minimum size for the RATD is obtained by setting the maximum angular velocity spun-up by RATs, $\Omega_{\rm RAT}$ (Eq. \ref{eq:omega_RAT}) to $\Omega_{\rm disr}$, which reads \citep{Hoang.2019nas,Hoang.2019}:
\bea
a_{\rm disr}\simeq 0.22\bar{\lambda}_{0.5}S_{\max,7}^{1/3.4}
(1+F_{\rm IR})^{1/1.7}\left(\frac{n_{1}T_{2}^{1/2}}{\gamma_{-1}U}\right)^{1/1.7}\mum,~~~\label{eq:adisr}
\ena
where $\bar{\lambda}_{0.5}=\bar{\lambda}/0.5\mum$ and $\gamma_{-1}=\gamma/0.1$. 

For strong radiation fields of $F_{\rm IR}>1$, the characteristic timescale for fast disruption of grains is defined as \citep{Hoang.2019nas,Hoang.2019}
\bea
t_{\rm disr}^{\rm fast}&&=\frac{I_{\|}\Omega_{\rm disr}}{\Gamma_{\rm RAT}}=\left(\frac{\Omega_{\rm disr}}{\Omega_{\rm RAT}}\right)\tau_{\rm damp},\nonumber\\
&& \simeq 10^{5} (\gamma U)^{-1}\bar{\lambda}_{0.5}^{1.7}\hat{\rho}^{1/2}S_{\max,7}^{1/2}a_{\eff,-5}^{-0.7} \yr~~~~\label{eq:tdisr1}
\ena
for $a_{\rm disr}<a_{\eff} \lesssim a_{\rm trans}$, and
\bea
t_{\rm disr}^{\rm fast}\simeq 7.4(\gamma U)^{-1}\bar{\lambda}_{0.5}^{-1}\hat{\rho}^{1/2}S_{\max,7}^{1/2}a_{\eff,-5}^{2}{~\yr}\label{eq:tdisr2}
\label{eq:tmin}
\ena
for $a_{\rm trans}<a_{\eff}<a_{\rm disr,max}$, where $a_{\rm trans}=\bar{\lambda}/2.5$ is the size at the transition between the flat and steep power-law of RAT efficiency  (\citealt{Hoang.2019nas}). 


\bibliography{ms.bbl}

\end{document}